\documentclass[11pt]{article}
\usepackage{jheppub}
\usepackage{mathtools,amssymb,amsthm}
\usepackage[utf8]{inputenc}
\usepackage{enumitem}
\usepackage{amsmath}
\usepackage{amssymb}
\usepackage[all]{xy}
\usepackage{xcolor}
\usepackage{simpler-wick}
\usepackage{physics}
\usepackage{tikz}
\usetikzlibrary{decorations.markings,arrows.meta}

\usepackage{babel}

\begin{document}
\global\long\def\aad{(a\tilde{a}+a^{\dagger}\tilde{a}^{\dagger})}%

\newcommand{\mbc}{\mathbb{C}}
\newcommand{\lam}{\lambda}
\newcommand{\Li}{\mathrm{Li}}
\newcommand{\ads}{\text{AdS}}
\newcommand{\mh}{\mathcal{H}}
\newcommand{\mbp}{\mathbb{P}}
\newcommand{\pd}{\partial}
\newcommand{\mc}{\mathcal{C}}
\newcommand{\me}{\mathcal{E}}

\global\long\def\ad{{\rm ad}}%

\global\long\def\bij{\langle ij\rangle}%

\global\long\def\df{\coloneqq}%

\global\long\def\bs{b_{\alpha}^{*}}%

\def\XD#1{{\color{blue}{#1}}}
\def\XDc#1{{\color{blue}{[XD: #1]}}}
\def\jx#1{{\color{purple}{#1}}}
\def\JX#1{{\color{purple}{[JX:#1]}}}

\global\long\def\dd{{\rm d}}%

\global\long\def\dg{{\rm {\rm \dot{\gamma}}}}%

\global\long\def\ddt{\frac{{\rm d^{2}}}{{\rm d}t^{2}}}%

\global\long\def\ddg{\nabla_{\dot{\gamma}}}%

\global\long\def\del{\mathcal{\delta}}%

\global\long\def\Del{\Delta}%

\global\long\def\dtau{\frac{\dd^{2}}{\dd\tau^{2}}}%

\global\long\def\ul{U(\Lambda)}%

\global\long\def\udl{U^{\dagger}(\Lambda)}%

\global\long\def\dl{D(\Lambda)}%

\global\long\def\da{\dagger}%

\global\long\def\id{{\rm id}}%

\global\long\def\ml{\mathcal{L}}%

\global\long\def\mm{\mathcal{\mathcal{M}}}%

\global\long\def\mf{\mathcal{\mathcal{F}}}%

\global\long\def\kpp{k^{\prime}}%

\global\long\def\lr{\leftrightarrow}%

\global\long\def\lf{\leftrightarrow}%

\global\long\def\ma{\mathcal{A}}%

\global\long\def\mb{\mathcal{B}}%

\global\long\def\md{\mathcal{D}}%

\global\long\def\mbr{\mathbb{R}}%

\global\long\def\mbz{\mathbb{Z}}%

\global\long\def\mh{\mathcal{\mathcal{H}}}%

\global\long\def\mi{\mathcal{\mathcal{I}}}%

\global\long\def\ms{\mathcal{\mathcal{\mathcal{S}}}}%

\global\long\def\mg{\mathcal{\mathcal{G}}}%

\global\long\def\mfa{\mathcal{\mathfrak{a}}}%

\global\long\def\mfb{\mathcal{\mathfrak{b}}}%

\global\long\def\mfb{\mathcal{\mathfrak{b}}}%

\global\long\def\mfg{\mathcal{\mathfrak{g}}}%

\global\long\def\mj{\mathcal{\mathcal{J}}}%

\global\long\def\mk{\mathcal{K}}%

\global\long\def\mmp{\mathcal{\mathcal{P}}}%

\global\long\def\mn{\mathcal{\mathcal{\mathcal{N}}}}%

\global\long\def\mq{\mathcal{\mathcal{Q}}}%

\global\long\def\mo{\mathcal{O}}%

\global\long\def\qq{\mathcal{\mathcal{\mathcal{\quad}}}}%

\global\long\def\ww{\wedge}%

\global\long\def\ka{\kappa}%

\global\long\def\nn{\nabla}%

\global\long\def\nb{\overline{\nabla}}%

\global\long\def\pathint{\langle x_{f},t_{f}|x_{i},t_{i}\rangle}%

\global\long\def\ppp{p^{\prime}}%

\global\long\def\qpp{q^{\prime}}%

\global\long\def\we{\wedge}%

\global\long\def\pp{\prime}%

\global\long\def\sq{\square}%

\global\long\def\vp{\varphi}%

\global\long\def\ti{\widetilde{}}%

\global\long\def\wg{\widetilde{g}}%

\global\long\def\te{\theta}%

\global\long\def\tr{{\rm tr}}%

\global\long\def\ta{{\rm \widetilde{\alpha}}}%

\global\long\def\sh{{\rm {\rm sh}}}%

\global\long\def\ch{{\rm ch}}%

\global\long\def\Si{{\rm {\rm \Sigma}}}%

\global\long\def\si{{\rm {\rm \sigma}}}%

\global\long\def\sch{{\rm {\rm Sch}}}%

\global\long\def\vol{{\rm {\rm {\rm Vol}}}}%

\global\long\def\reg{{\rm {\rm reg}}}%

\newcommand{\psid}{\Psi^{\Delta}_{\beta_L,\beta_R}}
\newcommand{\la}{\langle}
\newcommand{\ra}{\rangle}
\newcommand{\zb}{\bar{z}}
\newcommand{\be}
{\begin{equation}}
\newcommand{\mt}{\mathcal{T}}
\newcommand{\nt}{n_{{\rm tot}}}

\newcommand{\ee}{\end{equation}}
\newcommand{\tx}{\text}
\newcommand{\tb}{\textbf}

\title{A Semiclassical Entanglement Wedge in Double-Scaled SYK}

\author[a]{Xi Dong,}
\emailAdd{xidong@ucsb.edu}

\author[b]{Jiuci Xu}
\emailAdd{jiucixu@mit.edu}

\affiliation[a]{Department of Physics, University of California, Santa Barbara, CA 93106, USA}
\affiliation[b]{Center for Theoretical Physics - a Leinweber Institute,  \\ Massachusetts Institute of Technology, Cambridge, MA 02139, USA}

\abstract{We study the entanglement structure of partially entangled thermal states in
double-scaled SYK , allowing flavor degrees of freedom carried by matter insertions
to be in an arbitrary mixed state. In the semiclassical limit, followed by the heavy-probe
limit, our replica calculation gives a
Faulkner--Lewkowycz--Maldacena formula: the boundary entropy decomposes into an
`area' term (independent of the flavor state) and the bulk matter entropy in a corresponding entanglement wedge. Varying the Euclidean locations of the
operator insertions produces sharp transitions in the boundary entropy, which
we interpret as individual matter excitations crossing the boundary of the
wedge and hence as evidence for a sharply defined entanglement wedge. To derive this result, we
analyze the semiclassical behavior of general Euclidean $2n$-point functions and
give a probabilistic proof of  their factorization into sums over products of two-point functions.
We also give a complementary derivation from exact spectral representations,
using the semiclassical behavior of the quantum $6j$-symbol.
}

\maketitle

\section{Introduction}
A useful diagnostic of whether a microscopic quantum system admits an emergent
bulk geometry is whether its entanglement entropy exhibits the structure of an
entanglement wedge. In conventional holographic systems, this structure is
encoded in the RT/HRT \cite{Ryu_2006,Hubeny_2007} and Faulkner--Lewkowycz--Maldacena (FLM) \cite{Faulkner:2013ana} formulas: the boundary entropy is expressed
as the sum of an area term and the entropy of the bulk degrees of freedom
contained in the corresponding entanglement wedge. In this paper, we study whether
such a relation holds in the semiclassical limit of double-scaled SYK
(DSSYK)  \cite{Berkooz:2018jqr,Berkooz:2018qkz,Berkooz:2022mfk,Berkooz:2024lgq,Berkooz:2025ydg,Lin:2022rbf,Lin:2022nss,Rabinovici:2023yex,Goel:2023svz,Narovlansky:2023lfz,Okuyama:2023bch,Blommaert:2023opb,Blommaert:2023wad,Milekhin:2023bjv,Xu:2024hoc,Okuyama:2024yya,Almheiri:2024ayc,Almheiri:2024xtw,Milekhin:2024vbb,Aguilar-Gutierrez:2024oea,Okuyama:2025hsd,Sergio:2025Complexity-Dictionary,Belaey:2025ijg,Xu:2025Complexity,Cao:2025pir,Rajgadia:2026ask,Belaey:2025kiu,Mariani:2026jyp,Blommaert:2025avl,Schouten:2025tvn,Narovlansky:2025tpb,Goto:2026ipq}. In this model, semiclassical two-point functions admit a
fake-disk interpretation \cite{Lin_2023}, but this alone does not establish an FLM formula. Our entropy calculation establishes
an FLM formula in the heavy-probe limit, showing that these probes see a sharply defined entanglement wedge in the sense
described below.

Our main result is the following FLM formula for the class of states specified
below, in the semiclassical limit $\lambda\to 0$ with heavy probe operators:
\begin{equation} \label{eq:main-result}
    S_{\mathrm{bdy}}\left(\rho_R\right)=A+S\left(\widetilde{\rho}_r\right),
\end{equation}
where $\rho_R$ is the reduced density matrix\footnote{Throughout the paper, density matrices (such as $\rho, \rho_R$ and $\widetilde{\rho}$) are not necessarily properly normalized.} of the right subsystem for a state prepared by inserting $k$ matter operators from mutually independent matter species along the Euclidean boundary. Each operator creates a particle carrying a weight $\Delta_i$ and flavor $a_i$, with $i=1,2,\dots, k$:
\begin{equation} \label{eq:state-def2}
    |\Psi_{\vec \Delta, \vec a}(\vec \beta)\rangle =
    \begin{tikzpicture}[baseline=-3.5ex, scale=1.2]
     \draw[line width=0.8pt]
     (1,0)
     arc[
     start angle=0,
     end angle=-180,
     radius=1];
     \draw[thick,blue] (-40:1) -- ({cos(40)},0);
     \draw[thick, blue] (-60:1) -- ({cos(60)},0);
     \draw[thick, blue] (-135:1) -- ({cos(135)},0);
     \node at (-.15,-.35) {$\dots$} ;
     
     \node at (-160:1.2) {\tiny $\frac{\beta_1}{2}$};
     \node at (-120:1.2) {\tiny $\frac{\beta_2}{2}$};
     \node at (-90:1.2) {$\dots$};
     \node at (-50:1.2) {\tiny $\frac{\beta_k}{2}$};
     \node at (-20:1.3) {\tiny $\frac{\beta_{k+1}}{2}$};
     \node at ({cos(135)},0.15) {\tiny \textcolor{blue}{$(\Delta_1,a_1)$}};
     \node at ({cos(40)+.1},.15) {\tiny \textcolor{blue}{$(\Delta_{k},a_k)$} } ;
\end{tikzpicture}
\end{equation}
Consider a state where the flavor degrees of freedom are in a mixed state described by a density matrix $\widetilde{\rho}$:
\begin{equation}
    \rho=\sum_{\vec{a}_{1},\vec{a}_{2}}\widetilde{\rho}_{\vec{a}_{1},\vec{a}_{2}}|\Psi_{\vec \Delta, \vec{a}_1 }(\vec \beta)\ra\la\Psi_{\vec \Delta, \vec{a}_2}(\vec \beta)| , \quad \rho_R = \tr_L \rho ,
\end{equation}
where the flavor density matrix $\widetilde{\rho}$ acts on a Hilbert space
$\widetilde{\mh}^{(k)}$ formed by the product of $k$ single-insertion flavor
Hilbert spaces. The semiclassical result for the entropy of $\rho_R$
turns out to be best understood by factorizing $\widetilde{\mh}^{(k)}$ into two factors:
\begin{equation}
    \widetilde{\mathcal H}^{(k)}
    =\widetilde{\mathcal H}_{l}\otimes \widetilde{\mathcal H}_{r},
\end{equation}
where $\widetilde{\mh}_r$ contains the flavor degrees of freedom associated
with operator insertions at Euclidean locations closer to the right boundary
than to the left boundary. We use this $\widetilde{\mh}_r$ as an
operational definition of the bulk entanglement wedge of the right boundary for these probe particles. The
matter density matrix $\widetilde{\rho}_r$ appearing in \eqref{eq:main-result}
is then the reduction of $\widetilde{\rho}$ to $\widetilde{\mh}_r$.

The `area' term $A$ in \eqref{eq:main-result} is a number that depends on the Euclidean
separations $\vec \beta = (\beta_1,\dots, \beta_{k+1})$ and the matter weights
$\vec \Delta = (\Delta_1,\dots, \Delta_k)$. We will therefore consider fixed $\vec \beta$ and $\vec \Delta$, in which case $A$ is a constant that is independent of the state $\widetilde{\rho}$ chosen from $\widetilde \mh^{(k)}$.\footnote{We call $A$ an `area' term because it plays the role of the area contribution in the FLM formula, although we do not derive a geometric interpretation of $A$ here.}

In this sense, \eqref{eq:main-result} takes precisely the form of an FLM formula valid for all mixed states on the bulk Hilbert space defined here as $\widetilde \mh^{(k)}$ (also known as a code Hilbert space \cite{Almheiri_2015}), with the boundary entropy $S_{\mathrm{bdy}}(\rho_R)$ and bulk entropy $S(\widetilde{\rho}_r)$ differing by the `area' term $A$.
Therefore, flavor information carried by particles inside the right entanglement wedge $r$ (operationally defined as $\widetilde{\mh}_r$) can be reconstructed on the right boundary, whereas flavor information outside $r$ is inside the left entanglement wedge and can be reconstructed on the left boundary \cite{Jafferis_2016,Dong:2016eik,Harlow_2017}.
Thus, the FLM formula \eqref{eq:main-result} implies that semiclassical DSSYK
has a sharply defined entanglement wedge when probed by these heavy
operators.

Varying the Euclidean separations moves the matter insertions
through transition points at which the boundary entropy and reconstructibility can change sharply. We will therefore interpret this as the
corresponding particle crossing the entanglement-wedge boundary. Note that this does not require a local bulk description of the locations of the particles, as we defined the entanglement wedge operationally as $\widetilde{\mh}_r$ (or algebraically as its operator algebra).

For a single-particle insertion, this transition is illustrated by a simplified version of \eqref{eq:main-result}:
\begin{equation}\label{eq:intro-one-particle-transition}
    S_{{\rm bdy}}(\rho_R)
    =
    \begin{cases}
    A,
    & \beta_1<\beta_2,\\[4pt]
    A+S(\widetilde\rho),
    & \beta_1>\beta_2.
    \end{cases}
\end{equation}
Thus, the boundary entropy jumps by precisely the matter entropy when the
particle crosses the entanglement-wedge boundary.

The states defined in \eqref{eq:state-def} are partially entangled thermal
states (PETS). The entropy of such states has previously been
studied in the ordinary SYK model in \cite{Goel:2018ubv}. Our construction
follows the same logic but replaces the Schwarzian dynamics with exact DSSYK
results. This requires a careful analysis of semiclassical thermal $2n$-point
functions for arbitrary configurations of Wick-contracted matter chords. We
establish semiclassical factorization of these correlators: they reduce
to a sum over pairings, with each pairing contributing a product of $n$
two-point functions.\footnote{This is the generalized-free-field form of large-$N$ factorization familiar in holographic CFTs. In DSSYK, the double-scaled large-$N$ limit is not sufficient; the additional semiclassical limit $\lambda\to 0$ is required.} This property was derived for a class of uncrossed $2n$-point functions in
\cite{Goel:2023svz}
and supported more generally by a spectral argument in
\cite{Mukhametzhanov:2023tcg}. Here we give a
probabilistic proof for arbitrary fixed matter-chord diagrams,
independent of that spectral argument. Additionally, we
strengthen the spectral argument by bounding corrections to
the semiclassical $6j$-symbol\footnote{The underlying origin of this $6j$-symbol remains an active subject of investigation; see, for example, \cite{Berkooz:2022mfk,Verlinde:2024znh,Verlinde:2024zrh,Belaey:2025ijg,vanderHeijden:2025zkr,Schouten:2025tvn}.} approximation and making its application
to arbitrary $2n$-point functions explicit.
Related work has studied entanglement entropy of fixed-length states in DSSYK~\cite{Tang:2024entropy},
algebraic entropy in DSSYK including PETS~\cite{AguilarGutierrez:2025entropy},
and entanglement-wedge reconstruction in JT gravity~\cite{Gao:2024modular}.

The main result \eqref{eq:main-result} applies only in the semiclassical regime
of DSSYK with heavy probes. More precisely, we first take $\lambda\to 0$ and
then take $\Delta \to \infty$. The relation between this heavy-probe
entanglement wedge and the stringy horizon structure studied in
\cite{Gao:2025ohk} remains to be understood.

The remainder of the paper is organized as follows. In section
\ref{sec:PETS-def}, we define PETS beginning with the finite $N$ SYK model and
calculate their disorder-averaged entropy using the replica trick. In the
double-scaled limit, the replicated entropy reduces to exact DSSYK thermal
correlators. Taking the semiclassical and heavy-probe limits then isolates the
dominant saddle-point contribution and leads to \eqref{eq:main-result}. In
section \ref{sec:Correlation}, we evaluate the semiclassical limit of the DSSYK
partition function and the relevant thermal correlators. In particular, we
establish semiclassical factorization for general \(2n\)-point thermal correlators
and derive an explicit expression for the area term in
\eqref{eq:main-result}. In section~\ref{sec:factor}, we give a complementary explanation of factorization from the exact spectral representations. Appendix \ref{app:MB-6j} collects technical details
on the Mellin--Barnes representation of the
$\mathcal U_q (\mathfrak{su}(1,1))$ \(6j\)-symbol and its semiclassical limit.
Appendix \ref{app:disorder-averaged-partition-function} establishes the
convergence of the disorder-averaged partition functions and entropies.
Appendix \ref{app:Limitations} reviews the fake-disk interpretation and
discusses some of its limitations.

\section{Probing the entanglement wedge in double-scaled SYK} \label{sec:PETS-def}
In this section, we first briefly review the double-scaled limit of the SYK model. We then study the entropy of thermal states perturbed by matter operators carrying flavor degrees of freedom in a general mixed state. We show that, when probed by heavy matter insertions in the semiclassical regime of DSSYK, the resulting entropy exhibits a sharp transition as a function of the insertion location. This provides evidence for a sharply defined entanglement wedge in the bulk geometry.

\subsection{A brief review of double-scaled SYK}

The SYK model contains \(N\) Majorana fermions and is specified by the
Hamiltonian
\begin{equation} \label{eq:HN-def}
    H_N
    =
    i^{p/2}
    \sum_{|I|=p}J_I\Psi_I,
    \qquad
    \Psi_I=\psi_{i_1}\cdots\psi_{i_p},
\end{equation}
where \(I=(i_1,\ldots,i_p)\) is a subset of \(\{1,\ldots,N\}\).  We take
\(N\) and \(p\) to be even and use the convention
\(\{\psi_i,\psi_j\}=2\delta_{ij}\), so that \(\psi_i^2=1\) and the
microscopic Hilbert space \(\mathcal H_N\) has dimension \(2^{N/2}\).  The
coupling constants are sampled from a Gaussian ensemble with zero mean, normalized by
\begin{equation}
    \overline{J_IJ_J}
    =
    \delta_{IJ}\,\frac{\mathcal J^2}{\lambda \binom{N}{p}}, \quad \lambda \equiv \frac{2p^2}{N}.
\end{equation}
We set \(\mathcal J=1\). With this normalization, the second moment is
\begin{equation}
   \overline{\mathrm{Tr}H_N^{2}}= i^p\sum_{I,J}\overline{J_I J_J}\, \Tr (\Psi_I \Psi_J)= 
   \sum_{I,J}\delta_{IJ}\frac{1}{\lambda\binom{N}{p}}\Tr(\mathbf{1}_N)=\frac{1}{\lambda}\Tr(\mathbf{1}_N).
\end{equation}
Here \(\mathbf{1}_N\) is the identity operator on \(\mathcal H_N\), and we
have used \(\Psi_I^2=(-1)^{p/2}\).  It is often convenient to introduce the normalized
trace
\begin{equation} \label{eq:tr-def}
    \tr X_N :=2^{-N/2}\Tr X_N,
\end{equation}
where \(X_N\) is an operator acting on \(\mathcal H_N\).  Thus \(\Tr\)
denotes the Hilbert-space trace of the microscopic theory at fixed \(N\),
whereas \(\tr\) is normalized so that
\begin{equation}
    \tr \mathbf 1_{\mathcal{H}_N} =1, \quad \forall N \in 2\mathbb{Z}_{>0}.
\end{equation}
Given an unnormalized density matrix \(\rho\), one can define R\'enyi entropies
with respect to either trace:
\begin{equation} \label{eq:S-def}
     S^{(n)}_{\text{bdy}}(\rho) \equiv
     \frac{1}{1-n}\log\frac{\mathrm{Tr}(\rho^{n})}{(\mathrm{Tr}\rho)^{n}},
     \qquad
     S^{(n)}(\rho) \equiv
     \frac{1}{1-n}\log\frac{\mathrm{tr}(\rho^{n})}{(\mathrm{tr}\rho)^{n}},
\end{equation}
which differ only by the constant shift coming from
\(\Tr=2^{N/2}\tr\):
\begin{equation} \label{eq:S-relation1}
  S^{(n)}_{\text{bdy}}(\rho) = \frac{N}{2}\log 2 + S^{(n)}(\rho).
\end{equation}
Taking \(n\to 1\) gives the corresponding relation for von Neumann entropies,
\be\label{eq:S-relation2}
S_{\rm bdy}( \rho ) = \frac{N}{2}\log 2 + S(\rho). 
\ee
The relations \eqref{eq:S-relation1} and \eqref{eq:S-relation2} hold for each
member of the ensemble, and therefore also after disorder averaging.  In the
rest of the paper, the entropies $S^{(n)}$ and $S$ of microscopic and double-scaled SYK density matrices are defined with respect to the normalized trace
$\tr$. Bulk matter entropies use the ordinary trace on the flavor Hilbert space.

The double-scaled limit of SYK is obtained by taking
\begin{equation}
    N,p\to\infty,
    \qquad
    \lambda=\frac{2p^2}{N}\quad\text{fixed},
    \qquad
    q=e^{-\lambda}.
\end{equation}
We denote this limit by \(\lim_{\mathrm{DS}}\) throughout the paper.

The basic convergent observables are the disorder-averaged moments of the
Hamiltonian, defined using the normalized trace:
\begin{equation}
    m_k^{(N)}=
    \overline{\tr H_N^{k}}.
\end{equation}
Because the couplings are Gaussian, disorder averaging expands every even
moment as a sum over Wick pairings.  In the double-scaled limit, each pairing
is represented by a chord diagram: paired Hamiltonian insertions are connected
by a chord, and each crossing of two chords contributes a factor of \(q\).
Thus
\begin{equation} \label{eq:conv-moments}
    \lim_{\rm DS}m_{2k}^{(N)}
    =
    m_{2k}^{\rm chord}
    =
    \lambda^{-k}\sum_{\pi\in\operatorname{Pair}(2k)}q^{\operatorname{cr}(\pi)},
    \qquad
    \lim_{\rm DS}m_{2k+1}^{(N)}=0 .
\end{equation}
For instance, the fourth moment has three pairings:
\[
    \{(1,2),(3,4)\},\qquad
    \{(1,4),(2,3)\},\qquad
    \{(1,3),(2,4)\}.
\]
The first two are non-crossing, while the last has one crossing, so
\begin{equation}
    \lim_{\rm DS}\overline{\tr H_N^4}
    =
    \lambda^{-2}(1+1+q)
    =
    \lambda^{-2}(2+q) .
\end{equation}
This moment expansion can be represented diagrammatically by chord rules.  For the $k$-th moment, one places $k$ marked points on the boundary and sums over all contractions, with each contraction drawn as a chord connecting two boundary insertions.  The result is a sum over chord diagrams in which any pair of chords intersects at most once.  Each Hamiltonian chord contributes $\lambda^{-1}$, and each crossing contributes $q$:
\begin{equation}
   \lim_{\rm DS}\overline{\tr H_N^4} = \begin{tikzpicture}[baseline=-0.55ex, scale=0.75]
    \draw (0,0)  circle (1);
    \fill  (1,0) circle[radius=1.2 pt];
    \fill  (-1,0) circle[radius=1.2 pt];
    \fill  (0,-1) circle[radius=1.2 pt];
    \fill  (0,1) circle[radius=1.2 pt];
    \draw (-1,0)  .. controls (135:0.35) .. (0,1);
    \draw (0,-1) .. controls (-45:0.35) .. (1,0);
   \end{tikzpicture}  
   +
   \begin{tikzpicture}[baseline=-0.55ex, scale=0.75]
    \draw (0,0)  circle (1);
    \fill  (1,0) circle[radius=1.2 pt];
    \fill  (-1,0) circle[radius=1.2 pt];
    \fill  (0,-1) circle[radius=1.2 pt];
    \fill  (0,1) circle[radius=1.2 pt];
    \draw (-1,0)  .. controls (225:0.35) .. (0,-1);
    \draw (0,1) .. controls (45:0.35) .. (1,0);
   \end{tikzpicture} 
   +
   \begin{tikzpicture}[baseline=-0.55ex, scale=0.75]
    \draw (0,0)  circle (1);
    \fill  (1,0) circle[radius=1.2 pt];
    \fill  (-1,0) circle[radius=1.2 pt];
    \fill  (0,-1) circle[radius=1.2 pt];
    \fill  (0,1) circle[radius=1.2 pt];
    \draw (-1,0)  --  (1,0);
    \draw (0,-1)  -- (0,1);
   \end{tikzpicture} 
\end{equation}

The same convergence argument applies to matter operators of the form
\begin{equation} \label{eq:ON-def}
    O_N
    =
    i^{p'/2}\sum_{|I^\prime |=p'}K_{I^\prime}\Psi_{I^\prime},
\end{equation}
where the even length \(p'\) scales with \(p\).  More precisely, we keep
\(p'/p\to\Delta\) fixed and choose the normalization of the couplings
\(K_{I^\prime}\) so that \(\overline{\tr\, O_N^2}=1\).  The matter couplings are identically distributed Gaussian random variables with zero mean, independent of one another and of the Hamiltonian couplings $J_I$. Mixed moments of
\(H_N\) and \(O_N\) then have finite double-scaled limits represented by
Hamiltonian and matter chords.  In particular, a crossing between an
\(H\)-chord and a matter chord carries the weight \(q^\Delta\).

The double-scaled limit therefore produces limiting states and operators whose
moments are computed by chord diagrams.  Let \(\mathfrak A\) denote the
\(*\)-algebra freely generated by two elements \(H\) and \(O\).\footnote{In practice, one can introduce more generators for the algebra representing double-scaled operators of the form \eqref{eq:ON-def} with different species and weights. }  The
double-scaled limit of the normalized finite-\(N\) trace defines a tracial
state \(\omega\) on this algebra.  Explicitly, for a word
\(a=H^{k_1}OH^{k_2}\cdots OH^{k_n}\), let
\(a_N=H_N^{k_1}O_NH_N^{k_2}\cdots O_NH_N^{k_n}\).  Then
\begin{equation} \label{eq:omega-def}
    \omega(a) \equiv \lim_{\mathrm{DS}}\overline{\tr(a_N)}.
\end{equation}
The GNS construction for \((\mathfrak A,\omega)\) gives a von Neumann algebra
\(\mathcal A\) of type II\(_1\), which we call the double-scaled algebra~\cite{Lin:2022rbf,Xu:2024hoc}.  We
use the same notation \(\tr\) for its trace, since it is the limit of the
finite-\(N\) normalized traces in \eqref{eq:omega-def}.  Operators with a
subscript \(N\) act on the finite-\(N\) Hilbert space \(\mathcal H_N\), while
operators without the subscript denote their double-scaled limits.  Thus
\(\tr\) denotes the normalized trace \eqref{eq:tr-def} on finite-\(N\)
operators and the tracial state \eqref{eq:omega-def} on double-scaled
operators.

The convergence analysis \eqref{eq:conv-moments} also applies to the thermal partition function. In particular, we define
\begin{equation}
    Z_N(\beta) = \tr e^{-\beta H_N} = \sum_{k=0}^{\infty} \frac{(-\beta)^k}{k!} \tr(H_N^{k}).
\end{equation}
Appendix~\ref{app:disorder-averaged-partition-function} proves that this series converges uniformly in $N$ in the double-scaling limit for all finite values of $\lambda$ and $\beta$. Consequently, the double-scaled limit can be evaluated term by term:
\begin{equation}
    \begin{split}
            \lim_{\rm DS} \overline{Z_N(\beta)}
    & = \lim_{\mathrm{DS}} \sum_{k=0}^{\infty} \frac{(-\beta)^k}{k!} \overline{\tr H_N^k}  = \sum_{k=0}^\infty \lim_{\mathrm{DS}} \frac{(-\beta)^k}{k!}  \overline{\tr H_N^k} \\ 
    & = \sum_{k=0}^\infty \frac{(-\beta)^k}{k!}  \tr H^k   =  \tr e^{-\beta H}.
    \end{split}
\end{equation}
Here \(H\) is the double-scaled SYK Hamiltonian. Thus chord diagrams compute the
finite normalized-trace part of the microscopic theory.  Restoring the full
trace multiplies the answer by \(2^{N/2}\), so the logarithm gains the additive
term \((N/2)\log2\).

\subsection{Single-particle states and their entropy}
In this subsection, we study partially entangled states in the double-scaled limit of SYK. These are pure states of two SYK copies prepared by inserting a matter operator into a thermofield double state. We allow the flavor degrees of freedom to be in a general mixed state and use the matter insertion to probe the bulk entanglement wedge.

Now let us introduce the partially entangled state. Let $\mathcal{O}_{\Delta,a}$ be a matter chord operator of the form \eqref{eq:ON-def}. We have added a flavor index $a$ to allow different species of matter operators. It can be sandwiched between Euclidean evolution operators generated by the SYK Hamiltonian $H_N$, giving $e^{-\beta_L H_N/2} \mathcal{O}_{\Delta,a} e^{-\beta_R H_N/2}$.   We will assume the disorder coefficients corresponding to different flavors are independent. 

With the standard operator-state map, the operator can be viewed as a state in $\mathcal{H}_N^{(L)}\otimes \mathcal{H}_{N}^{(R)}$, which we denote as: 
\begin{equation}
     |\Psi_{\Delta, a}(\beta_L,\beta_R) \rangle
=
e^{-\beta_L H_N/2} \mathcal{O}_{\Delta,a} e^{-\beta_R H_N/2}
=\, \vcenter{\hbox{%
    \begin{tikzpicture}[
      x=1cm,
      y=1cm,
      line cap=round,
      line join=round
    ]
      \draw[line width=0.75pt]
        (-1.15,0.50)
        arc[
          start angle=180,
          end angle=360,
          x radius=1.15,
          y radius=1.05
        ];

      \draw[blue,line width=1.15pt]
        (0,0.42) -- (0,-0.55);

      \fill[blue]
        (0,-0.55) circle[radius=2.2pt];

      \node[anchor=east,inner sep=1pt]
        at (-0.93,-0.31)
        {\tiny $\frac{\beta_L}{2}$};

      \node[anchor=west,inner sep=1pt]
        at (0.93,-0.31)
        {\tiny $\frac{\beta_R}{2}$};

        \node at (0,0.6) {\tiny $(\Delta,a)$};
    \end{tikzpicture}
    }} %
\end{equation}
A density matrix that involves different flavors of matter insertions can be defined as: 
\begin{equation}
    \rho
    =
    \sum_{a,b=1}^{f}
    \tilde{\rho}_{ab}
    |\Psi_{\Delta,a}(\beta_L,\beta_R)\rangle
    \langle\Psi_{\Delta, b}(\beta_L,\beta_R)|,
\end{equation}
where $\tilde{\rho}$ is chosen to be a density matrix in the Hilbert space $\widetilde{\mathcal{H}}^{(1)}$ of flavor indices, and is normalized such that $\widetilde\Tr(\tilde \rho)= 1$. Here $\widetilde \Tr$ is the trace of the flavor Hilbert space $\widetilde{\mathcal{H}}^{(1)}$, whose dimension is assumed to be a constant number $f$. In the following calculation, we will sometimes keep explicit $\widetilde{\Tr}(\tilde{\rho})$ dependence for later convenience. 

We are interested in the boundary entropy \eqref{eq:S-def} of $\rho$ on the right subsystem $R$. By \eqref{eq:S-relation2}, we study the entropy, defined with respect to the normalized trace $\tr$, of the reduced density matrix
\begin{equation}
    \rho_R=\tr_L\rho,
\end{equation}
for which a useful diagrammatic representation is
\begin{equation} \label{eq:rhoR-diagram}
    \rho_R =\sum_{a,b} \tilde{\rho}_{ab} \,\, \begin{tikzpicture}[baseline=-0.55ex, scale=1.2]
         \draw[line width=0.75pt] 
         (10:1) 
         arc[ 
         start angle=10, 
         end angle=350, 
         radius=1];

         \draw[blue,thick] (0,-1) -- (0,-0.2);
         \draw[blue,thick] (0,1) -- (0,0.2);
         \fill[blue]  (0,1) circle[radius=1.5 pt];
         \fill[blue]  (0,-1) circle[radius=1.5 pt];
         
         \node  at (-.75,0) {\small $\beta_L$};
         \node at (50:1.3) {\small $\frac{\beta_R}{2}$};
         \node at (-50:1.3) {\small $\frac{\beta_R}{2}$};
         \node at (0.35,0.5) {\tiny $(\Delta,b)$};
         \node at (0.35,-0.5) {\tiny $(\Delta,a)$};
    \end{tikzpicture}
\end{equation}
At finite $N$, the normalization of $\rho_R$ is disorder dependent. This leads to several inequivalent notions of ensemble-averaged R\'enyi entropy---quenched, quasi-quenched, semi-quenched, and annealed---distinguished by the placement of the disorder average relative to the logarithm and normalization factors. Throughout this work, we use the semi-quenched entropy because it requires only a single $n$-replica calculation without additional replica limits, and its $n\to 1$ continuation can be expressed directly in terms of DSSYK thermal correlation functions. We expect this choice not to affect the leading result in the double-scaled limit.

Indeed, the additional replica constructions required by the other definitions involve disorder-averaged products of replicated traces and normalization factors. Their leading contributions are disconnected and therefore factorize into the same single-replica quantities that enter the semi-quenched entropy, whereas contractions connecting distinct replica components are suppressed in the double-scaled limit~\cite{Berkooz:2021multi}. Assuming that these corrections to the factorized answer remain subleading in the required replica limits, the quenched, quasi-quenched, semi-quenched, and annealed entropies therefore agree at leading order in the double-scaled SYK limit, differing only through finite-$N$ corrections that are invisible to the disk-level chord calculation.

Define the R\'enyi entropy by
\begin{equation}\label{eq:semi-quenched-renyi-def}
    S^{(n)} (\rho_R)
    =
    \frac{1}{1-n}
    \log
    \frac{\overline{\tr_R\rho_R^{n}}}
    {\overline{(\tr_R\rho_R)^{n}}} .
\end{equation}
The double-scaled limit converges and takes the form
\begin{equation} \label{eq:entropy-DS}
    \lim_{{\rm DS}} S^{(n)}(\rho_R)=  \frac{1}{1-n}\log\frac{Z_{2n}}{Z_{2}^{n}},
\end{equation}
where both $Z_{2n}$ and $Z_2$ are related to DSSYK thermal correlators. A proof of \eqref{eq:entropy-DS} for fixed integer $n>1$ is given in Appendix~\ref{subsec:entropy-convergence}. The denominator is a product of \(n\) identical one-copy amplitudes,
\begin{equation}
    \lim_{\rm DS}
    \overline{(\tr_R\rho_R)^n}
    =(\widetilde{\Tr}\tilde \rho)^n Z_2^{n} (\beta_L,\beta_R;\Delta) , \quad 
   Z_2 (\beta_L,\beta_R;\Delta) =Z(\beta) G_2^{\Delta}(\beta_L,\beta_R), 
\label{eq:DS-Z2}
\end{equation}
where $\beta=\beta_L+\beta_R$, and $Z(\beta)= \tr (e^{-\beta H})$ is the thermal partition function.  Equivalently, $Z_2$ is the unnormalized thermal two-point amplitude with fixed Euclidean boundary lengths:
\begin{equation}
    Z_2 (\beta_L,\beta_R;\Delta) =  \langle 0 | e^{-\beta_L H} \mathcal{O}_\Delta e^{-\beta_R H} \mathcal{O}_\Delta |0\rangle =  \, 
    \begin{tikzpicture}[baseline=-0.55ex, scale=0.9]
        \draw (0,0)  circle (1);
        \draw[thick,blue] (0,-1)--(0,1);
        \node at (-0.3,0) {\small $\Delta$};
        \node at (-1.3,0) {\small $\beta_L$};
        \node at (1.3,0) {\small $\beta_R$};
    \end{tikzpicture}\, ,
\end{equation}
with $\mathcal{O}_\Delta$ understood as a matter chord operator in the double-scaled algebra, carrying weight $\Delta$. 

The proof of \eqref{eq:DS-Z2} is the same uniform Wick-bound argument reviewed in
Appendix~\ref{app:disorder-averaged-partition-function}.  The random variable
\(\tr_R\rho_R\) is a finite linear combination of normalized Euclidean SYK
correlators.  Wick contractions connecting different copies in
\((\tr_R\rho_R)^n\) obey additional Majorana-index constraints and are therefore suppressed
in the double-scaled limit.  The disconnected contractions survive and give the
product of disorder-averaged one-copy correlators.  The same uniform bound allows the
double-scaled limit to pass through the thermal series, so the chord rules indeed
compute the normalization part of the semi-quenched entropy.

The numerator contains the genuinely replicated amplitude.  Before performing
the disorder average, the replicated trace $\tr (\rho_R^{n})$ is obtained by
taking $n$ copies of \eqref{eq:rhoR-diagram} and cyclically identifying their
right-boundary endpoints:
\begin{equation} \label{eq:rhon-diagram}
    \tr(\rho_R^{n}) = \begin{tikzpicture}[baseline=-0.55ex, scale=0.9]
         \draw[line width=0.75pt] 
         (10:1) 
         arc[ 
         start angle=10, 
         end angle=350, 
         radius=1];
         \draw[line width=0.75pt] 
         ($(10:1)+(2.5,0)$ )
         arc[ 
         start angle=10, 
         end angle=350, 
         radius=1];
         \node at (4.3,-.2) {$\cdots$};
         \draw[line width=0.75pt] 
         ($(10:1)+(6.0,0)$ )
         arc[ 
         start angle=10, 
         end angle=350, 
         radius=1];

         \draw[dashed] (350:1) .. controls (2.0,-0.2) .. ($(10:1)+(2.5,0)$);
         \draw[dashed] ($(350:1)+(2.5,0)$) .. controls ($(350:1)+(2.7,0)$) .. (4.5,0.1);
         \draw[dashed] ($(10:1)+(6,0)$) .. controls (6.2,-.1) .. (4.8,-.2) ; 
         \draw[dashed] (10:1) .. controls (1.5,2) and (8.5, 2.4) .. ($(350:1)+(6,0)$);

         \draw[blue,thick] (0,-1) -- (0,-0.2); 
         \draw[blue,thick] (0,1) -- (0,0.2);
         \draw[blue,thick] (2.5,-1) -- (2.5,-0.2); 
         \draw[blue,thick] (2.5,1) -- (2.5,0.2); 
         \draw[blue,thick] (6,-1) -- (6,-0.2); 
         \draw[blue,thick] (6,1) -- (6,0.2); 
    \end{tikzpicture}\, .
\end{equation}
Here the dashed lines indicate the cyclic endpoint identifications.  After the
disorder average, the blue matter insertions are paired by Wick contractions in
the bulk.  Thus the double-scaled numerator becomes a sum over Wick contractions
in the thermal \(2n\)-point function with this replica boundary condition.
Denoting the limit of
\eqref{eq:rhon-diagram} by $Z_{2n}$, we can write
\begin{equation} \label{eq:Zn}
Z_{2n}
=
\sum_{i_1,\ldots,i_n,j_1,\ldots,j_n}
\widetilde\rho_{i_1j_1}\widetilde\rho_{i_2j_2}\cdots
\widetilde\rho_{i_nj_n}
\sum_{\text{Wick contractions}}
\begin{tikzpicture}[baseline=-0.55ex, scale=1.2]
     \draw[line width=.8pt,dashed] 
     (-30:1) 
     arc[ 
     start angle=-30, 
     end angle=30, 
     radius=1];
      \draw[line width=0.8pt] 
      (30:1) 
      arc[ 
      start angle=30, 
      end angle=330, 
      radius=1];
    \fill[blue]  (160:1) circle[radius=1.5 pt];
    \fill[blue]  (200:1) circle[radius=1.5 pt];
    \fill[blue]  (130:1) circle[radius=1.5 pt];
    \fill[blue]  (230:1) circle[radius=1.5 pt];
    \fill[blue]  (100:1) circle[radius=1.5 pt];
    \fill[blue]  (260:1) circle[radius=1.5 pt];

    \node at (200:1.2) {\small $i_1$};
    \node at (230:1.2) {\small $j_2$};
    \node at (260:1.2) {\small $i_2$};
    \node at (100:1.25) {\small $j_n$};
    \node at (130:1.25) {\small $i_n$};
    \node at (160:1.25) {\small $j_1$};
    \node at (60:1.2) {\small $\dots$};
    \node at (-60:1.2) {\small $\dots$};
    \node at (0,0) {\small $\dots$};
    \node at (215:.8) {\tiny $\beta_R$};
    \node at (245:.8) {\tiny $\beta_L$};
    \node at (180:0.8) {\tiny $\beta_L$};
    \node at (145:.8) {\tiny $\beta_R$};
\end{tikzpicture}\, .
\end{equation}
The second sum runs over bulk Wick contractions compatible with the prescribed
cyclic ordering of the boundary flavor indices.

Let $\pi\in \operatorname{Pair}(2n)$ be a pairing of the \(2n\) boundary
endpoints.  To make the flavor contractions explicit, collect the labels around
the replicated boundary into the ordered vector
\begin{equation}
\boldsymbol i
=
(i_1,j_2,i_2,j_3,\ldots,i_{n-1},j_n,i_n,j_1).
\end{equation}
Then the sum over Wick contractions in \eqref{eq:Zn} can be written as
\begin{equation} \label{eq:cd-n}
\sum_{\text{Wick contractions}}
\begin{tikzpicture}[baseline=-0.55ex, scale=1.0]
     \draw[line width=.8pt,dashed] 
     (-30:1) 
     arc[ 
     start angle=-30, 
     end angle=30, 
     radius=1];
      \draw[line width=0.8pt] 
      (30:1) 
      arc[ 
      start angle=30, 
      end angle=330, 
      radius=1];
    \fill[blue]  (160:1) circle[radius=1.5 pt];
    \fill[blue]  (200:1) circle[radius=1.5 pt];
    \fill[blue]  (130:1) circle[radius=1.5 pt];
    \fill[blue]  (230:1) circle[radius=1.5 pt];
    \fill[blue]  (100:1) circle[radius=1.5 pt];
    \fill[blue]  (260:1) circle[radius=1.5 pt];

    \node at (200:1.2) {\small $i_1$};
    \node at (230:1.2) {\small $j_2$};
    \node at (260:1.2) {\small $i_2$};
    \node at (100:1.25) {\small $j_n$};
    \node at (130:1.25) {\small $i_n$};
    \node at (160:1.25) {\small $j_1$};
    \node at (60:1.2) {\small $\dots$};
    \node at (-60:1.2) {\small $\dots$};
    \node at (0,0) {\small $\dots$};
    \node at (215:.8) {\tiny $\beta_R$};
    \node at (245:.8) {\tiny $\beta_L$};
    \node at (180:0.8) {\tiny $\beta_L$};
    \node at (145:.8) {\tiny $\beta_R$};
\end{tikzpicture}
=
\sum_{\pi\in\operatorname{Pair}(2n)}
\left[
\prod_{(a,b)\in\pi}
\delta_{\boldsymbol i_a\boldsymbol i_b}
\right]
Z(n\beta) G_{2n}^{\pi}(\beta_L,\beta_R).
\end{equation}
Here \(G_{2n}^{\pi}(\beta_L,\beta_R)\) denotes the contribution to the normalized thermal \(2n\)-point
function from the Wick contraction specified by the pairing \(\pi\). Since different flavors share the same $\Delta$, $G_{2n}^\pi$ is independent of $\boldsymbol{i}$.  To see this, we note that at finite $N$, the $\boldsymbol{i}$ dependence is related to disorder averaging, and the Kronecker delta functions above implement the corresponding flavor
contractions.  After the flavor sum, each pairing produces a definite invariant
of the flavor density matrix.  We denote it by
\begin{equation}
\mathcal C_{\pi}(\widetilde\rho)
\equiv
\sum_{i_1,\ldots,i_n,j_1,\ldots,j_n}
\widetilde\rho_{i_1j_1}\cdots \widetilde\rho_{i_nj_n}
\prod_{(a,b)\in\pi}
\delta_{\boldsymbol i_a\boldsymbol i_b}.
\end{equation}
Thus the replicated trace has the compact form
\begin{equation} \label{eq:Z2n-result-exact}
Z_{2n}
=
\sum_{\pi\in\operatorname{Pair}(2n)}
\mathcal C_{\pi}(\widetilde\rho)
Z(n \beta) G_{2n}^{\pi}(\beta_L,\beta_R).
\end{equation}

For example, for \(n=2\) one finds
\begin{equation}
Z_4
=
(\widetilde{\Tr}\widetilde\rho)^2\,
\begin{tikzpicture}[baseline=-0.55ex, scale=0.9]
    \def\R{1.0}
    \draw (0,0) circle (\R);
    \draw (45:1) .. controls (0:0.3) .. (-45:1);
    \draw (135:1) .. controls (180:0.3) .. (225:1);
    \node at (90:1.28) {$\beta_R$};
    \node at (270:1.28) {$\beta_R$};
    \node at (180:0.72) {$\beta_L$};
    \node at (0:0.72) {$\beta_L$};
\end{tikzpicture}
\quad
+
\widetilde{\Tr}(\widetilde\rho^{\,2})\,
\begin{tikzpicture}[baseline=-0.55ex, scale=0.9]
    \def\R{1.0}
    \draw (0,0) circle (\R);
    \draw (45:1) .. controls (90:0.3) .. (135:1);
    \draw (225:1) .. controls (270:0.3) .. (315:1);
    \node at (90:1.28) {$\beta_R$};
    \node at (270:1.28) {$\beta_R$};
    \node at (180:0.72) {$\beta_L$};
    \node at (0:0.72) {$\beta_L$};
\end{tikzpicture}
+
\widetilde{\Tr}(\widetilde\rho\,\widetilde\rho^{T})\,
\begin{tikzpicture}[baseline=-0.55ex, scale=0.9]
    \def\R{1.0}
    \def\a{50}
    \draw (0,0) circle (\R);
    \draw (180-\a:\R) -- (-\a:\R);
    \draw (180+\a:\R) -- (\a:\R);
    \node at (90:1.28) {$\beta_R$};
    \node at (270:1.28) {$\beta_R$};
    \node at (180:0.72) {$\beta_L$};
    \node at (0:0.72) {$\beta_L$};
\end{tikzpicture}.
\end{equation}
Here \(\widetilde\rho^T\) denotes the transpose of
\(\widetilde\rho\) in the flavor basis used to define the boundary
contractions.

In the next section, we evaluate \eqref{eq:Z2n-result-exact} in the semiclassical
regime \(\lambda\to0\), with a heavy probe \(\Delta\gg1\). In this limit, the
sum over pairings is dominated by the geodesic configuration with minimal
effective length. The dominant pairing is determined by the Euclidean location
of the insertion, equivalently by the relative sizes of \(\beta_L\) and
\(\beta_R\). The
leading saddle contribution to \eqref{eq:Z2n-result-exact} is
\begin{equation} \label{eq:Z2n-saddle}
Z_{2n}^{*}
=
\begin{cases}
(\widetilde{\Tr}\widetilde\rho)^n\,
Z(n\beta) [G_2^\Delta(\beta_L,n\beta-\beta_L)]^n, 
& \beta_L<\beta_R, \\[4pt]
\widetilde{\Tr}(\widetilde\rho^{\,n})\,
Z(n\beta) [G_2^\Delta(n\beta-\beta_R,\beta_R)]^n, 
& \beta_L>\beta_R.
\end{cases}
\end{equation}
This saddle exhibits a sharp transition at \(\beta_L=\beta_R\).  Substituting  \eqref{eq:Z2n-saddle} into \eqref{eq:entropy-DS}, we obtain
\begin{equation} \label{eq:entropy-transition}
S^{(n)}(\rho_R)
=
\begin{cases}
\delta A^{(n)}, 
& \beta_L<\beta_R, \\[4pt]
\delta A^{(n)}
+
S^{(n)}(\widetilde\rho),
& \beta_L>\beta_R ,
\end{cases}
\end{equation}
where $\delta A^{(n)}$ is defined by
\begin{equation}
\label{eq:Sn-def}
\delta A^{(n)}(\beta_L,\beta_R;\Delta)
=
\frac{1}{1-n}
\log
\frac{
Z(n\beta)
\left[
G_2^{\Delta}\bigl(\beta_{\rm min},n\beta-\beta_{\rm min}\bigr)
\right]^n
}{
Z(\beta)^n
\left[
G_2^{\Delta}(\beta_L,\beta_R)
\right]^n
},
\end{equation}
with $\beta_{\min}\equiv \min(\beta_L,\beta_R)$.  Importantly, $\delta A^{(n)}$ is independent of the flavor density matrix $\widetilde{\rho}$. For fixed $\Delta$, $\beta_L$, and $\beta_R$, it is therefore a constant throughout the code subspace spanned by the matter-flavor Hilbert space. This suggests interpreting the $n\to1$ limit of $\delta A^{(n)}$ as the fluctuating contribution to the area operator, which acts as a multiple of the identity within this code subspace.

Taking $n\to1$ gives the
double-scaled von Neumann entropy:\footnote{Here and in the multi-particle generalization below, we assume that, away from transitions, corrections to the dominant saddle give a vanishing contribution to the von Neumann entropy in the limits considered here.}
\begin{equation} \label{eq:S2-tranSITION}
    S(\rho_R) = \begin{cases}\delta A , & \beta_L<\beta_R, \\ \delta A +S(\widetilde{\rho}), & \beta_L>\beta_R .\end{cases}
\end{equation}
Here $S(\tilde{\rho})$ is the von Neumann entropy of the flavor density matrix,
and $\delta A$ is the finite double-scaled fluctuation of the area term:
\begin{equation}\label{eq:deltaA-def}
\delta A = (1-\beta \partial_\beta) \log Z(\beta) - \beta \partial_{\beta} \log G_2^{\Delta}(\beta_{\rm min}, \beta-\beta_{\rm min}).
\end{equation}
Here and below, the $\partial_\beta$ derivative acting on the two-point function is taken with \(\beta_{\rm min}\) fixed.

Recall that the boundary entropy \eqref{eq:S-relation2} differs from $S(\rho)$ by a
constant shift \((N/2)\log 2\), which is naturally interpreted as the expectation
value of the leading area operator.  Introducing \(A=(N/2)\log 2+\delta A\),
we find from
\eqref{eq:Sn-def} that
\begin{equation} \label{eq:Sbdy-2pt}
    S_{\text{bdy}}(\rho_R)  = \begin{cases}  A, & \beta_L<\beta_R, \\  A+S(\widetilde{\rho}), & \beta_L>\beta_R .\end{cases}
\end{equation}
The transition in \eqref{eq:Sbdy-2pt} has the expected entanglement-wedge
interpretation.  When \(\beta_L<\beta_R\), the dominant saddle keeps the particle
outside the right entanglement wedge, so only the area term contributes.
When \(\beta_L>\beta_R\), the particle lies in the right entanglement wedge and
the entropy receives the additional flavor contribution \(S(\widetilde\rho)\).
This is the DSSYK analogue of the bulk matter entropy term in the FLM formula.

\subsection{Multi-particle states and the FLM formula}

The same logic extends directly to multi-particle states. Consider a state
prepared by perturbing the thermofield double state with $k$ matter insertions
along the Euclidean boundary, each carrying a weight $\Delta_i$ and flavor $a_i$, with $i=1,2,\dots, k$.
The slot index $i$ also distinguishes mutually independent matter species, even when their weights coincide. Wick contractions therefore connect only copies with the same slot and flavor labels. The state is
\begin{equation} \label{eq:state-def}
    |\Psi_{\vec \Delta, \vec a}(\vec \beta)\rangle =
    \begin{tikzpicture}[baseline=-3.5ex, scale=1.2]
     \draw[line width=0.8pt]
     (1,0)
     arc[
     start angle=0,
     end angle=-180,
     radius=1];
     \draw[thick,blue] (-40:1) -- ({cos(40)},0);
     \draw[thick, blue] (-60:1) -- ({cos(60)},0);
     \draw[thick, blue] (-135:1) -- ({cos(135)},0);
     \node at (-.15,-.35) {$\dots$} ;
     
     \node at (-160:1.2) {\tiny $\frac{\beta_1}{2}$};
     \node at (-120:1.2) {\tiny $\frac{\beta_2}{2}$};
     \node at (-90:1.2) {$\dots$};
     \node at (-50:1.2) {\tiny $\frac{\beta_k}{2}$};
     \node at (-20:1.3) {\tiny $\frac{\beta_{k+1}}{2}$};
     \node at ({cos(135)},0.15) {\tiny \textcolor{blue}{$(\Delta_1,a_1)$}};
     \node at ({cos(40)+.1},.15) {\tiny \textcolor{blue}{$(\Delta_{k},a_k)$} } ;
\end{tikzpicture}
\end{equation}
where $\vec \Delta=(\Delta_{1},\dots,\Delta_{k})$, $\vec \beta = (\beta_1,\dots,\beta_{k+1})$  and $\vec a = (a_1,a_2,\dots,a_k)$ collectively denote the matter
weights, Euclidean separations and flavor indices. The operator that corresponds to $\Psi_{\vec \Delta,\vec a}(\vec \beta)$   is:
\begin{equation}
    e^{-\frac{\beta_{1}}{2}H_{N}}\mathcal{O}_{\Delta_1, a_1}e^{-\frac{\beta_{2}}{2} H_{N}}\cdots\mathcal{O}_{\Del_{k},a_k}e^{-\frac{\beta_{k+1}}{2}H_{N}}.
\end{equation}
Each insertion slot occurs once in the bra and once in the ket. Independence of the matter species and flavors therefore gives, with the double-scaled inner product defined by the disorder-averaged trace in \eqref{eq:omega-def},
\begin{equation}
    \frac{\langle\Psi_{\vec\Delta,\vec b}(\vec\beta)|\Psi_{\vec\Delta,\vec a}(\vec\beta)\rangle}{\mathcal Z_\Psi}=\prod_{i=1}^{k}\delta_{a_i b_i},
\end{equation}
where $\mathcal Z_\Psi>0$ is the common squared norm, independent of the flavor labels. Dividing these states by $\sqrt{\mathcal Z_\Psi}$ therefore identifies the flavor Hilbert space isometrically with their span in the double-scaled Hilbert space.

A mixed state built from these multi-particle wavefunctions, together with its
right reduction, is defined by
\begin{equation}
    \rho=\sum_{\vec{a}_{1},\vec{a}_{2}}\widetilde{\rho}_{\vec{a}_{1},\vec{a}_{2}}|\Psi_{\vec \Del, \vec a_1}(\vec \beta)\ra\la\Psi_{\vec \Delta, \vec a_2}(\vec \beta)| , \quad \rho_R = \tr_L \rho .
\end{equation}
The flavor density matrix $\tilde{\rho}$ now acts on $k$ copies of the flavor
Hilbert space,
\begin{equation}
\widetilde{\mathcal{H}}^{(k)}=\bigotimes_{i=1}^k \widetilde{\mathcal{H}}(i) =\widetilde{\mh}_{l}\otimes\widetilde{\mathcal{H}}_{r},      
\end{equation}
where each $\widetilde{\mathcal{H}}(i)$ has dimension $f_i$ representing the number of different flavors carried by the $i$-th probe matter operator.  In the second equality, we factorize $\widetilde{\mathcal{H}}^{(k)}$ into left and right factors
according to the locations of the insertions relative to the midpoint of the Euclidean boundary; we assume that no insertion lies at this midpoint.
More explicitly, let $i_0$ be the largest integer in $\{0,\ldots,k\}$ such that
\begin{equation}
    \sum_{i=1}^{i_{0}}\frac{\beta_{i}}{2}<\frac{\beta}{4},
\end{equation}
where $\beta=\sum_{j=1}^{k+1}\beta_j$. We define the left and right flavor Hilbert spaces as
\begin{equation}
    \widetilde{\mh}_{l}=\otimes_{j=1}^{i_{0}}\widetilde{\mathcal{H}}(j),\quad\widetilde{\mh}_{r}=\otimes_{j=i_{0}+1}^{k}\widetilde{\mathcal{H}}(j).
\end{equation}
For the $i$-th operator insertion, introduce its Euclidean distances to the two boundaries,
\begin{equation}
    \beta_{i,L}=\sum_{i^{\pp}=1}^{i}\beta_{i^{\pp}},\quad\beta_{i,R}=\sum_{i^{\pp}=i+1}^{k+1}\beta_{i^{\pp}},\quad\beta_{i,{\rm min}}={\rm min}\{\beta_{i,L},\beta_{i,R}\}.
\end{equation}
In the semiclassical limit followed by the heavy-probe limit, the dominant saddle selects the
shorter geodesic segment for each insertion, as in the one-particle analysis.  Flavor indices in $\widetilde{\mathcal H}_l$ contract within each replica,
producing $\widetilde\rho_r=\operatorname{Tr}_{\widetilde{\mathcal H}_l}\widetilde\rho$, while those in $\widetilde{\mathcal H}_r$ contract cyclically between replicas.
The resulting flavor factor is $\widetilde{\Tr}(\widetilde\rho_r^{\,n})$.
Substituting this factor into the replicated amplitude and taking $n\to1$ gives
\begin{equation} \label{eq:main-result2}
    S_{{\rm bdy}}(\rho_{R})=A+S(\widetilde{\rho}_{r}),
\end{equation}
where the area term is
\begin{equation}
    A=\frac{N}{2}\log2+(1-\beta\partial_{\beta})\log Z(\beta)-\sum_{i=1}^{k}\beta\partial_{\beta}\log G_{2}^{\Del_i}(\beta_{i,{\rm min}},\beta-\beta_{i,{\rm min}}),
\end{equation}
and 
\begin{equation}
    \widetilde{\rho}_{r}=\mathrm{Tr}_{\widetilde{\mathcal H}_{l}}\widetilde{\rho}
\end{equation}
is the reduced density matrix of the flavor degrees of freedom carried by the
particles lying in the entanglement wedge of the right system.

\section{Semiclassical limit of Euclidean correlation functions}\label{sec:Correlation}

\subsection{Partition function}

The partition function of double-scaled SYK is given by
\begin{equation} \label{eq:Zb-def}
    Z(\beta)
\equiv \langle 0 | e^{-\beta H} |0 \rangle = 
\int_0^\pi \rho(\theta)\dd\theta \,e^{-\beta E(\theta)},
\end{equation}
where the energy $E$ and spectral density $\rho$ are parametrized by $\theta$ as
\begin{equation}\label{eq:E-def}
    E(\theta)=\frac{2\cos\theta}{\sqrt{\lambda(1-q)}},\quad \rho(\theta)  = \frac{\mathcal{N}_q}{
    \Gamma_q (\pm\frac{2i\theta}{\lambda})
    },
\end{equation}
with the numerical constant $\mathcal N_q =(2\pi)^{-1} (1-q)^2(q ; q)_{\infty}^3$ ensuring the unit normalization of $\rho(\theta)$.
The $q$-Pochhammer symbol and the $q$-Gamma function are defined by
\be\label{eq:qpochgamma-def}
(a;q)_\infty\equiv\prod_{n=0}^{\infty}(1-aq^n),
\qquad
\Gamma_q(z)\equiv (1-q)^{1-z}\frac{(q;q)_\infty}{(q^z;q)_\infty}.
\ee
In \eqref{eq:E-def} we have used the convenient notation $\Gamma_q(\pm i x):= \Gamma_q(ix)\Gamma_q(-ix)$. Recall the following reflection formula for the $q$-Gamma function \cite{Goel:2023svz,Okuyama:2025fhi,Goto:2026ipq}: 
\begin{equation} \label{eq:reflect-formula}
\frac{1}{\Gamma_{q}(\pm2i\theta/\lambda)}	=\sqrt{\frac{2\pi}{\lambda}}\frac{q^{-\frac{1}{8}}\tilde{q}^{\frac{1}{8}}\left(\tilde{q};\tilde{q}\right)_{\infty}}{(1-q)^{2}(q;q)_{\infty}^{3}}(2\sinh\frac{2\pi\theta}{\lambda})(2\sin\theta)(\tilde{q}e^{\pm4\pi\theta/\lambda};\tilde{q})_{\infty}e^{-2\theta^{2}/\lambda},
\end{equation}
where $\tilde q :=e^{-4\pi^2/\lambda}$.
For \(0<\theta<\pi\), an explicit application of \eqref{eq:reflect-formula} shows the following leading form of the spectral density 
\begin{equation}
    \rho(\theta)=\frac{e^{-\frac{\pi^{2}}{2\lambda}}}{\sqrt{2\pi\lambda}}(2\sin\theta)e^{-\frac{2\theta^{2}-2\pi\theta}{\lambda}}\left(1+\mo(\lambda)\right).
\end{equation}
At fixed $\beta$, the endpoint regions are exponentially suppressed relative to the interior saddle. Therefore, the integral can be put into the following form 
\begin{equation}\label{eq:Z-integral}
Z(\beta)
= 
\frac{e^{-\frac{\pi^2}{2\lambda}}}{\sqrt{2\pi \lambda}}
\int_0^\pi \dd\theta\,
\exp\left[
-\frac{1}{\lambda}f_\beta(\theta)
+g(\theta)
\right],    
\end{equation}
with 
\be
f_\beta(\theta)
=
2\theta^2-2\pi\theta+2\beta\cos\theta,
\qquad
g(\theta)=\log(2\sin\theta)-\frac{\beta}{2}\cos\theta.
\ee
The saddle point equation of $\theta$ determines it as a function of the inverse temperature $\beta$. Introduce a useful parameterization $\theta =\theta_\beta\equiv\frac{\pi }{2}(1+v)$. The saddle point equation becomes 
\begin{equation} \label{eq:saddle-v}
    f_\beta^{\prime} =0 \Rightarrow \cos \frac{\pi v}{2} = \frac{\pi v}{\beta}. 
\end{equation}
The integral \eqref{eq:Z-integral} can be evaluated up to one-loop order, leading to 
\begin{equation} \label{eq:Zb-result}
    Z(\beta)=\frac{\pi v}{\beta^{3/2}}\sqrt{\frac{2}{\frac{2}{\beta}+\sin\frac{\pi v}{2}}}\exp\left[\frac{2\pi v}{\lambda}\left(\tan\frac{\pi v}{2}-\frac{\pi v}{4}\right) +\frac{\pi v}{2}\tan\frac{\pi v}{2}\right],
\end{equation}
where we have neglected $\mathcal O(\lambda)$ corrections in presenting the semiclassical result. 

As a quick consistency check, note that at infinite temperature where $\beta=v=0$, the one-loop result \eqref{eq:Zb-result} reduces to $1$.

\subsection{Two-point function}\label{ssec:2pt-func}
The thermal two-point function in DSSYK is calculated by: 
\begin{equation} \label{eq:2pt-def}
    Z(\beta)G^{\Delta}_2 (\beta_1,\beta_2) = \, \int_{0}^{\pi}\prod_{i=1}^{2}e^{-\beta_i E_{i}}\rho(\theta_{i})\dd\theta_{i} \, \gamma_{\Del}^{2}(\te_{1},\te_{2}).
\end{equation}
It can be represented as:
\begin{equation}
    Z(\beta) G_2^{\Delta} (\beta_1,\beta_2)= \begin{tikzpicture}[baseline=-0.55ex, scale=0.9]
        \draw (0,0)  circle (1);
        \draw[blue] (-1,0) -- (1,0);
        \node at (0,0.2) {\small $\Delta$};
        \fill[blue]  (-1,0) circle[radius=1.5 pt];
        \fill[blue]  (1,0) circle[radius=1.5 pt];
        \node at (0,1.25) {\small $\beta_1$};
        \node at (0,-1.25) {\small $\beta_2$};
    \end{tikzpicture}
\end{equation}
where each boundary vertex contributes a factor $\gamma_\Delta$, defined by
\be
\gamma_\Delta(\theta_1,\theta_2)
=
\left[
\frac{(q^{2\Delta};q)_\infty}
{(q^\Delta e^{\pm i\theta_1\pm i\theta_2};q)_\infty}
\right]^{1/2} = \sqrt{\mathcal{N}_{\Delta} \frac{\Gamma_q\left(\Delta \pm \frac{i \theta_1}{\lambda} \pm \frac{i \theta_2}{\lambda}\right)}{\Gamma_q(2 \Delta)}}.
\ee
The normalization $\mathcal{N}_\Delta = (1-q)^{2\Delta -3 }(q;q)^{-3}_\infty$ ensures
\begin{equation}
    \int_0^\pi \rho(\theta_1) \rho(\theta_2) \dd \theta_1\dd \theta_2 \gamma_\Delta (\theta_1,\theta_2)^2 =1. 
\end{equation}
The only property we need is its semiclassical limit when the two spectral angles differ by order \(\lambda\). Set
\be
\theta_1=\theta+\frac{\lambda\nu}{2},
\qquad
\theta_2=\theta-\frac{\lambda\nu}{2},
\qquad
\kappa_\theta\equiv \frac{2\sin\theta}{\lambda}.
\ee
The corresponding change of variables has Jacobian
\be
\dd\theta_1\,\dd\theta_2=\lambda\,\dd\theta\,\dd\nu.
\ee
Now we use the fact that for fixed \(s\), when \(q\to1\) with \(q^z\) fixed on the unit circle away from one, we have
\be\label{eq:qgamma-shift}
\frac{\Gamma_q(z+s)}{\Gamma_q(z)}
=
(1-q)^{-s}
\frac{(q^z;q)_\infty}{(q^{z+s};q)_\infty}
=
\left(\frac{1-q^z}{1-q}\right)^s
\left[1+\mathcal O(\lambda)\right].
\ee
Taking \(s=\Delta\), \(z=\pm2i\theta/\lambda\) and multiplying the two signs, we find
\(\kappa_\theta^{2\Delta}[1+\mathcal O(\lambda)]\).
Then the two boundary vertices of a single matter chord combine into
\be \label{eq:gamma-approx}
\gamma_\Delta(\theta_1,\theta_2)
\gamma_\Delta(\theta_2,\theta_1)
=
\frac{\mathcal N_q}{\rho(\theta)}
\frac{\mathcal N_\Delta}{\Gamma(2\Delta)}
\kappa_\theta^{2\Delta}
\Gamma(\Delta\pm i\nu)
\left[1+\mathcal O(\lambda)\right].
\ee 
This is the basic boundary ingredient that will be used for analyzing higher-point thermal correlators.

As for the two-point function, the Boltzmann factor
expands as
\be \label{eq:e2-approx}
e^{-\beta_1E(\theta+\lambda\nu/2)-\beta_2E(\theta-\lambda\nu/2)}
=
e^{-\beta E(\theta)}
\exp\left[(\beta_1-\beta_2) \sin \theta \nu 
\right]
\left[1+\mathcal O(\lambda)\right].
\ee
Substituting \eqref{eq:gamma-approx} and \eqref{eq:e2-approx} into \eqref{eq:2pt-def} shows that the $\theta$-integral has the same leading saddle as $Z(\beta)$. After dividing by $Z(\beta)$, its leading contribution is obtained by evaluating the remaining amplitude at $\theta=\theta_\beta$, giving $G^{\Delta}_2$ as
\be\label{eq:G2-integral-final}
G_2^{\Delta}(\beta_1,\beta_2)
=
\frac{\lambda\mathcal N_q \mathcal N_\Delta}{\Gamma(2\Delta)}
\kappa_{\theta_\beta}^{2\Delta}
\int_{\mathbb R}\dd\nu\,
e^{(\beta_1-\beta_2)\sin\theta_\beta\,\nu}
\Gamma(\Delta\pm i\nu)
+\mathcal O(\lambda),
\ee
which evaluates to
\be \label{eq:2pt-result}
G^{\Delta}_2(\beta_1,\beta_2)=
\left[
\frac{\cos\frac{\pi v}{2}}
{\cos\left(\frac{\pi v}{2\beta}(\beta_1-\beta_2)\right)}
\right]^{2\Delta}
+\mathcal O(\lambda).
\ee

For fixed $\Delta>0$, the $\mathcal O(\lambda)$ errors in the two preceding expressions for $G_2^\Delta$ are locally uniform for $\beta_1,\beta_2>0$ and remain $\mathcal O(\lambda)$ after applying $\partial_{\beta_1}$ or $\partial_{\beta_2}$.\footnote{The saddle equation gives $\pi-|\beta_1-\beta_2|\sin\theta_\beta\geq\pi(1-v)>0$. In a fixed neighborhood of $\theta_\beta$, after including the Jacobian $\lambda$ and factoring out $\rho(\theta)e^{-\beta E(\theta)}$, the $q$-Pochhammer products defining $\rho$ and $\gamma_\Delta$, together with the expansion of the Boltzmann factor, bound the error in the remaining factor by $\mathcal O(\lambda)$ times an exponentially decaying function of $|\nu|$, for $|\nu|$ up to a fixed multiple of $\log(1/\lambda)$. The remaining exact factor is exponentially suppressed at larger $|\nu|$, up to fixed powers of $\lambda^{-1}$. Splitting the $\nu$-integral at a sufficiently large fixed multiple of $\log(1/\lambda)$ therefore gives an integrated $\mathcal O(\lambda)$ error. The contribution from outside this fixed neighborhood of $\theta_\beta$ is exponentially small relative to the saddle contribution. Applying the saddle expansion at $\theta_\beta$ to the two-point integral normalized by $Z(\beta)$, with error estimates valid also after applying $\partial_{\beta_1}$ or $\partial_{\beta_2}$, gives the stated bounds.}

\subsection{A probabilistic proof of semiclassical factorization}
In this section, we prove factorization in the semiclassical limit for a fixed matter-chord diagram, with fixed positive matter weights and fixed real nonnegative Euclidean intervals. Let $D$ denote a diagram for a $2n$-point function, with each boundary segment $i\in\{1,2,\dots,2n\}$ representing Euclidean evolution by $e^{-\beta_i H}$. We will show that 
\begin{equation} \label{eq:factorization-statement}
    G^{(D)}_{2n} =\prod_{i=1}^n G_2^{\Delta_i}\left(\frac{\beta+\tau_i}{2}, \frac{\beta-\tau_i}{2}\right)+\mathcal{O}(\lambda),
\end{equation}
where $\beta = \sum_{a=1}^{2n} \beta_{a}$ and $\tau_i$ is chosen so that the two arguments $(\beta\pm \tau_i)/2$ give the total Euclidean lengths of the boundary arcs on either side of the corresponding matter chord. For example, the diagram 
\begin{equation}
    D \, = \, \begin{tikzpicture}[baseline=-0.55ex, scale=0.9]
       \draw (0,0)  circle (1);
       \draw (0:1) -- (180:1);  
       \draw (60:1) -- (280:1);
       \draw (135:1) -- (315:1);
       \node at (165:1.25) {\small $\beta_1 $}; 
       \node at (100:1.25) {\small $\beta_2 $}; 
       \node at (35:1.25) {\small $\beta_3 $}; 
       \node at (-25:1.25) {\small $\beta_4 $}; 
       \node at (-55:1.25) {\small $\beta_5 $}; 
       \node at (-135:1.25) {\small $\beta_6 $}; 
       \node at (-0.5,-0.15) {\tiny $\Delta_3$};
       \node at (115:0.55) {\tiny $\Delta_1$};
       \node at (25:0.7) {\tiny $\Delta_2$};
    \end{tikzpicture}
\end{equation}
corresponds to the correlator 
\begin{equation}
    Z(\beta)\,G^{(D)}_{6} = \langle 0 | \wick{e^{-\beta_1 H} \c1 \mo_{\Delta_1} e^{-\beta_2 H} \c2 \mo_{\Delta_2} e^{-\beta_3 H} \c3 \mo_{\Delta_3} e^{-\beta_4 H} \c1  \mo_{\Delta_1} e^{-\beta_5 H} \c2 \mo_{\Delta_2} e^{-\beta_6 H} \c3 \mo_{\Delta_3}   } |0 \rangle \,.
\end{equation}
For a general $2n$-point function with Euclidean boundary intervals $\boldsymbol{\beta} = (\beta_1,\dots,\beta_{2n})$, we expand each evolution factor in powers of $H$. Let $\boldsymbol{k}=\left(k_1, \ldots, k_{2 n}\right)$ record the numbers of Hamiltonian insertions on the boundary arcs. Only terms with an even total number of insertions, $\sum_a k_a=2 m \in 2\mathbb{Z}$, survive the disorder average, so their expansion sign is $(-1)^{2 m}=1$. Each pairing $\pi \in \operatorname{Pair}(2 m)$ of these insertions carries the nonnegative weight
\begin{equation}
    w(\boldsymbol{k}, \pi)=\lambda^{-m} q^{\operatorname{cr}(\pi)} \prod_{a=1}^{2 n} \frac{\beta_a^{k_a}}{k_{a}!} \geq 0,
\end{equation}
where $\operatorname{cr}(\pi)$ counts crossings between Hamiltonian chords. To compute the thermal correlator, we first sum over pairings for a fixed distribution among the $2n$ boundary arcs, specified by $\boldsymbol{k}$ with a total of $2m$ insertions, including the appropriate matter-chord crossing factors. We then sum over $\boldsymbol{k}$ and finally over $m$. If the matter chords are erased, this sum reduces to the partition function:
\begin{equation}
    \sum_{m=0}^{\infty} \sum_{\substack{\boldsymbol{k}:\,k_a \geq 0 \\ \sum_a k_a=2 m}} \sum_{\pi \in \operatorname{Pair}(2 m)} w(\boldsymbol{k}, \pi)=\sum_{m=0}^{\infty} \frac{\beta^{2 m}}{(2 m)!} \lambda^{-m} \sum_{\pi \in \operatorname{Pair}(2 m)} q^{\mathrm{cr}(\pi)}=Z(\beta).
\end{equation}
The first equality follows from the multinomial identity. For every fixed $\lambda>0$, the sum converges absolutely, since
\begin{equation}
    1 \leq Z(\beta) \leq \sum_{m=0}^{\infty} \frac{\beta^{2 m}}{(2 m)!} \lambda^{-m}(2 m-1)!!=e^{\beta^2 /(2 \lambda)}.
\end{equation}
Thus, $w / Z(\beta)$ defines a probability measure on the pairs $(\boldsymbol{k}, \pi)$. We denote expectation with respect to this measure by $\mathbb{E}$, leaving its dependence on $\lambda$ and the boundary intervals implicit. This measure describes the chord configurations in the disorder-averaged sum; it does not introduce a further average over microscopic disorder.

For a fixed matter chord $i$, let $K_i(\boldsymbol{k}, \pi)$ count the Hamiltonian chords intersecting it, and set
\begin{equation}\label{eq:QD-def}
    X_i=q^{\Delta_i K_i}, \quad 0 \leq X_i \leq 1, \quad Q_D=\prod_{\substack{i<j \\ i, j \text { cross in } D}} q^{\Delta_i \Delta_j}.
\end{equation}
Restoring the matter-chord crossing factors multiplies the weight of each configuration contributing to $D$ by $Q_D\prod_i X_i$. The normalized correlator therefore takes the form
\begin{equation}
 G_{2n}^{(D)}=Q_D\,\mathbb{E}\!\left[\prod_{i=1}^{n}X_i\right].
 \label{eq:exact-expectation}
\end{equation}
This representation of $G_{2n}$ associates each matter chord with a random variable $X_i$ on a common probability space. No independence assumption is needed. 

Keeping only the crossing factor $X_i$ is equivalent to retaining a single matter chord. Its mean $\mu_i$ is therefore exactly the corresponding thermal two-point function:
\begin{equation}
    \mu_i \equiv \mathbb{E} X_i=G_2^{\Delta_i}\left(\frac{\beta+\tau_i}{2}, \frac{\beta-\tau_i}{2}\right).
\end{equation}
Thus, the factorization statement \eqref{eq:factorization-statement} reduces to proving 
\begin{equation} \label{eq:difference-to-prove}
    |G^{(D)}_{2n} - \prod_{i=1}^n \mu_i|  = |Q_D \mathbb{E}\prod_{i=1}^n X_i - \prod_{i=1}^n \mu_i| = \mo(\lambda).
\end{equation}
To control this difference, we first estimate the second moment. The doubled-weight expression $X_i^2 = q^{2\Delta_i K_i}$ corresponds to a matter chord at the same location with twice the original weight. Averaging this crossing factor with respect to the normalized thermal measure defined by $w$ gives
\begin{equation}
    \mathbb{E} X_i^2=G_2^{2 \Delta_i}\left(\frac{\beta+\tau_i}{2}, \frac{\beta-\tau_i}{2}\right).
\end{equation}
Together with the semi-classical two-point function \eqref{eq:2pt-result}, this shows that the variance of $X_i$ is of order $\mo(\lambda)$:
\begin{equation}
    V_i:=\operatorname{Var}\left(X_i\right)=G_2^{2 \Delta_i}\left(\frac{\beta+\tau_i}{2}, \frac{\beta-\tau_i}{2}\right)-\left[G_2^{\Delta_i}\left(\frac{\beta+\tau_i}{2}, \frac{\beta-\tau_i}{2}\right)\right]^2=\mathcal{O}(\lambda). \label{eq:variance}
\end{equation}
The key step toward \eqref{eq:difference-to-prove} is to bound the difference between the joint expectation and the product of the means in terms of the variances $V_i$:
\begin{equation} \label{eq:bound-V}
    \left|\mathbb{E} \prod_{i=1}^n X_i-\prod_{i=1}^n \mu_i\right| \leq \sum_{i<j} \sqrt{V_i V_j}.
\end{equation}
To prove this inequality, we use $\mathbb{E}(X_i-\mu_i)=0$ and a telescoping expansion to write
\begin{equation} \label{eq:telescope}
    \mathbb{E} \prod_{i=1}^n X_i-\prod_{i=1}^n \mu_i=\sum_{i=1}^n\left(\prod_{j>i} \mu_j\right) \mathbb{E}\left[\left(X_i-\mu_i\right)\left(\prod_{j<i} X_j-\prod_{j<i} \mu_j\right)\right].
\end{equation}
Since all factors lie in $[0,1]$, the difference of the partial products satisfies $\left|\prod_{j<i} X_j-\prod_{j<i} \mu_j\right| \leq \sum_{j<i}\left|X_j-\mu_j\right|$. Taking absolute values in \eqref{eq:telescope} and applying the Cauchy--Schwarz inequality to each pair of fluctuations gives \eqref{eq:bound-V}. 

Substituting \eqref{eq:bound-V} into \eqref{eq:difference-to-prove} and using $0<Q_D \leq 1$, we obtain the exact finite-$\lambda$ bound 
\begin{equation}
    \begin{aligned}\left|G_{2 n}^{(D)}-\prod_{i=1}^n \mu_i\right| & \leq Q_D \sum_{i<j} \sqrt{V_i V_j}+\left(1-Q_D\right) \prod_{i=1}^n \mu_i \\ & \leq \sum_{i<j} \sqrt{V_i V_j}+\lambda \sum_{\substack{i<j \\ i, j \text { cross in } D}} \Delta_i \Delta_j .\end{aligned}
\end{equation}
In the last step, we used $1-e^{-x}\le x$ for $x\ge0$.
For a fixed diagram, the number of matter chords is finite. By \eqref{eq:variance}, the first term on the last line is $\mathcal O(\lambda)$, while the second is explicitly proportional to the same small parameter. This proves \eqref{eq:factorization-statement}.

\subsection{The semiclassical area term}

We now evaluate the area term in the FLM formula \eqref{eq:main-result2} using the semiclassical results. In this subsection, we focus on single-particle states. The fluctuating part of the `area', defined in \eqref{eq:deltaA-def}, naturally separates into a matter-independent contribution and a matter-dependent contribution.

The matter-independent part is fixed entirely by the thermal partition function, and therefore depends only on the temperature.\footnote{See also \cite{Blommaert:2024hologram,Blommaert:2024entropic} for a derivation of the DSSYK thermal entropy from sine dilaton gravity.} Substituting \eqref{eq:Zb-result} into \eqref{eq:deltaA-def}, we find
\begin{equation}
    \begin{aligned}
\left(1-\beta \partial_\beta\right) \log Z(\beta)= & -\frac{\pi^2 v^2}{2 \lambda}+\frac{1}{2} \log \frac{\cos ^2 \frac{\pi v}{2}}{1+\frac{\pi v}{2} \tan \frac{\pi v}{2}} \\
& +\frac{\frac{\pi^2 v^2}{2}+\pi v\left(3\tan \frac{\pi v}{2}-\pi v\right)\left(1+\frac{\pi v}{2} \tan \frac{\pi v}{2}\right)}{4\left(1+\frac{\pi v}{2} \tan \frac{\pi v}{2}\right)^2}+\mathcal{O}(\lambda) .
\end{aligned}
\end{equation}
The leading $\mathcal{O}(\lambda^{-1})$ term is negative. This reflects the fact that moving away from the infinite-temperature state lowers the entropy.

We next turn to the matter-dependent part. This contribution is determined by the two-point function and is given by
\begin{equation} \label{eq:matter-contribution}
    -\beta\partial_{\beta}
    \log G_{2}^{\Delta}(\beta-\beta_{\rm min},\beta_{\rm min})
    =
 \frac{\Delta \pi v}{1+\frac{\pi v}{2} \tan \frac{\pi v}{2}}\left[\tan \frac{\pi v}{2}  -\tan \chi\left(1+\left(\frac{\pi v}{2}-\chi\right) \tan \frac{\pi v}{2}\right)\right]+\mathcal{O}(\lambda), 
\end{equation}
where
\begin{equation}
\chi=\frac{\pi v}{2} \frac{\left|\beta_L-\beta_R\right|}{\beta_L+\beta_R}, \quad \beta_{\min }=\min \left\{\beta_L, \beta_R\right\} .
\end{equation}
Since $0\leq \chi\leq \pi v/2$, one can verify that the expression in brackets in \eqref{eq:matter-contribution} is non-negative and is strictly positive when $\beta_{\rm min}>0$. Thus the insertion of a matter operator increases $\delta A$ at the displayed semiclassical order when $\Delta>0$ and $\beta_{\rm min}>0$, even before including the entropy associated with the flavor density matrix. In the infinite-temperature limit, this contribution vanishes.

This can be seen explicitly in the high-temperature limit. Taking $\beta\to 0$, or equivalently $v\to 0$, gives
\begin{equation}
-\beta \partial_\beta \log G_2^{\Delta}\left(\beta-\beta_{\min }, \beta_{\min }\right) \sim \Delta \frac{\pi^2 v^2}{2}\left(1-\frac{\left|\beta_L-\beta_R\right|}{\beta}\right)
\end{equation}
where $\beta=\beta_L+\beta_R$. Since
$|\beta_L-\beta_R|\leq \beta_L+\beta_R=\beta$, the factor in parentheses is non-negative.

Taking the low-temperature limit ($v\to1$) at fixed positive $\beta_{\min}/\beta$, we instead obtain
\begin{equation}
-\beta \partial_\beta \log G_2^{\Delta}\left(\beta-\beta_{\min }, \beta_{\min }\right) \sim 2 \Delta\left[1-\left(\frac{\pi}{2}-\chi\right) \tan \chi\right], \quad 0 \leq \chi \leq \frac{\pi}{2} .
\end{equation}
The expression in brackets is again non-negative throughout this interval. Therefore, at the displayed semiclassical order, the matter-dependent contribution to $\delta A$ is non-negative.

\section{Semiclassical factorization from exact spectral representations}\label{sec:factor}

In this section, we derive factorization of arbitrary $2n$-point thermal correlators directly from their exact integral representations in the semi-classical limit (see also \cite{Mukhametzhanov:2023tcg}). The factorization statement \eqref{eq:factorization-statement} was already established by the probabilistic argument in section~\ref{sec:Correlation}; the purpose here is to give an alternative derivation that makes its origin in the exact spectral representation explicit. We organize the integrals so that the semi-classical limit of the $\mathcal{U}_q (\mathfrak{su}(1,1)) $ $6j$-symbol enforces energy conservation at each internal vertex, yielding the factorized result. We begin with the crossed four-point function and then extend the argument to arbitrary matter-chord configurations.

\subsection{Crossed four-point function and the bulk crossing}

We next turn to the crossed four-point function, the simplest correlator that
is sensitive to a bulk crossing of two matter chords.  The four Euclidean
boundary segments have lengths \(\beta_a\), and the four intermediate
Hamiltonian energies are parametrized by \(\theta_a\).  In the crossed channel
the exact spectral representation is
\be \label{eq:G4-def}
Z(\beta) G_{4,\mathrm{cr}} (\{\beta_a\})
=
q^{\Delta_1\Delta_2}\int
\prod_{a=1}^{4}\rho(\theta_a)\dd\theta_a\,
e^{-\sum_{a=1}^{4}\beta_aE(\theta_a)}
\prod_{a=1}^{4}
\gamma_{a,a+1}^{\Delta_{a,a+1}}
\left\{
\begin{array}{ccc}
\Delta_1 & \theta_1 & \theta_4\\
\Delta_2 & \theta_2 & \theta_3
\end{array}
\right\}_q .
\ee
The factor $q^{\Delta_1\Delta_2}$ is the direct matter--matter crossing weight.
The braces denote the
\(\mathcal U_q(\mathfrak{su}(1,1))\) \(6j\)-symbol, whose Mellin--Barnes representation is reviewed in appendix~\ref{app:MB-6j}.
It
is the new ingredient relative to the two-point function and encodes the
interaction at the intersection of the two matter chords.  The corresponding
matter-chord diagram is
\begin{equation}
   Z(\beta) G_{4,\mathrm{cr}} = \, \begin{tikzpicture}[baseline=-0.55ex, scale=0.9]
        \draw (0,0)  circle (1);
        \draw[blue] (45:1) -- (225:1) ; 
        \draw[blue] (135:1) -- (315:1);
        \node at (-1.3, 0)  {\small $\beta_1$}; 
        \node at (1.25, 0)  {\small $\beta_3$};
        \node at (0, 1.25)  {\small $ \beta_2 $};  
        \node at (0, -1.25)  {\small $\beta_4 $}; 
        \fill[blue]  (45:1) circle[radius=1.25 pt];
        \fill[blue]  (135:1) circle[radius=1.25 pt];
        \fill[blue]  (225:1) circle[radius=1.25 pt];
        \fill[blue]  (315:1) circle[radius=1.25 pt];
        \node[blue] at (205:.65) {\tiny $\Delta_1$}; 
        \node[blue] at (335:.65) {\tiny $\Delta_2$};
    \end{tikzpicture}
\end{equation}
The four boundary arcs label the four thermal evolutions, while the blue lines
join alternating operator insertions.  As in the two-point function, every
boundary endpoint contributes
$\gamma^{\Delta_{a,a+1}}_{a,a+1}
\equiv\gamma_{\Delta_{a,a+1}}(\theta_a,\theta_{a+1})$.  Here
\(\Delta_{a,a+1}\) is the weight of the matter chord ending between regions
\(a\) and \(a+1\), with the cyclic identification \(a=5\equiv1\).

To take the semiclassical limit, we separate the common energy mode from the
relative energy fluctuations.  The thermal saddle forces all four spectral
parameters to lie within an \(\mo(\lambda)\) neighborhood of a common value,
so we write
\be
\theta_a=\theta+\lambda\alpha_a,
\qquad
a=1,\ldots,4 .
\ee
Only the combinations \(\theta+\lambda\alpha_a\) are physical: a common shift
of all \(\alpha_a\)'s can be absorbed into \(\theta\).  We fix this redundancy
by imposing \(\sum_{a=1}^{4}\alpha_a=0\).  The corresponding change of
variables has the exact Jacobian
\be \label{eq:change-var}
\prod_{a=1}^{4}\dd\theta_a
=
4\lambda^3\dd\theta
\prod_{a=1}^{4}\dd\alpha_a\,
\delta\!\left(\sum_{a=1}^{4}\alpha_a\right).
\ee
We use \(\alpha_{ab}\equiv\alpha_a-\alpha_b\) for boundary energy
differences, and the alternating combination
\begin{equation}
\omega\equiv\alpha_1-\alpha_2+\alpha_3-\alpha_4
\end{equation}
for the energy mismatch at the bulk crossing.  It is convenient to
parametrize the relative energies by the oriented differences
\begin{equation} \label{eq:nu-def}
\nu_a=\alpha_{a+1}-\alpha_a,
\qquad a=1,\ldots,4,
\end{equation}
with cyclic indices.  They obey
\begin{equation}
\sum_{a=1}^{4}\nu_a=0,
\qquad
\omega=\nu_2+\nu_4=-(\nu_1+\nu_3).
\end{equation}
The constrained transformation from the \(\alpha_a\)'s to the \(\nu_a\)'s
has determinant \(4\), so \eqref{eq:change-var} becomes
\begin{equation}\label{eq:4pt-var-change}
\prod_{a=1}^{4}\dd\theta_a
=
\lambda^3\dd\theta
\prod_{a=1}^{4}\dd\nu_a\,
\delta\!\left(\sum_{a=1}^{4}\nu_a\right).
\end{equation}

The finite part of the four endpoint factors is symmetric under
\(\nu_a\to-\nu_a\).  The two dominant residues of the \(6j\)-symbol, however,
carry different finite Gamma-function phases.  Combining each residue with
the endpoint factors gives the leading integrand
\begin{equation}  \label{eq:6j-simple}
    \prod_{a=1}^4 \gamma_{a, a+1}^{\Delta_{a, a+1}}\left\{\begin{array}{lll}
\Delta_1 & \theta_1 & \theta_4 \\
\Delta_2 & \theta_2 & \theta_3
\end{array}\right\}_q \xrightarrow{\lambda \rightarrow 0} \frac{\mathcal{N}_q^2}{2 \pi \lambda \rho(\theta)^3}\left[\mathcal{S}_\theta(i \omega) \Psi_{\Delta_1 \Delta_2}^{+}\left(\nu_i\right)+\mathcal{S}_\theta(-i \omega) \Psi_{\Delta_1 \Delta_2}^{-}\left(\nu_i\right)\right],
\end{equation}
where 
\begin{align}
\mathcal{S}_\theta(i \omega) & \equiv \kappa_\theta^{i \omega} \Gamma(i \omega), \qquad \kappa_\theta=\frac{2 \sin \theta}{\lambda}, \qquad \omega=-(\nu_1+\nu_3),  \\
\Psi_{\Delta_1 \Delta_2}^{+}\left(\nu_i\right) & \equiv \left[ \prod_{i=1}^2 \frac{\mathcal{N}_{\Delta_i}}{\Gamma\left(2 \Delta_i\right)} \kappa_\theta^{2 \Delta_i}\right] \Gamma\left(\Delta_2+i \nu_1\right) \Gamma\left(\Delta_1-i \nu_2\right) \Gamma\left(\Delta_2+i \nu_3\right) \Gamma\left(\Delta_1-i \nu_4\right),
\end{align}
and $\Psi^-$ is the complex conjugate of $\Psi^+$. We leave a detailed derivation of \eqref{eq:6j-simple} to appendix~\ref{app:MB-6j}.

The density and power counting are unchanged.  The four spectral densities supply \(\rho(\theta)^4\), while \eqref{eq:6j-simple} supplies
\(\rho(\theta)^{-3}\).  The common \(\theta\)-integral therefore has the same leading saddle as the
partition-function integral and localizes at \(\theta=\theta_\beta\), producing
the factor \(Z(\beta)\) on the left-hand side of \eqref{eq:G4-def}.  The
relative-energy Jacobian contributes \(\lambda^3\), and the crossing contributes
\(\lambda^{-1}\), leaving one factor of \(\lambda\) for each complete matter
chord.

We now integrate \eqref{eq:6j-simple} against the measure \eqref{eq:4pt-var-change} and extract its leading behavior. The delta function eliminates one of the $\nu$-variables, which we take to be $\nu_4$, leaving a three-dimensional integral over $\nu_{1,2,3}$. The remaining integrand is therefore evaluated at
\begin{equation}
\nu_4=-\nu_1-\nu_2-\nu_3.
\end{equation}
Viewed as a function of complex $\nu_1$, the first branch of \eqref{eq:6j-simple} has poles from $\Gamma(i\omega)$ in the lower half-plane, including its boundary, and poles from $\Psi^+$ in the upper half-plane:
\begin{equation} \label{eq:three-familiy}
\nu_1 = - \nu_3 - i m,  \quad \nu_1= i(\Delta_2+m), 
\quad
\nu_1=-\nu_2-\nu_3+i(\Delta_1+m),
\quad m=0,1,\ldots .
\end{equation}
The second branch has the complex-conjugate pole locations.

At $\omega=-(\nu_1+\nu_3)=0$, the delta function in \eqref{eq:4pt-var-change} further implies $\nu_2=-\nu_4$, and the $\nu$-dependent factors in either waveform combine to give
\begin{equation} \label{eq:factorized-integrand}
\Gamma(\Delta_1\pm i\nu_2)
\Gamma(\Delta_2\pm i\nu_3).
\end{equation}
Consequently, the simple poles of $\Gamma(i\omega)$ and $\Gamma(-i\omega)$ cancel in the complete two-branch integrand. In particular, the full integrand \eqref{eq:6j-simple} is regular at $\omega=0$. This observation allows us to evaluate the original integral using the following contour prescription.

Choose $0<\epsilon<\min\{1,\Delta_1,\Delta_2\}$. For strictly positive Euclidean separations, first shift the $\nu_1$ contour from $\mathbb R$ to $\mathbb R+i\epsilon$, without crossing any pole of the complete integrand. We then separate the two branches and shift the contour for the first branch to $\mathbb R-i\sigma$ and that for the second branch to $\mathbb R+i\sigma$, with fixed $1<\sigma<2$. The first deformation crosses the $m=0,1$ poles of $\Gamma(i\omega)$, while the second crosses only the $m=1$ pole of $\Gamma(-i\omega)$. After factoring out the common normalization in $\Psi^\pm$, the $m=1$ residues contribute $O(\lambda)$, and the displaced-contour integrals contribute $O(\lambda^\sigma)$. The leading contribution is therefore the residue at $\omega=0$ from the first branch.

The dominant contribution to the $\nu_1$ integral can therefore be summarized by the following distributional replacement:
\begin{equation} \label{eq:replacement}
    \begin{split}
        \frac{1}{2\pi}\delta\!\left(\sum_{a=1}^4 \nu_a\right) & 
\left[\mathcal{S}_\theta(i \omega) \Psi_{\Delta_1 \Delta_2}^{+}\left(\nu_i\right)+\mathcal{S}_\theta(-i \omega) \Psi_{\Delta_1 \Delta_2}^{-}\left(\nu_i\right)
\right] \\ 
& \xrightarrow{\lambda\to 0}
\left[\prod_{i=1}^{2}\frac{\mathcal N_{\Delta_i}}{\Gamma(2\Delta_i)}\kappa_\theta^{2\Delta_i}\right]
\delta(\nu_1+\nu_3)\delta(\nu_2+\nu_4)
\Gamma(\Delta_1 \pm i \nu_2)
\Gamma(\Delta_2\pm i\nu_1).
    \end{split}
\end{equation}
The two delta functions set \(\nu_4=-\nu_2\) and \(\nu_3=-\nu_1\).  Together
with \(\sum_a\alpha_a=0\), they give
\begin{equation}
(\alpha_1,\alpha_2,\alpha_3,\alpha_4)
=
\frac{1}{2}
(-\nu_1-\nu_2,\,
 \nu_1-\nu_2,\,
 \nu_1+\nu_2,\,
-\nu_1+\nu_2).
\end{equation}
The relative part of the Boltzmann exponent is therefore
\begin{equation}
    2 \sum_{a=1}^4 \beta_a \alpha_a=\left(\beta_2+\beta_3-\beta_1-\beta_4\right) \nu_1 + \left(\beta_3+\beta_4-\beta_1-\beta_2\right) \nu_2 .
\end{equation}
Using the two-point integral in section~\ref{ssec:2pt-func}, we obtain
\begin{align}
G_{4,\mathrm{cr}}(\{\beta_a\})
&=
\left[
\frac{\lambda\mathcal N_q\mathcal N_{\Delta_1}}
{\Gamma(2\Delta_1)}
\kappa_{\theta_\beta}^{2\Delta_1}
\int_{\mathbb R}\dd\nu_2\,
e^{(\beta_3+\beta_4-\beta_1-\beta_2)
\sin\theta_\beta\,\nu_2}
\Gamma(\Delta_1\pm i\nu_2)
\right]
\nonumber\\
&\quad\times
\left[
\frac{\lambda\mathcal N_q\mathcal N_{\Delta_2}}
{\Gamma(2\Delta_2)}
\kappa_{\theta_\beta}^{2\Delta_2}
\int_{\mathbb R}\dd\nu_1\,
e^{(\beta_2+\beta_3-\beta_1-\beta_4)
\sin\theta_\beta\,\nu_1}
\Gamma(\Delta_2\pm i\nu_1)
\right]
+\mathcal{O}(\lambda)
\nonumber\\
&=
G_2^{\Delta_1}(\beta_3+\beta_4,\beta_1+\beta_2)\,
G_2^{\Delta_2}(\beta_2+\beta_3,\beta_1+\beta_4)
+\mathcal{O}(\lambda).
\label{eq:factor-4}
\end{align}
Thus the only leading effect of the bulk crossing is to conserve the relative
energy along each complete matter chord.

As a side remark, the $\alpha_a$ variables may be interpreted as energy flows assigned to the boundary segments. In this picture, the $\nu$ variables defined in \eqref{eq:nu-def} represent the energy transferred from the bulk matter lines into the boundary:
\begin{equation} \label{eq:face-to-line}
    \begin{tikzpicture}[baseline=-0.55ex, scale=0.9]
         \draw[line width=0.75pt] 
         (-60:1) 
         arc[ 
         start angle=-60, 
         end angle=60, 
         radius=1];
         \draw[blue, line width=0.75pt] (-0.3,0) -- (1,0);
          \path[ 
          postaction={
          decorate, 
          decoration={ 
          markings,
          mark=at position 0.5 
          with { 
          \arrow{Stealth[length=4pt,width=5pt]} 
          } 
          } 
          } 
          ] 
          (-60:1) arc[start angle=-60,end angle=0,radius=1];
           \path[ 
           postaction={
           decorate, 
           decoration={ 
           markings,
           mark=at position 0.5 
           with { 
           \arrow{Stealth[length=4pt,width=5pt]} 
           } 
           } 
           } 
           ] 
           (0:1) arc[start angle=0,end angle=60,radius=1];
            \path[ 
            postaction={
            decorate, 
            decoration={ 
            markings,
            mark=at position 0.5 
            with { 
            \arrow{Stealth[length=4pt,width=5pt]} 
            } 
            } 
            } 
            ] 
            (-0.3,0) -- (1,0);
            \node at (-30:1.4) {\small $\alpha_i$};
            \node at (30:1.45) {\small $ \alpha_{i+1}$};  
            \node at (0.3,0.3) {\small $\nu_i$}; 
    \end{tikzpicture}
\end{equation}
The leading pole then imposes the condition \eqref{eq:replacement}, which admits a simple interpretation: the energy carried by each matter line is conserved as it passes through the bulk crossings. This condition leads to the following parametrization of the energy assigned to the individual segments of the crossed diagram:
\begin{equation} \label{eq:assign-rule}
    \begin{tikzpicture}[baseline=-0.55ex, scale=1.1]
        \draw (0,0)  circle (1);
        \draw (-1,0) -- (1,0);
        \draw (0,-1) -- (0,1);
         \path[ 
         postaction={
         decorate, 
         decoration={ 
         markings,
         mark=at position 0.35 
         with { 
         \arrow{Stealth[length=4pt,width=5pt]} 
         } 
         } 
         } 
         ] 
         (-1,0) -- (1,0);
          \path[ 
          postaction={
          decorate, 
          decoration={ 
          markings,
          mark=at position 0.35 
          with { 
          \arrow{Stealth[length=4pt,width=5pt]} 
          } 
          } 
          } 
          ] 
          (0,-1) -- (0,1);
          \node at (0.3,-0.6) {\small $ \nu_1 $}; 
          \node at (-0.6,0.3) {\small $\nu_2$ };
          \node at (45:1.5) {\small $\frac{\nu_1-\nu_2}{2}$ };
          \node at (135:1.5) {\small $ -\frac{\nu_1+\nu_2}{2}$}; 
          \node at (-45:1.5) {\small $ \frac{\nu_1+\nu_2}{2}$}; 
          \node at (-135:1.5) {\small $ \frac{\nu_2-\nu_1}{2}$}; 
           \path[ 
           postaction={
           decorate, 
           decoration={ 
           markings,
           mark=at position 0.5 
           with { 
           \arrow{Stealth[length=4pt,width=5pt]} 
           } 
           } 
           } 
           ] 
           (90:1) arc[start angle=90,end angle=0,radius=1];
            \path[ 
            postaction={
            decorate, 
            decoration={ 
            markings,
            mark=at position 0.5 
            with { 
            \arrow{Stealth[length=4pt,width=5pt]} 
            } 
            } 
            } 
            ] 
            (0:1) arc[start angle=0,end angle=-90,radius=1];
             \path[ 
             postaction={
             decorate, 
             decoration={ 
             markings,
             mark=at position 0.5 
             with { 
             \arrow{Stealth[length=4pt,width=5pt]} 
             } 
             } 
             } 
             ] 
             (-90:1) arc[start angle=-90,end angle=-180,radius=1];
              \path[ 
              postaction={
              decorate, 
              decoration={ 
              markings,
              mark=at position 0.5 
              with { 
              \arrow{Stealth[length=4pt,width=5pt]} 
              } 
              } 
              } 
              ] 
              (180:1) arc[start angle=180,end angle=90,radius=1];
    \end{tikzpicture}
\end{equation}  
where the orientations are chosen so that energy is conserved at every boundary and bulk vertex, with the total incoming energy flow balancing the total outgoing flow. This provides a useful graphical interpretation of the parametrization of the boundary segments in terms of the $\nu$ variables and explains how it reproduces the correct time configuration in the factorized integral \eqref{eq:factor-4}. In the next section, we extend this assignment to a general $2n$-point function.

We conclude that, in the semiclassical limit, the $\mathcal{U}_q(\mathfrak{su}(1,1))$ $6j$-symbol reduces to the universal form \eqref{eq:6j-simple}, which closely resembles the Dray--'t Hooft $S$-matrix recently reviewed in \cite{Lam_2018}. The leading contribution arises from the zero-exchange-energy pole and causes the crossed four-point function to factorize into the corresponding product of two-point functions.

\subsection{General \(2n\)-point function}

We now apply the preceding residue analysis to a general diagram $D$ representing a $2n$-point thermal correlator with a fixed contraction pattern of boundary insertions. Let $F$ be the set of faces of $D$, and let $E$ be the set of chord segments obtained by cutting every matter chord at its crossings. We denote the set of bulk crossings by $V_{\mathrm{int}}$ and the set of boundary endpoints by $V_{\rm ext}$. We also use these symbols for the numbers of elements in the corresponding sets. For a $2n$-point function, $|V_{\mathrm{ext}}|=2n$. Boundary arcs are not included in $E$. We label the boundary endpoints cyclically by $a=1,\ldots,2n$, and let $\beta_a$ be the length of the arc from endpoint $a$ to endpoint $a+1$, with $2n+1$ identified with $1$. Thus
\be
\beta=\sum_{a=1}^{2n}\beta_a .
\ee
Assign a spectral parameter \(\theta_f\) to each face. If \(f(a)\) denotes the
face adjacent to the boundary arc \(a\), the spectral representation takes the
schematic form
\be\label{eq:Z2n-def}
Z(\beta)G_{2n}^{(D)}
=
Q_D\int \prod_{f\in F}\rho(\theta_f)\dd\theta_f\,
\prod_{a=1}^{2n}e^{-\beta_aE(\theta_{f(a)})}
\prod_{a=1}^{V_{\rm ext}}
\gamma_{a,a+1}^{\Delta_{a,a+1}}
\prod_{v\in V_{\rm int}}\mathcal R_v ,
\ee
where $Q_D$ is the matter--matter crossing factor defined in Eq.~\eqref{eq:QD-def} and satisfies $Q_D=1+\mathcal O(\lambda)$ for a fixed diagram and fixed matter weights.
Here, we integrate over the spectral parameters $\theta_f$ associated with all
faces, while only the faces adjacent to the boundary carry Boltzmann weights
$e^{-\beta_aE(\theta_{f(a)})}$. The factor $\gamma_{a,a+1}$ is the boundary
vertex coupling the two faces that meet at the corresponding matter endpoint,
whereas $\mathcal R_v$ is the quantum $6j$-symbol associated with the four
faces surrounding the crossing $v$. In what follows, we will need only the
semiclassical limits of these factors.

As before, we separate the common thermal mode from the relative face
energies:
\be
\theta_f=\theta+\lambda\alpha_f,
\qquad
\sum_{f\in F}\alpha_f=0.
\ee
Up to an overall power of \(\lambda\), which is absorbed into the normalization
of the relative-energy measure, this gives
\be \label{eq:measure-general}
\prod_{f\in F}\rho(\theta_f)\dd\theta_f
\propto 
\lambda^{F-1}\rho(\theta)^F\dd\theta
\prod_{f\in F}\dd\alpha_f\,
\delta\!\left(\sum_{f\in F}\alpha_f\right)
\left[1+\mathcal O(\lambda)\right],
\ee
with the constant numeric prefactor fixed in the following discussion. The Boltzmann factor must be expanded at the same stage. Since
\be
E(\theta+\lambda\alpha_f)
=E(\theta)-2\sin\theta\,\alpha_f+\mathcal O(\lambda),
\ee
we find
\be \label{eq:Boltzmann-general}
\prod_{a=1}^{2n}e^{-\beta_aE(\theta_{f(a)})}
=
e^{-\beta E(\theta)}
\exp\!\left[
2\sin\theta\sum_{a=1}^{2n}\beta_a\alpha_{f(a)}
\right]
\left[1+\mathcal O(\lambda)\right].
\ee
This expression records which Euclidean boundary interval is adjacent to each
face. As we will see below, it reduces to the familiar difference between the
two thermal intervals associated with each complete matter chord.

The spectral density $\rho$ in each endpoint factor is evaluated at the mean of its two adjacent face angles. Since, for fixed $a$,
\be
\frac{\rho(\theta+\lambda a)}{\rho(\theta)}
=e^{-2(2\theta-\pi)a}[1+\mathcal O(\lambda)],
\ee
the shifts of these means contribute at order one. The external vertices therefore contribute
\be \label{eq:vertex-general}
\begin{aligned}
\prod_{a=1}^{V_{\rm ext}}
\gamma_{a,a+1}^{\Delta_{a,a+1}}
&=
\frac{\mathcal{N}_q^{n}}{\rho(\theta)^{n}}
\exp\!\left[
\frac{2\theta-\pi}{2}
\sum_{a=1}^{2n}\bigl(\alpha_{f(a-1)}+\alpha_{f(a)}\bigr)
\right]\\
&\quad\times
\prod_{i=1}^{n}
\left[
\frac{\mathcal N_{\Delta_i}}{\Gamma(2\Delta_i)}
\kappa_\theta^{2\Delta_i}
\right]
\prod_{a=1}^{V_{\rm ext}}
\left[
\Gamma\!\left(\Delta_{a,a+1}\pm i\alpha_{a,a+1}\right)
\right]^{1/2}
\left[1+\mathcal O(\lambda)\right],
\end{aligned}
\ee
where \(\alpha_{a,a+1}\) is the oriented difference between the fluctuations
of the two faces adjacent to endpoint \(a\), and
\(\Gamma(\Delta\pm ix)\equiv
\Gamma(\Delta+ix)\Gamma(\Delta-ix)\).

At a bulk vertex \(v\in V_{\mathrm{int}}\), label the four surrounding faces
cyclically by \(f_1, f_2,f_3,f_4 \in F\). We define the exchange energy at $v$
by
\be
\begin{tikzpicture}[baseline=-0.55ex, scale=1.1]
    \draw (45:1) -- (225:1);
    \draw (135:1)--(-45:1);
      \fill[blue]  (0,0) circle[radius=1.2 pt];
      \node at (.3,0) {\small $v$}; 
      \node at (.8,0) {\small $f_3 $}; 
      \node at (-.8,0) {\small $f_1 $}; 
      \node at (0,.8) {\small $ f_2 $}; 
      \node at (0,-.8) {\small $f_4 $}; 
\end{tikzpicture}
\quad \longrightarrow\quad \omega_v=\alpha_{f_1}-\alpha_{f_2}+\alpha_{f_3} -\alpha_{f_4}.
\ee
In the semiclassical limit, the corresponding $6j$-symbol reduces, according to \eqref{eq:6j-symbol-limit}, to an expression involving the scattering $S$-matrix $\mathcal{S}(i\omega_v)$. As in the crossed four-point function analysis, the dominant contribution to the relative-energy integral arises from the pole at $\omega_v=0$. Evaluating the integral by taking the residue at this pole is equivalent to replacing $\mathcal{R}_v$ by the constraint
\begin{equation} \label{eq:bulk-crossing-general}
\mathcal R_v
\xrightarrow[\text{Dominant Residue}]{\lambda\to 0}
\frac{1}{\lambda\rho(\theta_v)}\,\delta(\omega_v)=
\frac{1}{\lambda\rho(\theta)}
\exp\!\left[
\frac{2\theta-\pi}{2}\sum_{j=1}^{4}\alpha_{f_j}
\right]\delta(\omega_v)\,[1+\mathcal O(\lambda)].
\end{equation}
Here $\theta_v=\frac14\sum_{j=1}^{4}\theta_{f_j}=\theta+\frac{\lambda}{4}\sum_{j=1}^{4}\alpha_{f_j}$ is the mean of the four adjacent face angles.

For fixed $n$, the number of crossings is finite. We assume that each $\mathcal R_v$ is integrated against functions with sufficient regularity in $\omega_v$ as $\lambda\to0$ so that the distributional replacement in \eqref{eq:bulk-crossing-general} can be taken inside the remaining integrals over the relative-energy variables in \eqref{eq:measure-general}. This uniformity assumption is natural for positive Euclidean boundary separations, since the endpoint factors decay exponentially at large relative energy. It then follows that, at leading order, the product of crossing factors imposes
\begin{equation} \label{eq:bulk-crossing-replacement-all}
\prod_{v\in V_{\rm int}} \mathcal{R}_v
\longrightarrow
\left(
\frac{1}{\lambda \rho(\theta)}
\right)^{|V_{\rm int}|}
\exp\!\left[
\frac{2\theta-\pi}{2}
\sum_{v\in V_{\rm int}}\sum_{j=1}^{4}\alpha_{f_j}
\right]
\prod_{v\in V_{\rm int}}\delta(\omega_v).
\end{equation}
For each $v$, the inner sum runs over its four adjacent faces.
We will show below that these constraints are independent and derive their consequences explicitly.

We now evaluate \eqref{eq:Z2n-def} by using
\eqref{eq:measure-general}, \eqref{eq:Boltzmann-general},
\eqref{eq:vertex-general}, and \eqref{eq:bulk-crossing-replacement-all}. We first
isolate the terms of order $\exp(\mathcal{O}(\lambda^{-1}))$, which determine the saddle point. These terms arise from the
Boltzmann factor and the spectral density $\rho$. The
total power of $\rho$ is
\be
\rho(\theta)^{F-V_{\rm int}-V_{\rm ext}/2}.
\ee
Because the diagram is drawn on a disk and \(E\) counts only chord segments,
the Euler relation for the disk gives
\be \label{eq:Euler-general}
F-E+V_{\rm int}=1.
\ee
The four-valency of each bulk crossing and the one-valency of each boundary
endpoint further imply
\be \label{eq:structure-general}
4V_{\rm int}+V_{\rm ext}=2E.
\ee
Together, these relations give
\be \label{eq:Euler-2}
F=n+V_{\rm int}+1.
\ee
The face measures, endpoint factors, and crossing factors therefore leave
one power of \(\rho(\theta)\), while their net power of \(\lambda\) is
\(\lambda^{F-1-V_{\rm int}}=\lambda^n\).  Thus the common
\(\theta\)-integral has the same leading saddle as the partition-function integral \eqref{eq:Zb-def}, and one factor of
\(\lambda\) remains for each complete matter chord.

We now show directly that the crossing constraints are independent and
leave one relative-energy variable for each complete matter chord.

To see this directly, let \(e\) be a chord segment, with \(f_L(e)\) and
\(f_R(e)\) denoting the faces on its two sides. Left and right are defined
relative to an arbitrarily chosen orientation of the segment; in the
illustration below, the segment is oriented to the right. We then define the
discrete gradient
\be \label{eq:line-assignment}
\begin{tikzpicture}[baseline=-0.55ex, scale=1.1]
    \draw (-2,0) -- (2,0);
    \draw (-1.2,-1) -- (-1.2,1);
    \draw (1.2,-1) -- (1.2,1);
     \path[ 
     postaction={
     decorate, 
     decoration={ 
     markings,
     mark=at position 0.5 
     with { 
     \arrow{Stealth[length=4pt,width=5pt]} 
     } 
     } 
     } 
     ] 
     (-2,0) -- (2,0);
     \node at (.5,0.2) {\small $ \nu_e$}; 
     \node at (-.2,0.8) {\small $f_L(e) $}; 
     \node at (-.2,-.8) {\small $ f_R(e) $}; 
\end{tikzpicture} 
\quad \longrightarrow\quad \nu_e=\alpha_{f_L(e)}-\alpha_{f_R(e)}.
\ee
Around $v$, denote the four face variables by $\alpha_{1,2,3,4}$. Using the
same convention as in \eqref{eq:line-assignment}, we assign
$\nu_v^{(1,2,3,4)}$ to the four segments that intersect at $v$:
\be \label{eq:nu-v-def}
\nu_v^{(1)}=\alpha_1-\alpha_2,
\quad
\nu_v^{(2)}=\alpha_2-\alpha_3,
\quad
\nu_v^{(3)}=\alpha_3-\alpha_4,
\quad
\nu_v^{(4)}=\alpha_4-\alpha_1.
\ee
These variables satisfy
\be
\nu_v^{(1)}+\nu_v^{(2)}+\nu_v^{(3)}+\nu_v^{(4)}=0, \quad \omega_v = \nu_v^{(1)}+\nu_v^{(3)}.
\ee
Together with the first relation, the condition $\omega_v = 0$ implies
\be \label{eq:vertex-conservation-general}
\nu_v^{(1)}+\nu_v^{(3)}=0,
\qquad
\nu_v^{(2)}+\nu_v^{(4)}=0.
\ee

When both strands are oriented through a crossing,
\eqref{eq:vertex-conservation-general} says that the jump of
\(\alpha_f\) is unchanged along either strand.  It therefore assigns one
conserved variable \(\nu_i\) to all segments of the same complete chord.

Choose an orientation for each complete matter chord \(i\), and define
\(\sigma_i(f)=+1\) when the face \(f\) lies on its left and
\(\sigma_i(f)=-1\) when it lies on its right.  The solutions of the crossing constraints are
\begin{equation}\label{eq:alpha-general-solution}
\alpha_f
=
\frac{1}{2}
\sum_{i=1}^{n}\left[
\sigma_i(f)-\frac1F\sum_{g\in F}\sigma_i(g)
\right]\nu_i .
\end{equation}
Indeed, the four faces around a crossing of chords \(i\) and \(j\)
carry, up to cyclic order, the four sign pairs
\((+,+),(+,-),(-,-),(-,+)\), so
\eqref{eq:alpha-general-solution} obeys \(\omega_v=0\).
Conversely, \eqref{eq:vertex-conservation-general} says that the jump
of \(\alpha_f\) is constant along every complete chord.  After subtracting
the right-hand side of \eqref{eq:alpha-general-solution}, no jump remains
across any chord segment.  Since the face-adjacency graph is connected,
the difference is a common constant, which vanishes by the condition $\sum_f\alpha_f=0$.

This construction also proves that the \(V_{\rm int}\) crossing
conditions are independent.  Since \(F=n+V_{\rm int}+1\), imposing them
and removing the common shift leaves exactly \(n\) variables, one
\(\nu_i\) for each complete chord.  There is no additional
diagram-dependent Jacobian.  To see this, add the chords one at a time.
A chord crossing \(k\) existing chords introduces \(k+1\) segment jumps;
replacing them by one conserved jump and the \(k\) successive exchanged energies
\(\omega_v\) is triangular with unit determinant.  Midpoint and jump
variables for each newly split face also have unit determinant.

For each matter chord define
\begin{equation}\label{eq:tau-general}
\tau_i
=
\sum_{a=1}^{2n}
\beta_a\,\sigma_i(f(a)).
\end{equation}
The two boundary arcs between the endpoints of chord \(i\) then have
lengths \((\beta+\tau_i)/2\) and \((\beta-\tau_i)/2\).  Substituting
\eqref{eq:alpha-general-solution} into \eqref{eq:Boltzmann-general} gives
\begin{equation}
2\sum_{a=1}^{2n}\beta_a\alpha_{f(a)}
=
\sum_{i=1}^{n}\tau_i\nu_i
-\frac{\beta}{F}\sum_{g\in F}\sum_{i=1}^{n}\sigma_i(g)\nu_i.
\end{equation}
At either endpoint of chord \(i\), the adjacent faces differ only in
the sign of \(\sigma_i\).  Its two endpoint Gamma-function factors therefore combine
into \(\Gamma(\Delta_i\pm i\nu_i)\).

Applying the Euler relation to the two regions separated by each complete chord shows that the additional endpoint and crossing factors multiply to
\begin{equation}
\exp\!\left[
\frac{2\theta-\pi}{F}
\sum_{g\in F}\sum_{i=1}^{n}\sigma_i(g)\nu_i
\right].
\end{equation}
Combining them with the additional Boltzmann term gives
\begin{equation}
\exp\!\left[
\frac{2\theta-\pi-\beta\sin\theta}{F}
\sum_{g\in F}\sum_{i=1}^{n}\sigma_i(g)\nu_i
\right],
\end{equation}
which equals one at the thermal saddle, where $2\theta_\beta-\pi=\beta\sin\theta_\beta$.

We therefore obtain
\begin{equation}
\begin{aligned}
G_{2n}^{(D)}
&=
\prod_{i=1}^{n}
\left[
\frac{\lambda\mathcal N_q\mathcal N_{\Delta_i}}
{\Gamma(2\Delta_i)}
\kappa_{\theta_\beta}^{2\Delta_i}
\int_{\mathbb R}\dd\nu_i\,
e^{\tau_i\sin\theta_\beta\,\nu_i}
\Gamma(\Delta_i\pm i\nu_i)
\right]
+\mathcal{O}(\lambda)
\\
&=
\prod_{i=1}^{n}
G_2^{\Delta_i}
\left(
\frac{\beta+\tau_i}{2},
\frac{\beta-\tau_i}{2}
\right)
+\mathcal{O}(\lambda).
\end{aligned}
\end{equation}
Thus every fixed pairing contributes the product of the corresponding
two-point functions.  Summing over pairings gives the
generalized-free-field Wick expansion of the Euclidean \(2n\)-point
function in the semiclassical limit.

\section{Discussion and outlook}

In this work, we studied the entanglement entropy of partially entangled thermal states in double-scaled SYK, allowing the matter insertions to carry an arbitrary flavor density matrix $\widetilde{\rho}$. Our main result is that, in the semiclassical regime followed by the heavy-probe limit, the boundary entropy takes the form
\begin{equation}
S_{\mathrm{bdy}}(\rho_R) =A+S(\widetilde{\rho}_r).
\end{equation}
Here $A$ is independent of the flavor density matrix and agrees with the area-like contribution obtained for the corresponding flavorless PETS. The density matrix $\widetilde{\rho}_r$ is the reduced density matrix of the matter degrees of freedom associated with insertions lying closer to the right Euclidean boundary. We interpret these degrees of freedom as belonging to the entanglement wedge of the right boundary theory. The second term therefore admits the interpretation of a bulk matter entropy. In this sense, the microscopic DSSYK calculation realizes the characteristic structure of the FLM formula.

An important feature of the calculation is that the wedge assignment is determined directly by the dominant microscopic Wick-contraction pattern. As an insertion is moved across a transition point, its flavor index changes from being contracted within a single replica to being contracted between adjacent replicas, producing a sharp transition in the boundary entropy. This provides a microscopic account of how the bulk matter entropy enters the boundary entropy, without assuming an entanglement wedge or the FLM formula at the outset.

A central technical ingredient in our analysis is the semiclassical factorization of general DSSYK thermal correlation functions. For diagrams without matter-matter crossings, this factorization follows relatively directly from the saddle-point representation of the two-point functions. We gave a probabilistic proof of this factorization for contributions to these correlators from arbitrary fixed matter-chord diagrams, including crossed ones. This proof reduces the factorization of these contributions to the two-point function by associating each matter chord with a bounded random variable whose variance vanishes in the semiclassical limit.

We also gave a complementary explanation of this factorization for general crossed diagrams by analyzing the semiclassical behavior of the quantum $6j$-symbol in an appropriate Mellin--Barnes representation. At leading order, the pole conditions enforce conservation of the energy parameters flowing along each matter chord, and the resulting expression factorizes into a product of semiclassical two-point functions. The graph-theoretic formulation of this argument applies to general $2n$-point functions and demonstrates that the emergence of the FLM-like entropy structure is not restricted to a particular ordering of the insertions.

Several aspects of this construction merit further investigation. First, it would be useful to identify the area-like term $A$ more directly with a gravitational observable. A derivation of $A$ from the effective bulk theory, including its one-loop contribution and its dependence on the insertion locations, would sharpen the connection with gravitational generalized entropy. It may also clarify whether the fluctuating part of $A$ can be promoted to a nontrivial operator when the code subspace is enlarged beyond the matter flavor degrees of freedom considered here.

Second, the present results are obtained by taking
\begin{equation}
\lambda\to0\quad(\Delta_i>0\text{ fixed}),\qquad\text{then}\qquad\Delta_i\to\infty,
\end{equation}
with the insertion locations and flavor dimensions held fixed.
The first limit ensures semiclassical factorization of general DSSYK thermal correlators, and the second makes the competing Wick contractions sharply distinguishable. Matter backreaction is suppressed when $\lambda\Delta_i\ll1$. It would be interesting to relax these assumptions. At large but finite $\Delta_i$, the sharp wedge transition is smoothed over a parametrically small region of insertion space. When $\lambda\Delta_i$ is held finite, matter crossings contribute at leading order and the insertions are expected to backreact on the semiclassical geometry. Understanding this regime may provide a microscopic model of quantum extremal surfaces in the presence of appreciable matter backreaction.

A related question is whether the factorization derived here admits a systematic expansion in $\lambda$. Such corrections should encode interactions among the semiclassical matter propagators, as well as fluctuations of the emergent geometry. The exact quantum-group structure of DSSYK provides a concrete framework in which these effects may be tractable. It would be valuable to compare the resulting corrections with loop effects computed directly in the bulk description.

Another natural direction is to study the full family of Rényi entropies rather than only the von Neumann limit. Away from $n=1$, different replica contraction patterns need not be governed by the same saddle, and additional transitions may occur as functions of $n$, the insertion positions, and the pattern of flavor entanglement. The resulting Rényi spectrum could probe more detailed features of the entanglement-wedge transition and may provide a microscopic realization of replica-wormhole saddles in DSSYK.

Finally, the present results suggest a natural algebraic interpretation. The double-scaled algebra admits a tracial type-$\mathrm{II}_1$ representation, while the additional flavor degrees of freedom generate a finite-dimensional matter algebra. It would be interesting to characterize the inclusion of the wedge algebra into the full boundary algebra and to formulate the entropy relation in terms of conditional expectations, relative entropy, and modular theory.

Taken together, our results demonstrate that the characteristic separation between an area-like contribution and a bulk matter entropy can emerge directly from a microscopic boundary calculation. DSSYK therefore provides a useful solvable setting in which geometric, algebraic, and microscopic descriptions of entropy can be compared within a single framework.

\section*{Acknowledgment}
We thank Sergio E. Aguilar-Gutierrez, Ahmed Almheiri,  Henry Lin, Hong Liu,   Alexey Milekhin, Vladimir Narovlansky, Geoffrey R. Penington, Douglas Stanford, Elisa Tabor, and Herman Verlinde for helpful discussions. This material is based upon work supported by the U.S. Department of Energy, Office of Science, Office of High Energy Physics, under Award Number DE-SC0011702, by funds from the University of California, and by the John Templeton Foundation under Grant Number 63670.

\appendix

\section{Mellin--Barnes representation of the quantum $R$-matrix} \label{app:MB-6j}
In this appendix we review the Mellin--Barnes representation of the quantum $6j$-symbol and identify its dominant poles in the semiclassical limit.  We show that their leading contribution reduces to the universal Dray--'t Hooft scattering $S$-matrix in \eqref{eq:6j-limit}\footnote{See \cite{Lam_2018} for a similar discussion on the classical SL$(2,\mathbb{R})$ 6j symbol.}.  More importantly, the exact residue representation gives systematic control over corrections to this result to arbitrary perturbative order in $\lambda$.  The two subdominant pole families are exponentially suppressed, whereas the descendants of the two dominant poles enter at progressively higher powers of $\lambda$.  The eikonal kernel \eqref{eq:6j-limit} is therefore the unique leading term of a controlled semiclassical expansion, and all corrections to it vanish in the strict $\lambda\to0$ limit.

The Mellin--Barnes representation for the quantum $6j$-symbol, given in eq.~(6.3) of~\cite{Berkooz:2018jqr}, reads:
\be\label{eq:MB-integral}
\left\{ \begin{array}{lll}
\Delta_{1} & \te_{1} & \te_{4}\\
\Delta_{2} & \te_{2} & \te_{3}
\end{array}\right\} _{q}=N_{R}e^{i\vp}\int_{\mathcal{C}}\frac{\dd u}{2\pi i}q^{u-\Del_{2}}e^{i\te_{34}}f_{q}(u).
\ee
We use the following notation hereafter:
\be
\theta_{ij}= \theta_i -\theta_j, \quad \theta_{i+j}= \theta_i + \theta_j.
\ee
The prefactor $N_R$ is given by:
\be \label{eq:NR-def}
N_{R}= \frac{1}{2\pi(1-q)\mathcal{N}_{q}}\frac{\Gamma_{q}\left(1-(\Del_{1}-\Del_{2}+i\frac{\te_{13}}{\lambda})\right)\Gamma_{q}\left(\Del_{1}-\Del_{2}+i\frac{\te_{13}}{\lambda}\right)}{\Gamma\left(1-(\Del_{1}-\Del_{2}+i\frac{\te_{13}}{\lambda})\right)\Gamma\left(\Del_{1}-\Del_{2}+i\frac{\te_{13}}{\lambda}\right)},
\ee
and the phase factor $e^{i\vp}$ is given by:\footnote{The square root is chosen continuously from coincident real spectral angles, where $e^{i\varphi}=1$.}
\be
e^{i \varphi}=\sqrt{\frac{\Gamma_q\left(\Delta_1-\frac{i \theta_1 \pm i \theta_4}{\lambda}\right) \Gamma_q\left(\Delta_1+\frac{i \theta_3 \pm i \theta_2}{\lambda}\right) \Gamma_q\left(\Delta_2+\frac{i \theta_1 \pm i \theta_2}{\lambda}\right) \Gamma_q\left(\Delta_2-\frac{i \theta_3 \pm i \theta_4}{\lambda}\right)}{\Gamma_q\left(\Delta_1+\frac{i \theta_1 \pm i \theta_4}{\lambda}\right) \Gamma_q\left(\Delta_1-\frac{i \theta_3 \pm i \theta_2}{\lambda}\right) \Gamma_q\left(\Delta_2-\frac{i \theta_1 \pm i \theta_2}{\lambda}\right) \Gamma_q\left(\Delta_2+\frac{i \theta_3 \pm i \theta_4}{\lambda}\right)}}.
\ee 
The integrand $f_q(u)$ is defined via 
\be
f_{q}(u)=\frac{\Gamma\left(u+1-A_{2}\right)\Gamma\left(u+1-A_{1}\right)}{\Gamma_{q}\left(u+1-A_{2}\right)\Gamma_{q}\left(u+1-A_{1}\right)}\prod_{i=1}^{4}\Gamma_{q}\left(a_{i}+u\right)\frac{\Gamma\left(A_{1}-u\right)\Gamma\left(A_{2}-u\right)}{\Gamma_{q}\left(B_{1}+u\right)\Gamma_{q}\left(B_{2}+u\right)},
\ee
where the parameters in the argument of the gamma functions are
\be \label{eq:parameter-value}
\begin{aligned}
a_1 & =-i \frac{\theta_{1+2+3-4}}{\lambda}, \quad a_2=-i \frac{\theta_{1+3-2-4}}{\lambda}, \\
a_3 & =2 i \frac{\theta_4}{\lambda}, \quad a_4=0, \\
A_1 & =\Delta_1+\frac{i \theta_{14}}{\lambda}, \quad A_2=\Delta_2+\frac{i \theta_{34}}{\lambda}, \\
B_1 & =\Delta_1-\frac{i \theta_{14}}{\lambda}, \quad B_2=\Delta_2-\frac{i \theta_{34}}{\lambda}.
\end{aligned}
\ee
Note that, for real values of $\theta_i$, the parameters $a_i$ are purely imaginary, whereas the parameters $A_{i},B_i$ have positive real parts. The function $f_q(u)$ has six families of poles. The four families, defined by $-a_i -n+2\pi i\ell/\lambda,\, i=1,2,3,4,\,  n=0,1,\dots,\, \ell\in\mathbb Z$, begin on the imaginary axis and extend into the left half-plane. The other two families, labeled by $A_j+n,\, j=1,2,\, n=0,1,2\dots $, begin at real part $\Delta_{1}$ or $\Delta_2$ and extend into the right half-plane. The contour $\mathcal{C}$ is defined to separate these two classes of poles. In practice, the integral can be evaluated as
\be
\int_{\mathcal{C}} \frac{\dd u}{2 \pi i} = \int_{c-i \infty}^{c+i\infty} \frac{\dd u}{2\pi i} ,\quad 0< c < \min\{\Delta_1,\Delta_2\},
\ee
where we focus on the case with positive matter weights. When one of the matter weights vanishes, the $6j$-symbol simplifies so that the crossed four-point function reduces to a two-point function, as shown in \cite{Berkooz:2018jqr}.

\subsection{Closing the contour to the left half-plane}

In the mathematical literature~\cite{koekoek1996askeyscheme}, one usually closes the contour in the right half-plane and picks up the two corresponding families of residues, yielding a sum of two basic hypergeometric ${}_4\phi_3$ series. For our purposes, however, it is more convenient to close the contour to the left. In this form, the structure and magnitude of the residues are easier to control, and the semiclassical analysis becomes more transparent.

To justify this contour deformation, we must show that the contribution from the closing contour in the left half-plane vanishes. This is nontrivial because the factor $q^u$ in the integrand of~\eqref{eq:MB-integral} grows exponentially as ${\rm Re}\,u\to -\infty$. We must therefore show that the remaining factor $f_q(u)$ decays sufficiently rapidly along the chosen contour. To this end, we study $f_q(u)$ on the rectangular contour defined by
\be
\begin{aligned}
& \gamma_1: u=x+i T, \quad-u_0 \leq x \leq c, \\
& \gamma_2: u=-u_0+i t, \quad-T \leq t \leq T, \\
& \gamma_3: u=x-i T, \quad-u_0 \leq x \leq c,
\end{aligned}
\ee
where $u_0, T>0$, and $\gamma_1$ and $\gamma_3$ have large imaginary part $\pm T$, while $\gamma_2$ lies at large negative real part $-u_0$.

Using the reflection formula for $\Gamma$-functions, we may rewrite part of $f_q(u)$ as
\be
\prod_{i=1,2}\Gamma(u+1-A_{i})\Gamma(A_{i}-u)=\frac{\pi^{2}}{\sin\pi(A_{1}-u)\sin\pi(A_{2}-u)}.
\ee
Together with the definition of $\Gamma_q$, this gives
\be \label{eq:f-simpler}
f_{q}(u)=\frac{\pi^{2}(1-q)^{2}}{\sin\pi(A_{1}-u)\sin\pi(A_{2}-u)}\prod_{i=1}^{4}\frac{\left(q^{\beta_{i}+u};q\right)_{\infty}}{\left(q^{\alpha_{i}+u};q\right)_{\infty}},
\ee
where
\be
\alpha_{i}=a_{i},\quad i=1,2,3,4,
\qquad
\beta_{1,2}=1-A_{1,2},
\qquad
\beta_{3,4}=B_{1,2},
\ee
and these parameters satisfy
\be
2+\sum_{i=1}^{4}\alpha_{i}=\sum_{i=1}^{4}\beta_{i},
\ee
as follows from~\eqref{eq:parameter-value}.

Since the $q$-Pochhammer symbol is a product of factors of the form $(1-q^x)$, and $0<q=e^{-\lambda}<1$, the triangle inequality implies
\be
|1-q^x|\le |1+q^{{\rm Re}\,x}|.
\ee
This leads to the useful bound
\be \label{eq:ineq-1}
\left|(q^{x};q)_{\infty}\right|\le (-q^{{\rm Re}\,x};q)_{\infty}.
\ee
For the $q$-Pochhammer symbols in the denominator of $f_q(u)$, one must in addition avoid poles when choosing the contour. A convenient estimate is
\be\label{eq:inequ-2}
\frac{1}{|(q^{x};q)_{\infty}|}
=
\left|\frac{(q^{x^{*}};q)_{\infty}}{\left(q^{x},q^{x^{*}};q\right)_{\infty}}\right|
\le
\frac{(-q^{{\rm Re}\,x};q)_{\infty}}{(q^{{\rm Re}\,x\pm i{\rm Im}\,x};q)_{\infty}},
\ee
where $x^*$ denotes the complex conjugate of $x$. The denominator here is
\be
(q^{{\rm Re}\,x\pm i{\rm Im}\,x};q)_{\infty}
=
\prod_{n=0}^{\infty}
\left(1+q^{2 ( {\rm Re}\,x+n) }-2q^{n+{\rm Re}\,x}\cos(\lambda\,{\rm Im}\,x)\right),
\ee
which is nonzero for generic values of ${\rm Im}\,x$.

Applying~\eqref{eq:ineq-1} and~\eqref{eq:inequ-2} to~\eqref{eq:f-simpler}, we obtain
\be
\left|\prod_{i=1}^{4}\frac{(q^{u+\beta_{i}};q)_{\infty}}{(q^{u+\alpha_{i}};q)_{\infty}}\right|
\le
\prod_{i=1}^{4}
\frac{\left(-q^{{\rm Re}(u+\alpha_{i})},-q^{{\rm Re}(u+\beta_{i})};q\right)_{\infty}}
{\left(q^{{\rm Re}(u+\alpha_{i})\pm i{\rm Im}(u+\alpha_{i})};q\right)_{\infty}}.
\ee

We now estimate the integrand of~\eqref{eq:MB-integral} along the contour. For $u\in\gamma_1,\gamma_3$, one finds
\be
\left|q^{u-\Del_2}\prod_{i=1}^{4}\frac{(q^{u+\beta_{i}};q)_{\infty}}{(q^{u+\alpha_{i}};q)_{\infty}}\right|
\le
q^{-u_{0}-\Del_2}
\frac{\left(-q^{-u_{0}},-q^{-u_{0}+\rho_{\Del}};q\right)_{\infty}^{4}}
{\prod_{i=1}^{4}\left(q^{{\rm Re}(u)\pm i{\rm Im}(u+\alpha_i)};q\right)_{\infty}},
\ee
where we have used ${\rm Re}\,\alpha_i=0$ and introduced
\be
\rho_{\Del}
=
\min_{i}\{{\rm Re}\,\beta_i\}
=
\min\{1-\Del_{1},1-\Del_{2},\Del_{1},\Del_{2}\}.
\ee
For fixed $u_0$, this bound is periodic in $T$. We therefore take $T=T_0+ 2 n\pi/\lambda$, choosing $T_0$ to avoid denominator zeros on both horizontal segments.

The decisive suppression instead comes from the sine factors in~\eqref{eq:f-simpler}. Indeed, for ${\rm Im}\,u=\pm T$,
\be
\frac{1}{\prod_{i=1,2}|\sin\pi(A_i-u)|}
\le C_1\,e^{-2\pi T},
\ee
where $C_1$ is a bounded function of ${\rm Re}\,u\in[-u_0,c]$. Therefore, along $\gamma_1$ and $\gamma_3$, the integrand is bounded by a bounded function of ${\rm Re}\,u\in[-u_0,c]$ times the exponentially decaying factor $e^{-2\pi T}$.  For any fixed $u_0$, one may thus choose $T$ large enough that the contributions from the horizontal segments become arbitrarily small, and they vanish in the limit $T\to\infty$.

For the contour $\gamma_2$, a more careful estimate of the ratio of $q$-Pochhammer symbols is required, since an individual factor $(q^u;q)_\infty$ can become exponentially large when ${\rm Re}\,u$ is large and negative. To obtain a useful bound in this regime, we use the decomposition
\be \label{eq:etac-def}
\left(q^{u+c};q\right)_{\infty}=(q^{u+c};q)_{N_{c}}(q^{\eta_c };q)_{\infty},
\ee
where $(a;q)_N\equiv\prod_{n=0}^{N-1}(1-aq^n)$, $c$ denotes one of the parameters $\alpha_i$ or $\beta_i$, and
\be
\eta_c=u+c+N_c.
\ee
We choose the integer $N_c$ so that the real part of $\eta_c$ lies in the interval $(0,1)$:
\be
0<\rho_c={\rm Re}(\eta_c)<1.
\ee
This decomposition separates off the dominant behavior at large negative ${\rm Re}\,u$, which is encoded in $N_c$, from a bounded remainder depending on $\eta_c$.

The second factor in~\eqref{eq:etac-def} can then be bounded as
\be \label{eq:bound-eta-1}
(q^{\rho_c};q)_\infty
\le
\left|(q^{\eta_c};q)_\infty\right|
\le
(-q^{\rho_c};q)_\infty.
\ee
To estimate the first factor in~\eqref{eq:etac-def}, we use
\be
\left(q^{u+c};q\right)_{N_c}
=
\left(-q^{u+c}\right)^{N_c}q^{\binom{N_c}{2}}\left(q^{1-\eta_c};q\right)_{N_c},
\ee
which makes the exponentially large dependence on $N_c$ explicit. This gives the bound
\be \label{eq:bound-eta-2}
q^{\binom{N_c}{2}+N_c\,{\rm Re}(u+c)}(q^{1-\rho_c};q)_\infty
\le
\left|\left(q^{u+c};q\right)_{N_c}\right|
\le
q^{\binom{N_c}{2}+N_c\,{\rm Re}(u+c)}(-q^{1-\rho_c};q)_\infty.
\ee
Combining~\eqref{eq:bound-eta-1} and~\eqref{eq:bound-eta-2}, we obtain
\be
q^{\binom{N_c}{2}+N_c\,{\rm Re}(u+c)}(q^{1-\rho_c},q^{\rho_c};q)_\infty
\le
\left|\left(q^{u+c};q\right)_\infty\right|
\le
q^{\binom{N_c}{2}+N_c\,{\rm Re}(u+c)}(-q^{1-\rho_c},-q^{\rho_c};q)_\infty,
\ee
which isolates the large exponential dependence on $N_c$, and hence on ${\rm Re}\,u$. Applying this estimate to $f_q(u)$, we find
\be
\prod_{i=1}^{4}\left|\frac{(q^{u+\beta_i};q)_\infty}{(q^{u+\alpha_i};q)_\infty}\right|
\le
q^E
\prod_{i=1}^{4}
\frac{(-q^{\rho_{\beta_i}},-q^{1-\rho_{\beta_i}};q)_\infty}
{(q^{\rho_{\alpha_i}},q^{1-\rho_{\alpha_i}};q)_\infty},
\ee
where
\be
E=
\sum_{i=1}^{4}
\left[
\binom{N_{\beta_i}}{2}+N_{\beta_i}{\rm Re}(u+\beta_i)
-
\binom{N_{\alpha_i}}{2}-N_{\alpha_i}{\rm Re}(u+\alpha_i)
\right].
\ee
Along $\gamma_2$, where ${\rm Re}\,u=-u_0$, we may use the identity
\be
\frac{N_c(N_c-1)}{2}+{\rm Re}(u+c)N_c
=
-\frac{1}{2}\left(u_0-{\rm Re}\,c\right)^2
-\frac{1}{2}\left(u_0-{\rm Re}\,c\right)
+\frac{1}{2}\rho_c(\rho_c-1),
\ee
from which it follows that
\be
E=2u_0+\mathcal O(1), \qquad u_0\gg 1.
\ee
Therefore,
\be
|q^{u-\Delta_2}f_q(u)|
\le
C_2\,q^{u_0+\mathcal O(1)}
\prod_{i=1}^4 \frac{\left(-q^{\rho_{\beta_i}},-q^{1-\rho_{\beta_i}} ; q\right)_{\infty}}{\left(q^{\rho_{\alpha_i}}, q^{1-\rho_{\alpha_i}} ; q\right)_{\infty}},
\qquad \forall\,u\in\gamma_2,
\ee
where $C_2$ is an upper bound for the prefactor multiplying the $q$-Pochhammer ratio in~\eqref{eq:f-simpler}. We choose the vertical segment so that the fractional parts $\rho_{\alpha_i},\rho_{\beta_i}$ stay in a compact subinterval of $(0,1)$. Along such a sequence of contours, the remaining $q$-Pochhammer factors are uniformly bounded. Since $\beta_j=1-A_j$ for $j=1,2$, this condition keeps $\Delta_j+u_0$ away from integers. Retaining the inverse sine factors in~\eqref{eq:f-simpler} gives
\be
|q^{u-\Delta_2}f_q(u)|
\le
C_3 q^{u_0}e^{-2\pi|t|},
\qquad u=-u_0+it,
\ee
and hence
\be
\left|\int_{\gamma_2}\frac{\dd u}{2\pi i}\,
q^{u-\Delta_2}f_q(u)\right|
\le
C_4q^{u_0}.
\ee
For fixed $q$ and external parameters, $C_3,C_4$ are independent of $T$ and $u_0$ along this sequence. Taking $T\to\infty$ along the chosen sequence at fixed $u_0$, and then $u_0\to\infty$, makes all three closing integrals vanish. This justifies closing the original Mellin--Barnes contour to the left and summing the enclosed residues in this order.

\subsection{The phase factor in the \(q\to 1\) limit}

In this subsection we analyze the phase $e^{i\varphi}$ of the quantum \(6j\)-symbol in the semi-classical limit in~\eqref{eq:MB-integral}. The main point is that, in the $\lambda \to 0^+$ limit with $q=e^{-\lambda}$,  the leading contribution to the phase reduces to a simple power of $\kappa_\theta = \frac{2\sin\theta}{\lambda}$. This phase will play an important role later.

We start from the exact expression of the phase factor
\begin{equation}
\begin{aligned}
e^{i\varphi}
&=
\sqrt{
\frac{
\Gamma_q\!\left(\Delta_1-\frac{2 i \theta}{\lambda}-i \alpha_{1+4}\right)
\Gamma_q\!\left(\Delta_1+\frac{2 i \theta}{\lambda}+i \alpha_{3+2}\right)
\Gamma_q\!\left(\Delta_2+\frac{2 i \theta}{\lambda}+i \alpha_{1+2}\right)
\Gamma_q\!\left(\Delta_2-\frac{2 i \theta}{\lambda}-i \alpha_{3+4}\right)
}{
\Gamma_q\!\left(\Delta_1+\frac{2 i \theta}{\lambda}+i \alpha_{1+4}\right)
\Gamma_q\!\left(\Delta_1-\frac{2 i \theta}{\lambda}-i \alpha_{3+2}\right)
\Gamma_q\!\left(\Delta_2-\frac{2 i \theta}{\lambda}-i \alpha_{1+2}\right)
\Gamma_q\!\left(\Delta_2+\frac{2 i \theta}{\lambda}+i \alpha_{3+4}\right)
}
}\\
&\qquad\times
\sqrt{
\frac{
\Gamma_q\!\left(\Delta_1-i \alpha_{14}\right)
\Gamma_q\!\left(\Delta_1+i \alpha_{32}\right)
\Gamma_q\!\left(\Delta_2+i \alpha_{12}\right)
\Gamma_q\!\left(\Delta_2-i \alpha_{34}\right)
}{
\Gamma_q\!\left(\Delta_1+i \alpha_{14}\right)
\Gamma_q\!\left(\Delta_1-i \alpha_{32}\right)
\Gamma_q\!\left(\Delta_2-i \alpha_{12}\right)
\Gamma_q\!\left(\Delta_2+i \alpha_{34}\right)
}
}.
\end{aligned}
\label{eq:phase-factor-start}
\end{equation}
The second square root is independent of $\theta/\lambda$, hence contributes only an \(\mo(\lambda^0)\) phase in the limit of interest. The nontrivial \(\lambda\)-dependence is entirely encoded in the first square root. We now extract this leading dependence by studying ratios of \(\Gamma_q\)-functions whose arguments contain the large imaginary pieces \(\pm 2i\theta/\lambda\).

We use the definition \eqref{eq:qpochgamma-def} to derive
\be
\sqrt{\frac{\Gamma_{q}(x+\frac{2i\theta}{\lambda}+i\alpha)\Gamma_{q}(x-\frac{2i\theta}{\lambda}-i\beta)}{\Gamma_{q}(x-\frac{2i\theta}{\lambda}-i\alpha)\Gamma_{q}(x+\frac{2i\theta}{\lambda}+i\beta)}}=\left(\frac{2\sin\theta}{\lambda}\right)^{i\left(\alpha-\beta\right)}\left[1+\mo(\lambda)\right],
\label{eq:qgamma-phase-ratio}
\ee
where \(x,\alpha,\beta\) are held fixed as \(\lambda\to0\), and \(0<\theta<\pi\) is kept away from the endpoints. To see this, write each \(\Gamma_q\) in terms of \(q\)-Pochhammer symbols. The factors \((q;q)_\infty\) cancel, while the explicit powers of \(1-q\) give \((1-q)^{-2i(\alpha-\beta)}\). The remaining \(q\)-Pochhammer ratio is estimated by:
\be
\log (a e^{-i\lambda \alpha};e^{-\lambda})_\infty
-
\log (a e^{-i\lambda \beta};e^{-\lambda})_\infty
=
-i(\alpha-\beta)\log(1-a)+\mo(\lambda),
\ee
with \(a=e^{\mp 2i\theta}\). Combining the two signs gives
\be
(1-e^{-2i\theta})(1-e^{2i\theta})=4\sin^2\theta,
\ee
which proves~\eqref{eq:qgamma-phase-ratio}. The choice of branch is the one
obtained by continuing from \(0<\theta<\pi\).

We can now apply~\eqref{eq:qgamma-phase-ratio} to the first square root in~\eqref{eq:phase-factor-start}. The two pairs of \(\Gamma_q\)-functions give the exponents
\be
\alpha_{2+3}-\alpha_{1+4},
\qquad
\alpha_{1+2}-\alpha_{3+4}.
\ee
Since
\be
\alpha_{2+3}-\alpha_{1+4}+\alpha_{1+2}-\alpha_{3+4}=2(\alpha_{2}-\alpha_{4}),
\ee
we obtain
\be
e^{i\vp}	=\kappa_\theta^{2i\alpha_{24}}e^{i\chi}+\mo\left(\lambda\right),
\label{eq:phase-factor-limit}
\ee
where the remaining \(\lambda^0\) phase is obtained by replacing the \(\Gamma_q\)-functions in the second square root by ordinary \(\Gamma\)-functions:
\be
e^{i\chi}\equiv\sqrt{\frac{\Gamma\left(\Delta_{1}-i\alpha_{14}\right)\Gamma\left(\Delta_{1}+i\alpha_{32}\right)\Gamma\left(\Delta_{2}+i\alpha_{12}\right)\Gamma\left(\Delta_{2}-i\alpha_{34}\right)}{\Gamma\left(\Delta_{1}+i\alpha_{14}\right)\Gamma\left(\Delta_{1}-i\alpha_{32}\right)\Gamma\left(\Delta_{2}-i\alpha_{12}\right)\Gamma\left(\Delta_{2}+i\alpha_{34}\right)}}.
\ee

\subsection{Dominant poles and all-orders semiclassical control}

We now identify which pole families of the Mellin--Barnes integrand survive at
leading order as \(\lambda\to0\).  After closing the contour to the left, the
possible residues come from the four numerator \(q\)-Gamma functions in
\begin{equation} \label{eq:fq-def}
    f_{q}(u)=\frac{\Gamma\left(u+1-A_{2}\right)\Gamma\left(u+1-A_{1}\right)}{\Gamma_{q}\left(u+1-A_{2}\right)\Gamma_{q}\left(u+1-A_{1}\right)}\prod_{i=1}^{4}\Gamma_{q}\left(a_{i}+u\right)\frac{\Gamma\left(A_{1}-u\right)\Gamma\left(A_{2}-u\right)}{\Gamma_{q}\left(B_{1}+u\right)\Gamma_{q}\left(B_{2}+u\right)},
\end{equation}
namely
\be
-a_{i}-n+\frac{2\pi i\ell}{\lambda},\quad n=0,1,2,\dots,\quad \ell\in\mathbb Z,
\ee
with \(i=1,2,3,4\).  We first consider the poles with $\ell=0$; the contributions from poles with $\ell\ne0$ are estimated below. At fixed \(n\), all four families have the same local
residue from the \(q\)-Gamma function that is at the pole:
\be
{\rm Res}(\Gamma_{q}(a_{i}+u),\quad u=-a_{i}-n)=\frac{1-q}{\lambda}\frac{(-1)^{n}q^{n(n+1)/2}}{[n]_{q}!},
\ee
where \([n]_{q}!\equiv\prod_{m=1}^{n}\frac{1-q^{m}}{1-q}\).
Moving to the next pole in a given family does not introduce an additional
power of \(\lambda\):
\be
\frac{{\rm Res}(\Gamma_{q}(a_{i}+u),\,u=-a_{i}-n-1)}{{\rm Res}(\Gamma_{q}(a_{i}+u),\,u=-a_{i}-n)}=-\frac{q^{n+1}(1-q)}{1-q^{n+1}}\simeq\mo(\lambda^{0})
\ee
for \(n=\mo(\lambda^0)\).  Thus the residue of $\Gamma_q(a_i+u)$ does not
itself suppress descendants by powers of $\lambda$. Any such suppression
comes from the remaining factors in the complete residue.

The \(q\)-Gamma recursion first shows that lowering a large-imaginary
argument of order $\mathcal{O}(\lam^{-1})$ by one unit costs a factor \(1-q\simeq\lambda\):
\begin{equation}
\frac{\Gamma_q(\frac{i\alpha}{\lambda}+u-1)}
{\Gamma_q(\frac{i\alpha}{\lambda}+u)}
=
\frac{1-q}{1-e^{-i\alpha}q^{u-1}}
=
\mathcal O(\lambda).
\end{equation}
To compare different pole families, we  also
need the absolute large-argument behavior.  For fixed \(x\) and
\(0<\alpha<2\pi\), expanding the products in
\eqref{eq:qpochgamma-def} gives
\begin{equation}
  \log\left|\Gamma_q\!\left(x+\frac{i\alpha}{\lambda}\right)\right|  =
-\frac{\alpha(2\pi-\alpha)}{4\lambda}
+\mathcal O(\log\lambda).
\label{eq:qgamma-large-argument}
\end{equation}
Phases outside this interval are understood modulo \(2\pi\).  Thus a
\(q\)-Gamma function with \(e^{-i\alpha}\neq1\) is exponentially suppressed
as \(\lambda\to0\).

By contrast, the ratios \(\Gamma/\Gamma_q\) in the prefactor of
\eqref{eq:fq-def} change by factors of order one under the corresponding shifts.  Indeed,
\be
\frac{\Gamma(A+u+n+1)}{\Gamma_{q}(B+u+n+1)}
\bigg/
\frac{\Gamma(A+u+n)}{\Gamma_{q}(B+u+n)}
=
\frac{(1-q)(A+u+n)}{1-q^{B+u+n}}.
\ee
At the semiclassical pole locations with
\begin{equation}
{\rm Im}\,u=\frac{\theta}{\lambda},
\end{equation}
one finds
\be
\frac{(1-q)(A+u+n)}{1-q^{B+u+n}}
=
\frac{i\theta}{1-e^{-i\theta}}
+\mathcal O(\lambda)
=
\mathcal O(\lambda^0),
\ee
provided the denominator is not tuned to a pole.  This estimate controls
fixed integer shifts of these ratios within a pole family.

The residue at a pole in the \(k\)-th family is
\begin{equation}
    \begin{split}
        {\rm Res}(f_{q}(u),u & =-a_{k}-n)=\frac{1-q}{\lambda}\frac{(-1)^{n}q^{n(n+1)/2}}{[n]_{q}!}\prod_{i=1}^{2}\frac{\Gamma(1-A_{i}-a_{k}-n)}{\Gamma_{q}(1-A_{i}-a_{k}-n)}\\
 & \quad\times\frac{\Gamma\left(A_{1}+a_{k}+n\right)\Gamma\left(A_{2}+a_{k}+n\right)}{\Gamma_{q}\left(B_{1}-a_{k}-n\right)\Gamma_{q}\left(B_{2}-a_{k}-n\right)}\prod_{j=1,j\not=k}^{4}\Gamma_{q}(a_{j}-a_{k}-n).
    \end{split}
    \label{eq:f-residue}
\end{equation}
For the separate $a_1$ and $a_3$ residue estimates below, we assume fixed $\theta\ne\pi/2$, as holds at the thermal saddle for fixed $0<\beta<\infty$.
In the scaling \(\theta_i=\theta+\lambda\alpha_i\), the
parameters \(a_2\) and \(a_4\) are finite, whereas \(a_1\) and \(a_3\) contain
the large pieces \(\mp 2i\theta/\lambda\).  The summed bound
\eqref{eq:remote-summed-bound} establishes that the $a_1$ and $a_3$
families are exponentially suppressed.

Consequently, the only potentially leading residue families are
\be
u=-a_{2}-n\quad\text{or}\quad u=-a_{4}-n,
\qquad n=0,1,2,\dots .
\ee
There is one further simplification inside each of these two families.  Consider
the \(a_2\) family and denote its residue by
\(\mathcal R^{(2)}_n={\rm Res}(f_q(u),u=-a_2-n)\).  Comparing the \(n\)-th
residue to the leading one gives
\begin{align}
\frac{\mathcal R^{(2)}_n}{\mathcal R^{(2)}_0}
&=
\frac{(-1)^n q^{n(n+1)/2}}{[n]_q!}
\prod_{i=1}^{2}\Biggl[
\frac{\Gamma(1-A_i-a_2-n)}{\Gamma(1-A_i-a_2)}
\frac{\Gamma_q(1-A_i-a_2)}{\Gamma_q(1-A_i-a_2-n)}
\nonumber\\
&\quad\times
\frac{\Gamma(A_i+a_2+n)}{\Gamma(A_i+a_2)}
\frac{\Gamma_q(B_i-a_2)}{\Gamma_q(B_i-a_2-n)}\Biggr]
\prod_{j\neq2}
\frac{\Gamma_q(a_j-a_2-n)}{\Gamma_q(a_j-a_2)} .
\label{eq:a2-residue-ratio}
\end{align}
The prefactor, the ordinary Gamma ratios, and the $q$-Gamma function ratios involving the
finite argument \(a_4-a_2=-a_2\) are all of order \(\lambda^0\) for fixed
\(n\).
Suppression instead comes from the two large-imaginary $q$-Gamma ratios, for which
\begin{equation}
\frac{\Gamma_q(a_j-a_2-n)}{\Gamma_q(a_j-a_2)}
=
\prod_{m=1}^{n}
\frac{1-q}{1-q^{a_j-a_2-m}}
=
\mathcal O(\lambda^n),
\qquad j=1,3,
\label{eq:a2-large-descendant-suppression}
\end{equation}
which follows from the fact that \(q^{a_j-a_2}\) stays away from one.  Hence, for each fixed \(n\),
\begin{equation}
\frac{\mathcal R^{(2)}_n}{\mathcal R^{(2)}_0}
=
\mathcal O(\lambda^{2n}),
\qquad n\geq1.
\end{equation}
The same argument applies to the \(-a_4-n\) family, with \(a_2\) and \(a_4\)
interchanged.  Therefore only the \(n=0\) poles \(u=-a_2\) and \(u=-a_4\)
contribute at leading semiclassical order.

The residue hierarchy also controls the corrections to this result beyond
leading order.  For generic fixed external energies,
the two dominant families have the order-by-order structure
\begin{equation}
\frac{\mathcal R^{(k)}_n}{\mathcal R^{(k)}_0}
=
\lambda^{2n}
\left(c^{(k)}_{n,0}+c^{(k)}_{n,1}\lambda+c^{(k)}_{n,2}\lambda^2+\cdots\right),
\qquad k=2,4,
\label{eq:residue-all-orders-power-counting}
\end{equation}
where the coefficients are finite functions of the fixed kinematic variables.
They can be obtained systematically by expanding the exact $q$-Gamma ratios
appearing in the residues.  Consequently, at any prescribed relative order
$\lambda^K$, only the finitely many descendants with
$n\leq\lfloor K/2\rfloor$ can contribute.  The bound \eqref{eq:uniform-all-orders-tail}
controls the large-$n$ tail, so it does not modify this
fixed-order perturbative counting.

The $a_1$ and $a_3$ families are even smaller.  With their residues denoted
analogously, the summed bound \eqref{eq:remote-summed-bound} implies,
for each fixed $n$,
\begin{equation}
\frac{|\mathcal R^{(k)}_n|}
{|\mathcal R^{(2)}_0|+|\mathcal R^{(4)}_0|}
=
\mathcal O\!\left(e^{-c_\theta/\lambda}\right),
\qquad k=1,3, \qquad c_\theta>0,
\end{equation}
when $\theta$ is kept away from the endpoints.  These contributions are
beyond all orders in perturbation theory.  Thus the Mellin--Barnes residue
sum does not merely recover the eikonal answer at leading order: it organizes
all finite-$\lambda$ corrections into a controlled asymptotic expansion.  The corrections from the two $n=0$ residues and their
prefactors can be computed successively, while the $n\geq1$ descendants begin
at the orders fixed by \eqref{eq:residue-all-orders-power-counting}.  Every
such correction is parametrically suppressed and may be consistently ignored
in the strict semiclassical limit.  This power counting is not uniform in
endpoint or additional double-scaling regimes, where
$q^{a_j-a_k}\to1$ and the assumption used in
\eqref{eq:a2-large-descendant-suppression} must be reconsidered. 

Since the finite differences such as \(\alpha_{14}\) are
held fixed and do not scale with \(\lambda\), the corresponding \(q\)-Gamma
functions can be replaced by ordinary Gamma functions at this order. It is important, however, not to discard the large-imaginary \(q\)-Gamma
functions that remain inside the two dominant residues.  Using the definition of density $\rho$, we write  
\be
\Gamma_q\left(\frac{2i\theta}{\lambda}\right)
\Gamma_q\left(-\frac{2i\theta}{\lambda}\right)
=\frac{\mathcal N_q}{\rho(\theta)}.
\ee
With \(\sum_a\alpha_a=0\), the relevant parameters become
\be
a_2=-i\omega,\qquad a_4=0,\qquad
a_1=-\frac{2i\theta}{\lambda}+2i\alpha_4,\qquad
a_3=\frac{2i\theta}{\lambda}+2i\alpha_4,
\ee
where \(\omega=\alpha_1-\alpha_2+\alpha_3-\alpha_4\).  The large pair left at each
leading pole is therefore
\be
\begin{aligned}
u=-a_4=0:\qquad
\Gamma_q(a_1)\Gamma_q(a_3)
&=
\frac{\mathcal N_q}{\rho(\theta)}\,
\kappa_\theta^{4i\alpha_4}
\left[1+\mo(\lambda)\right],
\\
u=-a_2:\qquad
\Gamma_q(a_1-a_2)\Gamma_q(a_3-a_2)
&=
\frac{\mathcal{N}_q}{\rho(\theta)}\,
\kappa_\theta^{-4i\alpha_2}
\left[1+\mo(\lambda)\right].
\end{aligned}
\label{eq:residue-large-pairs}
\ee
Combining these powers with the large phase in
\eqref{eq:phase-factor-limit}, and using
\(\omega=-2(\alpha_2+\alpha_4)\), gives
\(\kappa_\theta^{-i\omega}\) at \(u=-a_4\) and
\(\kappa_\theta^{i\omega}\) at \(u=-a_2\).  The remaining finite \(q\)-Gamma ratios have nontrivial
\(\mathcal O(\lambda^0)\) limits.  At \(n=0\), they are
\begin{align}
u=-a_2:\quad&
\frac{
\Gamma(\Delta_1-i\alpha_{32})
\Gamma(\Delta_2-i\alpha_{12})
}{
\Gamma(\Delta_1+i\alpha_{32})
\Gamma(\Delta_2+i\alpha_{12})
}  \equiv e^{i\vartheta_+},\\
u=-a_4:\quad&
\frac{
\Gamma(\Delta_1+i\alpha_{14})
\Gamma(\Delta_2+i\alpha_{34})
}{
\Gamma(\Delta_1-i\alpha_{14})
\Gamma(\Delta_2-i\alpha_{34})
} \equiv e^{i\vartheta_-},
\label{eq:finite-residue-ratios}
\end{align}
where the gamma ratios are phase factors; therefore,  $\vartheta_{\pm}$ thus defined are real.  
The remaining factor \(q^{u-\Delta_2}e^{i\theta_{34}}\) in
\eqref{eq:MB-integral} is \(1+\mathcal O(\lambda)\) at both dominant poles,
while the normalization \eqref{eq:NR-def} becomes
\begin{equation}
N_R=
\frac{1}{2\pi \lambda\mathcal N_q}
\left[1+\mathcal O(\lambda)\right].
\end{equation}
The leading \(6j\)-symbol is therefore
\begin{align}
&\left\{
\begin{array}{ccc}
\Delta_1 & \theta_1 & \theta_4\\
\Delta_2 & \theta_2 & \theta_3
\end{array}
\right\}_q  \xrightarrow{\lambda \to 0} 
\frac{1}{2\pi\lambda\rho(\theta)}
\Bigg[ e^{i\chi + i \vartheta_+}
\kappa_\theta^{i\omega}\Gamma(i\omega)+  e^{i\chi + i \vartheta_- }\kappa_\theta^{-i\omega}\Gamma(-i\omega)
\Bigg].
\label{eq:6j-symbol-limit}
\end{align}
For real relative energies, the two complete branch phases are complex
conjugates; they are not a common factor away from \(\omega=0\).

As in section~\ref{ssec:2pt-func}, the four external vertices have the
symmetric semiclassical limit
\begin{align}
\prod_{a=1}^{4}\gamma_{a,a+1}^{\Delta_{a,a+1}}
&=
\frac{\mathcal N_q^2}{\rho(\theta)^2}
\prod_{i=1}^{2}
\left[
\frac{\mathcal N_{\Delta_i}}{\Gamma(2\Delta_i)}
\kappa_\theta^{2\Delta_i}
\right]
\nonumber\\
&\quad\times
\left[
\Gamma(\Delta_2\pm i\nu_1)
\Gamma(\Delta_1\pm i\nu_2)
\Gamma(\Delta_2\pm i\nu_3)
\Gamma(\Delta_1\pm i\nu_4)
\right]^{1/2}
\left[1+O(\lambda)\right].
\label{eq:vertex-limit}
\end{align}
Keeping this symmetric form avoids absorbing the finite phase into a chiral
endpoint product.  Multiplying \eqref{eq:6j-symbol-limit} by
\eqref{eq:vertex-limit} gives
\begin{equation} \label{eq:6j-limit}
    \prod_{a=1}^{4}\gamma_{a,a+1}^{\Delta_{a,a+1}}\left\{ \begin{array}{lll}
\Delta_{1} & \theta_{1} & \theta_{4}\\
\Delta_{2} & \theta_{2} & \theta_{3}
\end{array}\right\} _{q}\xrightarrow{\lambda\to0}\frac{\mathcal{N}_{q}^{2}}{2\pi\lambda\rho(\theta)^{3}}\left[\mathcal{S}_{\theta}(i\omega)\Psi_{\Del_{1}\Del_{2}}^{+}\left(\nu_{i}\right)+\mathcal{S}_{\theta}(-i\omega)\Psi_{\Del_{1}\Del_{2}}^{-}(\nu_{i})\right],
\end{equation}
where we have defined the scattering $S$-matrix $\mathcal{S}_\theta$ and waveform as:
\begin{align}
\mathcal{S}_\theta(i \omega) & \equiv \kappa_\theta^{i \omega} \Gamma(i \omega), \qquad \kappa_\theta=\frac{2 \sin \theta}{\lambda}, \qquad \omega=-(\nu_1+\nu_3),  \\
\Psi_{\Delta_1 \Delta_2}^{+}\left(\nu_i\right) & \equiv \left[ \prod_{i=1}^2 \frac{\mathcal{N}_{\Delta_i}}{\Gamma\left(2 \Delta_i\right)} \kappa_\theta^{2 \Delta_i}\right] \Gamma\left(\Delta_2+i \nu_1\right) \Gamma\left(\Delta_1-i \nu_2\right) \Gamma\left(\Delta_2+i \nu_3\right) \Gamma\left(\Delta_1-i \nu_4\right),
\end{align}
and $\Psi^{-}_{\Delta_1\Delta_2}= \overline{\Psi^+_{\Delta_1\Delta_2}}$.

Equation~\eqref{eq:6j-limit} is therefore the leading term of the exact
residue expansion rather than an independent eikonal ansatz.  At any fixed
perturbative order, the subleading terms are determined by a finite set of
descendant residues together with the corresponding asymptotic expansion of
the two dominant residues and the external vertices.  This makes manifest
that finite-$\lambda$ corrections are computable to arbitrary
order, while none survives at leading order in the semiclassical limit. Furthermore, in the next subsection we bound the sum of the subleading residues and the error in the leading semiclassical approximation, thereby justifying the semiclassical results \eqref{eq:6j-symbol-limit} and \eqref{eq:6j-limit}.

Note that $\omega=0$  sets
$\nu_4=-\nu_2$ and $\nu_3=-\nu_1$, so the $\nu$-dependent factors in either waveform in \eqref{eq:6j-limit} combine to give
$\Gamma(\Delta_1\pm i\nu_2)\Gamma(\Delta_2\pm i\nu_1)$.

\subsection{Bound on corrections}
\label{ssec:paired-bounds}

In this section, we justify the replacement rule \eqref{eq:6j-symbol-limit} by controlling the corrections in the semi-classical regime. Specifically, we show that the normalized difference between the integral of the exact $6j$-symbol and that of its semi-classical approximation vanishes in the semi-classical limit. 

To make the statement precise, let 
\begin{equation}
    F_q(u)=N_R q^{u-\Delta_2} e^{i \varphi+i \theta_{34}} f_q(u), \quad K(\omega, \lambda)=\int_{\mathcal{C}} \frac{d u}{2 \pi i} F_q(u),
\end{equation}
where $K(\omega,\lambda)$ is the exact quantum $6j$-symbol. The allowed range of $\omega$ is determined by requiring all energy parameters $\theta_a = \theta_\beta + \lambda \alpha_a$ to lie in $(0,\pi)$. We denote this range by $D_\lambda$:
\begin{equation} \label{eq:D-def-1}
    D_{\lambda}=\left\{ \omega\in\mathbb{R}:0<\theta+\lambda\alpha_{a}(\omega)<\pi\text{ for every }a\right\} .
\end{equation}
This interval can be written as $D_{\lambda}=(\omega_{-},\omega_{+})$, where 
\begin{equation} 
\begin{aligned} & \omega_{-}=\max\left(-\frac{4\theta}{\lambda}+2|\alpha_{13}|,-\frac{4(\pi-\theta)}{\lambda}+2|\alpha_{24}|\right),\\
 & \omega_{+}=\min\left(\frac{4(\pi-\theta)}{\lambda}-2|\alpha_{13}|,\frac{4\theta}{\lambda}-2|\alpha_{24}|\right).
\end{aligned}
\end{equation}
The condition in \eqref{eq:D-def-1} requires $\omega_- \leq \omega_+$. In the special case $\alpha_{13}=\alpha_{24}=0$, the interval $D_\lambda$ reduces to 
\begin{equation}
    D_{\lambda}=\left(-\frac{4d}{\lambda},\frac{4d}{\lambda}\right),\quad d=\min(\theta,\pi-\theta).
\end{equation}

We denote the individual residue contributions of $F_q(u)$ and their pairings by
\begin{equation} \label{eq:PT-def}
 \begin{split}
 T_{n,\ell}^{(k)} (\omega,\lambda)
 &=\mathop{\rm Res}_{u=-a_k-n+2\pi i\ell/\lambda}F_q(u),\\
 P_{n,\ell}(\omega,\lambda)&=T_{n,\ell}^{(2)}+T_{n,\ell}^{(4)},
 \quad P_n=P_{n,0},\quad n\ge0,\quad\ell\in\mathbb Z,
 \end{split}
\end{equation}
where the dependence on $\omega$ is understood with the remaining independent parameters held fixed.
At $\omega=0$, the $a_2$ and $a_4$ pole families coalesce, and $P_{n,\ell}$ becomes the residue of the resulting double pole. Equivalently, the paired contribution is obtained by integrating $F_q$ around a small contour enclosing both poles. Although the individual residues $T^{2, 4}$ may diverge at $\omega=0$, their singular parts cancel, leaving $P_{n,\ell}$ finite.  

Because of these cancellations, the contributions from subleading poles $P_n, n\geq 1$ can exceed the leading contribution $P_0$. Thus, $|P_0|$ alone cannot bound the remaining residues for all values of $\omega$. To control the descendant residues even when the leading pair cancels, we introduce 
\begin{equation} \label{eq:R-def}
   \mathcal  R_\lambda(z, \omega) \equiv z(z-i \omega) F_q(z),
\end{equation}
and set $R_\lambda (\omega) \equiv  \mathcal R_\lambda(i\omega, \omega)$. We will prove the following three bounds on the remaining residue contributions for $|\omega|\leq b\log\frac{1}{\lambda}$, with a fixed constant $b>0$ and sufficiently small $\lambda$: 
\begin{equation}
     \sum_{n \geq N}\left|P_n\right| \leq C_{1, N} \lambda^{2 N}(1+|\omega|)^{k_N}\left(\left|P_0\right|+\left|R_\lambda\right|\right),  \quad k_N >0\label{eq:bound-1}
\end{equation}
\begin{equation}
   \sum_{n \geq 0, \ell \neq 0}\left|P_{n, \ell}\right| \leq C_2 e^{-c_2 / \lambda}\left(\left|P_0\right|+\left|R_\lambda\right|\right), \quad 0<c_2<4 \pi^2 ,  \label{eq:bound-2}
\end{equation}
\begin{equation}
 \sum_{k=1,3} \sum_{n, \ell}\left|T_{n, \ell}^{(k)}\right| \leq C_3 e^{-c_3 / \lambda}\left(\left|P_0\right|+\left|R_\lambda\right|\right), \quad 0<c_3<2 d^2 . \label{eq:bound-3}
\end{equation}
The positive constants $c_2$, $c_3$, $C_{1,N}$, $C_2$, and $C_3$ are independent of $\omega$ and $\lambda$. Throughout the proof, constants denoted by $C$ may change from line to line, but are independent of sufficiently small $\lambda$ and of $\omega$ in the stated range.

For $N=1$, the first bound shows that the descendant contributions from the $(k,\ell)= (2,0), (4,0)$ families are suppressed by $\lambda^2$, up to the displayed polynomial factor. Bounds 2 and 3 show that the other pole families are non-perturbatively suppressed. 

Using these three bounds, we show that, on the interval $I_\lambda = (-b\log\frac{1}{\lambda}, b \log \frac{1}{\lambda})$ with fixed $b>0$ and for $|\mathrm{Re} \tau| <\frac{\pi}{2}$,  
\begin{equation} \label{eq:integral-bound}
\lim_{\lambda\to0}\frac{1}{\mathcal{N}}\int_{I_{\lambda}}\mathrm d\omega\,|e^{\tau\omega}||K(\omega,\lambda)-P^{{\rm sc}}_{0}(\omega,\lambda)|=0,
\end{equation}
where $\mathcal{N}= 1/(2\pi \lambda \rho(\theta))$ is the prefactor in the semi-classical expression \eqref{eq:6j-symbol-limit}. This prefactor makes explicit the normalization of the leading-order integral of $P^{\rm sc}_0$. The bound states that the normalized difference between the exact and semi-classical integrals vanishes on a frequency interval whose width grows logarithmically in $\lambda^{-1}$. 

To control the remaining range of $\omega$, we also require the following bound on the large $\omega$ tail, with $\theta=\theta_\beta$ given by the saddle point equation \eqref{eq:saddle-v} of the canonical ensemble at fixed $0<v<1, |\operatorname{Re} \tau|<\pi(1+v) / 4$,~\footnote{This range of $\tau$ is sufficient for the Euclidean Boltzmann weight at each crossing. As an illustration, in the crossed four-point function, holding $\alpha_{13}$ and $\alpha_{24}$ fixed, the coefficient of $\omega$ in the relative Boltzmann exponent is $\tau=\frac12\sin\theta_\beta(\beta_1-\beta_2+\beta_3-\beta_4)$, which satisfies $|\tau|\leq\frac12\beta\sin\theta_\beta=\pi v/2<\pi(1+v)/4$.} and sufficiently large fixed $b$:
\begin{equation} \label{eq:bound-tail}
    \lim_{\lambda\to0}\frac{1}{\mathcal{N}}\int_{D_\lambda \backslash I_{\lambda}} \dd \omega |e^{\tau\omega}||K(\omega,\lambda)|=0.
\end{equation}
This states that the normalized integral over the physical range of $\omega$ outside $I_\lambda$ vanishes in the semi-classical limit. The three bounds \eqref{eq:bound-1}--\eqref{eq:bound-3} are combined in subsequent discussion to establish the integrand replacement \eqref{eq:6j-symbol-limit} and \eqref{eq:6j-limit}.

For $|\omega| \geq B \geq 1$,  the tail of the semiclassical expression $P_0^{\rm sc}$ is controlled separately by the Gamma-function decay; the real branch phases in \eqref{eq:6j-symbol-limit} and the ordinary Gamma-function asymptotics give
\begin{equation}
    \frac{1}{\mathcal{N}} \int_{\mathbb{R} \backslash I_\lambda} d \omega\left|e^{\tau \omega} P_0^{\mathrm{sc}}(\omega)\right| \leq C e^{-\delta_\eta B}=C \lambda^{b \delta_\eta} \longrightarrow 0.
\end{equation}
Since $I_\lambda \subset D_\lambda$ for sufficiently small $\lambda$, we may combine this bound with \eqref{eq:integral-bound} and \eqref{eq:bound-tail}. After we extend the exact kernel by zero outside $D_\lambda$, the triangle inequality yields
\begin{equation}
    \lim_{\lambda \to 0}  \frac{1}{\mathcal{N}} \int_{\mathbb{R}} d \omega\left|e^{\tau \omega}\right|\left|\mathbf{1}_{D_\lambda}(\omega) K(\omega, \lambda)-P_0^{\mathrm{sc}}(\omega)\right|  =0
\end{equation}
Here $\mathcal{C}$ is evaluated at $B=b \log (1 / \lambda)$. Equation \eqref{eq:C-limit} holds for every fixed $b$, so choosing $b$ large enough for the tail does not affect the growing-interval estimate.

In the discussion below, we always assume $\omega \in I_\lambda = (- b \log(1/\lambda), b \log(1/\lambda) ) $ for some fixed $b>0$ for the proof of \eqref{eq:bound-1}--\eqref{eq:bound-3}, and thus \eqref{eq:integral-bound}. The bound on the large $\omega$ tail treats the rest of $D_\lambda$.

\subsubsection{Proof of bound 1}
We begin with \eqref{eq:bound-1}. With $\beta=(1-A_1,1-A_2,B_1,B_2)$, an integer shift of $F_q(z)$ by $n$ gives the recursion relation
\begin{equation}  
 F_q(z-n)=h_n(z)F_q(z),\quad
 h_n(z)\equiv \prod_{m=1}^{n} r_m(z), \quad 
 r_m(z)\equiv q \prod_{j=1}^{4} \frac{(1-q^{m-z-\beta_j})}
 {(1-q^{m-z-a_j})},
 \label{eq:residue-recursion}
\end{equation}
where $h_0 =1$. This recursion relates the $n$-th paired residue $P_n$ to $P_0$ through
\begin{equation} \label{eq:Pn-relation}
    \begin{split}
        P_{n}	&=h_{n}(0)T_{0,0}^{(4)}+h_{n}(i\omega)T_{0,0}^{(2)} \\ 
	& =h_n(0) P_0+\frac{h_n(i \omega)-h_n(0)}{i \omega} R_\lambda ,
    \end{split}
\end{equation}
where we have used \eqref{eq:R-def}. To establish \eqref{eq:bound-1}, we control the factors involving $h_n$ that multiply the leading residue and $R_\lambda(\omega)$. We work along the segment $z=i s$, with $s \in [0, \omega]$ when $\omega\geq 0$ or $[\omega,0]$ when $\omega<0$. Throughout the proofs of bounds 1 and 2, $\sup_s$ denotes the supremum over this interval. Since $h_n$ is a product of $r_m(z)$, we first establish the bounds 
\begin{equation}
 |r_m(is)| \le q, \quad |r^\prime_m (i s) | \leq \frac{C_r q}{m} (1+|\omega|)^3, 
 \quad m\ge1 , \label{eq:bound-r}
\end{equation}
where $C_r >0$ is independent of $s$. We prove \eqref{eq:bound-r} below. Assuming these estimates for now, the product definition of $h_n$ in \eqref{eq:residue-recursion} gives 
    \begin{align} \label{eq:bound-h}
 |h_n(is)|&\le q^{n-M}|h_M(is)|,\\
 |h_n'(is)|&\le q^{n-M}
 \left(|h_M'(is)|+C_r (1+|\omega|)^3 |h_M(is)|\sum_{m=M+1}^{n}\frac1m\right).
\end{align}
For fixed $n$, there is a constant $C_{h,n}>0$, independent of $s,\omega,\lambda$, such that
\begin{equation}
 |h_n(is)|+|h_n'(is)|\le C_{h,n}\lambda^{2n}(1+|\omega|)^{4n}.
 \label{eq:order-lambda}
\end{equation}
We separate the term $n=N$ and apply \eqref{eq:bound-h} with $M=N+1$ to the remaining terms. Summing the geometric factor and the harmonic sum gives
\begin{equation}
\sum_{n\ge N+1}\sup_s\bigl(|h_n(is)|+|h_n'(is)|\bigr)\le \widetilde C_{h,N}\lambda^{2N+1}(1+|\omega|)^{4N+7}[1+\log(1/\lambda)].
\end{equation}
Here $\widetilde C_{h,N}>0$ is independent of $\omega$ and $\lambda$.
Relative to $\lambda^{2N}(1+|\omega|)^{4N}$, the additional factor is bounded by $C\lambda[1+b\log(1/\lambda)]^7[1+\log(1/\lambda)]$, which tends to zero. Adding the $n=N$ term therefore gives
\begin{equation}
 \sum_{n=N}^{\infty}\sup_s
 \bigl(|h_n(is)|+|h_n'(is)|\bigr)
 \le C_{1,N} \lambda^{2N} (1+|\omega|)^{4N}.
 \label{eq:uniform-all-orders-tail}
\end{equation}
Here $C_{1,N}$ is independent of $s$ and $\lambda$ for sufficiently small $\lambda$. This estimate controls both $|h_n(is)|$ and $| h^\prime_n (is)|$. In particular, for $\omega \not=0$, the divided difference can be written and bounded as
\begin{equation} \label{eq:uniform-bound-h-differnece}
     \frac{h_n(i\omega)-h_n(0)}{i\omega}
 =\int_0^1 h_n'(i\tau\omega)\,d\tau \quad 
 \Rightarrow \quad 
 \left|\frac{h_n(i\omega)-h_n(0)}{i\omega}\right|
 \le\sup_s|h_n'(is)|.
\end{equation}
At $\omega=0$, the integral reduces to $h_n'(0)$, so the bound extends continuously to the coincident-pole limit. Thus \eqref{eq:uniform-all-orders-tail} controls
both coefficients in \eqref{eq:Pn-relation}, giving
\begin{equation} \label{eq:bound-P-intermediate}
\sum_{n \geq N}\left|P_n\right| \leq C_{1, N} \lambda^{2 N}(1+|\omega|)^{k_N}\left(\left|P_0\right|+\left|R_\lambda\right|\right), \quad |\omega| \leq b \log\frac{1}{\lambda},
\end{equation}
for $k_N =4N$. This proves \eqref{eq:bound-1}.

\paragraph{Proof of \eqref{eq:bound-r}.}
To prove \eqref{eq:bound-r}, we first bound $|r_m(is)|$ by estimating its numerator and denominator separately.

Choose $s$-independent  $B_\ast>0$ such that
\begin{equation}
|is+\beta_r|\leq B_\ast,
\qquad r=1,\ldots,4,
\end{equation}
for the fixed external kinematic parameters and uniformly for $s$ between $0$ and $\omega$.  This is possible because the shifts entering $\beta_r$ remain finite on the interval under consideration. There exists a constant $b_1$ such that $B_* \leq b_1 (1+ |\omega|)$.  Each factor in the numerator of \eqref{eq:residue-recursion} can then be bounded as
\begin{equation} \label{eq:bound-r-numerator}
\begin{aligned}
|1-q^{m-is-\beta_r}|
&\leq 1-q^m+q^m|e^{\lambda(is+\beta_r)}-1|\\
&\leq t_m+\lambda q^m|is+\beta_r|e^{\lambda|is+\beta_r|}\\
&\leq t_m+ \mathcal C_0\lambda q^m,
\end{aligned}
\end{equation}
where
\begin{equation}
t_m=1-q^m, \quad \mathcal{C}_0 \equiv B_* e^{\lambda B_*}. 
\end{equation}
Since $\lambda B_*\to0$ uniformly for $\omega\in I_\lambda$, $e^{\lambda B_*}\le2$ for sufficiently small $\lambda$. We may therefore choose $b_2=2b_1$, independent of $\lambda$ and $\omega$, so that
\begin{equation} \label{eq:C0-bound}
    \mathcal{C}_0 \leq  b_2 (1+|\omega|), \qquad |\omega|\leq b \log(\frac{1}{\lambda}).
\end{equation}

Multiplying the four factors therefore gives the bound
\begin{equation}
\prod_{r=1}^4 |1-q^{m-is-\beta_r}|
\leq
\left(t_m+\mathcal C_0\lambda q^m\right)^4.
\end{equation}
For the denominator, we use the identity
\begin{equation}
|1-q^m e^{i\varphi}|^2
=
t_m^2+4q^m\sin^2(\varphi/2). 
\end{equation}
Among the four factors $1-q^{m-z-a_j}$, the parameters $a_2$ and $a_4$ remain finite as $\lambda\to0$. Their corresponding factors therefore obey the elementary lower bound
\begin{equation}\label{eq:bound24}
\prod_{j=2,4}\left|1-q^{m-is-a_j}\right|
\geq t_m^2.
\end{equation}
The remaining factors associated with $a_1$ and $a_3$ contain imaginary parts of order $\lambda^{-1}$. Writing the corresponding phases as $\varphi_{1,3}$, we have
\begin{equation}
\varphi_{1,3}
=
\pm 2\theta-\lambda (s+2\alpha_{4})
\end{equation}
For fixed $\theta\in(0,\pi)$, there exists a constant $\delta>0$ such that, for sufficiently small $\lambda$,
\begin{equation}
\left|\sin\left(\varphi_j/2\right)\right|
\geq
\sin\theta-\frac{\lambda}{2}|s+2\alpha_4|
\geq
\frac{\delta}{2}.
\end{equation}
We may therefore choose $c_0=\delta^2$, which gives
\begin{equation} \label{eq:bound13}
\prod_{j=1,3}\left|1-q^{m-is-a_j}\right|
\geq
t_m^2+c_0q^m.
\end{equation}
Combining \eqref{eq:bound13} and \eqref{eq:bound24}, we obtain
\begin{equation} \label{eq:bound-r-denomiantor}
\prod_{j=1}^4\left|1-q^{m-is-a_j}\right|
\geq
t_m^2\left(t_m^2+c_0q^m\right).
\end{equation}
Substituting the numerator bound \eqref{eq:bound-r-numerator} and the denominator bound \eqref{eq:bound-r-denomiantor} into the definition of $r_m$ in \eqref{eq:residue-recursion}, we find
\begin{equation} \label{eq:bound-r-proved1}
\left|r_m(is)\right|
\leq
q\,
\frac{\left(t_m+\mathcal C_0\lambda q^m\right)^4}
{t_m^2\left(t_m^2+c_0q^m\right)}
\leq q,
\qquad
m\geq1,
\end{equation}
for sufficiently small $\lambda$. The second inequality follows by expanding the numerator and using
\begin{equation}
\frac{\lambda q^m}{t_m}
=
\frac{\lambda}{e^{\lambda m}-1}
\leq
\frac{1}{m}
\leq1.
\end{equation}
Differentiating the product defining the multiplier with respect to $s$ and applying the same estimates gives
\begin{equation}
|r_m'(is)|
\leq
\frac{C_r q}{m} (1+|\omega|)^3,
\label{eq:bound-r-proved2}
\end{equation}
where $C_r>0$ depends only on $b_2$ and $c_0$ and is independent of $m$,  $\lambda$ and $\omega$ for sufficiently small $\lambda$. The bounds \eqref{eq:bound-r-proved1} and \eqref{eq:bound-r-proved2} together establish \eqref{eq:bound-r}.

\subsubsection{Proof of bound 2}

We next turn to bounds $2$ and $3$, which control the exponentially suppressed contributions. We begin with the periodic images of the dominant pole families. 

To prove \eqref{eq:bound-2}, we examine how the integrand changes under a shift by an imaginary period. The reflected expression
\eqref{eq:f-simpler} gives
\begin{equation}
 F_q(u+2\pi i\ell/\lambda)=W_\ell(u)F_q(u),\quad
 W_\ell(u)=\prod_{r=1}^{2}
 \frac{\sin\pi(A_r-u)}
 {\sin\pi(A_r-u-2\pi i\ell/\lambda)},
 \label{eq:residue-image-relation}
\end{equation}
with $W_0 =1$. The required bound on $W_\ell$ and its derivative is 
\begin{equation}
 \sup_s
 \bigl(|W_\ell(is)|+|W_\ell'(is)|\bigr)
 \le  C_W e^{-4\pi^2|\ell|/\lambda + 2 \pi |\omega|},\qquad \ell\ne0,
 \label{eq:image-multiplier-bound}
\end{equation}
for some $\omega$-independent constant $C_W >0$.  We prove this inequality below. Since $W_\ell$ is invariant under integer shifts of $u$, we can define $H_{n,\ell}=W_\ell h_n$ and apply \eqref{eq:residue-recursion} to obtain 
\begin{equation}
 P_{n,\ell}=H_{n,\ell}(0)P_0+
 \frac{H_{n,\ell}(i\omega)-H_{n,\ell}(0)}{i\omega}R_\lambda.
 \label{eq:double-pole-images}
\end{equation}
This is the analogue of \eqref{eq:Pn-relation} for the imaginary images. At $\omega=0$, the divided difference reduces to $H_{n,\ell}'(0)$, so the formula also covers the double-pole residue.

Using \eqref{eq:uniform-all-orders-tail} with $N=1$, together with
$h_0=1$ and $h_0'=0$, gives
\begin{equation} \label{eq:n0-h-bound}
     \sum_{n=0}^{\infty}\sup_s
 \bigl(|h_n(is)|+|h_n'(is)|\bigr)\le 1 + C_{1,1} \lambda^{2} (1+|\omega|)^4.
\end{equation}
Since 
\begin{equation}
    C_{1,1}\lambda^2(1+|\omega|)^{k_1} \leq C_h \lambda^2(1+b \log (1 / \lambda))^{k_1} \longrightarrow 0,
\end{equation}
there exists $\lambda_0 >0$ such that 
\begin{equation} \label{eq:lambda0-def}
    C_{1,1} \lambda^2(1+b \log (1 / \lambda))^{k_1} \leq 1 \quad\left(0<\lambda<\lambda_0\right) .
\end{equation}
Therefore for $\lambda\in(0,\lambda_0)$, the right hand side of \eqref{eq:n0-h-bound} is no more than 2. 

As in the preceding proof, we control the coefficients in \eqref{eq:double-pole-images} by uniform estimates on $|H_{n,\ell}(is)|$ and the corresponding derivative for $s$ between $0$ and $\omega$. Combining \eqref{eq:n0-h-bound} and \eqref{eq:image-multiplier-bound} gives 
\begin{equation}
\sum_{n \geq 0, \ell \neq 0}\left|P_{n, \ell}\right| \leq 8 C_W e^{-4 \pi^2 / \lambda+2 \pi|\omega|}\left(\left|P_0\right|+\left|R_\lambda\right|\right) .
 \label{eq:one-image-bound}
\end{equation}
For any fixed $0<c_2<4\pi^2$, sufficiently small $\lambda$ satisfies $2\pi b\lambda\log(1/\lambda)\le4\pi^2-c_2$. Thus the exponent in \eqref{eq:one-image-bound} is at most $-c_2/\lambda$ for $\omega\in I_\lambda$. Taking $C_2=8C_W$ therefore proves \eqref{eq:bound-2}.

\paragraph{Proof of \eqref{eq:image-multiplier-bound}.}
Along the segment $u=is$ between $0$ and $i\omega$, $|\operatorname{Im}A_r-s|\le|\omega|/2+C$.
Each numerator sine in \eqref{eq:residue-image-relation} and its derivative are therefore bounded by $Ce^{\pi|\omega|/2}$.
For each denominator, the identity
\begin{equation} \label{eq:sinh-square}
     |\sin(x+iy)|^2=\sin^2x+\sinh^2y
\end{equation}
gives a lower bound $C^{-1}e^{2\pi^2|\ell|/\lambda-\pi|\omega|/2}$ for sufficiently small $\lambda$. Its derivative is bounded by a constant times its modulus.
Applying these bounds to both sine ratios and their derivatives gives \eqref{eq:image-multiplier-bound}. The argument remains valid when a numerator sine vanishes.

\subsubsection{Proof of bound 3}
To prove \eqref{eq:bound-3}, we show that neither the growth along the $k=1,3$ families nor the sum over their imaginary images overcomes their exponential suppression relative to the dominant residues.
For each $k=1,3$, choose the image whose poles lie closest to the real axis and denote its residues by $\widetilde T_n^{(k)}$. We use the following three inequalities, which are proved below:
\begin{align}
 |\widetilde T_0^{(k)}|
 &\le C\lambda^{-p}e^{-2\pi d/\lambda+C(1+|\omega|)}|R_\lambda|,
 \qquad p\ge0,\label{eq:remote-initial-bound}\\
 |\widetilde T_n^{(k)}|
 &\le Ce^{C(1+|\omega|)}q^n|\widetilde T_0^{(k)}|
 \prod_{m=1}^{n}
 \frac{|1-e^{2i\theta}q^m|^2}{(1-q^m)^2},
 \label{eq:remote-descendant}\\
 \sum_{\ell\in\mathbb Z}|T_{n,\ell}^{(k)}|
 &\le Ce^{C(1+|\omega|)}|\widetilde T_n^{(k)}|.
 \label{eq:remote-image-sum}
\end{align}
Here $d$ is the distance to the nearest endpoint that enters \eqref{eq:bound-3}. The positive constants $C$ may differ between the three estimates but are independent of $n$, $\omega\in I_\lambda$, and sufficiently small $\lambda$.

The product in
\eqref{eq:remote-descendant} is bounded by
\begin{align}
 \log\prod_{m=1}^{\infty}
 \frac{|1-e^{2i\theta}q^m|^2}{(1-q^m)^2}
 &=2\sum_{n=1}^{\infty}
 \frac{1-\cos(2n\theta)}{n(e^{\lambda n}-1)}\nonumber\\
 &\le\frac{2}{\lambda}\sum_{n=1}^{\infty}
       \frac{1-\cos(2n\theta)}{n^2}
 =\frac{2d(\pi-d)}{\lambda}.
 \label{eq:bound-product}
\end{align}
The growth of the descendant product is therefore dominated by the initial $e^{-2\pi d/\lambda}$ suppression. Since $\sum_{n\ge0}q^n=(1-q)^{-1}\le C/\lambda$,
\begin{equation}
 \begin{aligned}
 \sum_{k=1,3}\sum_{n=0}^{\infty}\sum_{\ell\in\mathbb Z}
 |T_{n,\ell}^{(k)}|
 &\le \widetilde{C}_3 \lambda^{-p-1}e^{-2d^2/\lambda+C(1+|\omega|)}|R_\lambda|\\
 &\le C_3 e^{-c_3/\lambda}\bigl(|P_0|+|R_\lambda|\bigr),
 \qquad 0<c_3<2d^2.
 \end{aligned}
 \label{eq:remote-summed-bound}
\end{equation}
For the last inequality, we use $e^{C|\omega|}\le\lambda^{-Cb}$ for $\omega\in I_\lambda$. The resulting fixed power of $\lambda^{-1}$ can be absorbed into any fixed rate $0<c_3<2d^2$, proving \eqref{eq:bound-3}.

\paragraph{Proof of \eqref{eq:remote-initial-bound}, \eqref{eq:remote-descendant} and \eqref{eq:remote-image-sum}.}
We now establish these three inequalities in turn. For \eqref{eq:remote-initial-bound}, we evaluate the residue at the leading pole closest to the real axis. Its imaginary part has magnitude
$2d/\lambda+\mathcal O(1+|\omega|)$. The two inverse sine factors contribute
$-4\pi d/\lambda$ to the logarithm of the residue magnitude, while
the $q$-Pochhammer quotient contributes $2d^2/\lambda$. The common prefactor cancels in the ratio. The large-argument estimates give, uniformly for $\omega\in I_\lambda$,
\begin{equation}
 \frac{|\widetilde T_0^{(k)}|}{|R_\lambda|}
 \le C\lambda^{-p}e^{-2\pi d/\lambda+C(1+|\omega|)}.
\end{equation}
The phases of the $q$-Pochhammer arguments shift by $\mathcal O(\lambda(1+|\omega|))$, changing the leading logarithms of the products by at most $C(1+|\omega|)$.
The ratio extends continuously to $\omega=0$, since $R_\lambda(0)\ne0$. This proves \eqref{eq:remote-initial-bound}.

For \eqref{eq:remote-descendant}, fix $k=1$ or $3$ and evaluate
\eqref{eq:residue-recursion} at $z=-a_k$:
\begin{equation}
 r_m(-a_k)
 =q\frac{\prod_{r=1}^{4}(1-q^{m+a_k-\beta_r})}
          {\prod_{j=1}^{4}(1-q^{m+a_k-a_j})}.
\end{equation}
Here $k$ labels the chosen pole family, while $j$ labels the
denominator factors. We first compare the four numerator factors
with the two denominator factors indexed by $j=2,4$.

Each of these six factors has the form
$1-q^m e^{\pm2i\theta}e^{\lambda\zeta}$. Each shift $\zeta$ satisfies $|\zeta|\le C(1+|\omega|)$. Since $\lambda(1+|\omega|)\to0$ uniformly for $\omega\in I_\lambda$, retaining $e^{\lambda|\zeta|}$ gives
\begin{equation}
 \left|
 \frac{1-q^m e^{\pm2i\theta}e^{\lambda\zeta}}
      {1-q^m e^{\pm2i\theta}}-1
 \right|
 \le
 \frac{q^m|e^{\lambda\zeta}-1|}
      {|1-q^m e^{\pm2i\theta}|}
 \le \frac{\lambda|\zeta|e^{\lambda|\zeta|}}{\sin\theta}q^m
 \le C\lambda(1+|\omega|)q^m,
\end{equation}
where we have used $|1-q^m e^{\pm2i\theta}|\ge\sin\theta>0$.
Thus each modulus differs from
$|1-e^{2i\theta}q^m|$ by a relative correction bounded by
$C\lambda(1+|\omega|)q^m$, uniformly in $m$. Consequently,
\begin{equation}
 \frac{\prod_{r=1}^{4}|1-q^{m+a_k-\beta_r}|}
      {\prod_{j=2,4}|1-q^{m+a_k-a_j}|}
 \le
 |1-e^{2i\theta}q^m|^2
 \frac{(1+C\lambda(1+|\omega|)q^m)^4}{(1-C\lambda(1+|\omega|)q^m)^2}
 \le
 e^{8C\lambda(1+|\omega|)q^m}|1-e^{2i\theta}q^m|^2.
\end{equation}

Of the remaining denominator factors, the one with $j=k$ equals
$1-q^m$. The other has modulus at least $1-q^m$, since all
$a_j$ are purely imaginary. Absorbing $8C$ into a new constant $C$, we obtain
\begin{equation}
 |r_m(-a_k)|
 \le q e^{C\lambda(1+|\omega|)q^m}
       \frac{|1-e^{2i\theta}q^m|^2}{(1-q^m)^2}.
\end{equation}
The same estimate applies at any periodic image because $r_m(z)$
is periodic under $z\mapsto z+2\pi i\ell/\lambda$. Multiplying
over $m$ proves \eqref{eq:remote-descendant}, since
\begin{equation}
 \prod_{m=1}^{n}e^{C\lambda(1+|\omega|)q^m}
 \le \exp\left(C\lambda(1+|\omega|)\frac{q}{1-q}\right)
 =\exp\left(\frac{C\lambda(1+|\omega|)}{e^\lambda-1}\right)
 \le e^{C(1+|\omega|)}.
\end{equation}

Finally, the poles in a fixed family are
\begin{equation}
 u_{n,\ell}=-a_k-n+\frac{2\pi i\ell}{\lambda}.
\end{equation}
Changing $\ell$ moves the pole parallel to the imaginary axis.
The sine factors are controlled by its distance
$|\operatorname{Im}u_{n,\ell}|$ from the real axis.
Choose $\ell_*$ to minimize this distance, so that
$\widetilde T_n^{(k)}=T_{n,\ell_*}^{(k)}$. Using
\eqref{eq:residue-image-relation} and \eqref{eq:sinh-square}, we find
\begin{equation}
 |T_{n,\ell}^{(k)}|
 \le Ce^{C(1+|\omega|)}|\widetilde T_n^{(k)}|
 \exp\left[
 -2\pi\left(
 |\operatorname{Im}u_{n,\ell}|
 -|\operatorname{Im}u_{n,\ell_*}|
 \right)\right].
\end{equation}
Since $|\operatorname{Im}A_r|\le C(1+|\omega|)$, the bounds on the two sine ratios contribute at most the factor $Ce^{C(1+|\omega|)}$.
Since $|\operatorname{Im}u_{n,\ell_*}|\le\pi/\lambda$,
\begin{equation}
 |\operatorname{Im}u_{n,\ell}|
 -|\operatorname{Im}u_{n,\ell_*}|
 \ge \frac{2\pi}{\lambda}
       \bigl(|\ell-\ell_*|-1\bigr).
\end{equation}
Thus, the residues at the two neighboring images satisfying $|\ell-\ell_*|=1$ are bounded by $Ce^{C(1+|\omega|)}$ times the closest-pole residue. More distant images carry the same prefactor and additional powers of $e^{-4\pi^2/\lambda}$. Summing these contributions proves \eqref{eq:remote-image-sum}.

\subsubsection{Integrated error and tail bounds}

To prove \eqref{eq:integral-bound}, we first establish a pointwise estimate on a bounded interval $I=(-B,B)\subset \mathbb{R}$, keeping track of its width. For Euclidean time $\tau$ satisfying $|\mathrm{Re}(\tau)|= \eta <\pi/2$, the estimate is
\begin{equation} \label{eq:bound-pointwise-I}
    |e^{\tau\omega}||K(\omega,\lambda)-P^{{\rm sc}}_{0}(\omega,\lambda)|\leq\lambda\mathcal{C}\mathcal{N}e^{-\delta_{\eta}|\omega|},\quad\omega\in I,
\end{equation}
where $\delta_\eta = \frac{\pi}{2}-\eta$ and the coefficient $\mathcal{C}=\mathcal{C}(\lambda,B,L)$, with $L=\log(\kappa_\theta)$, satisfies 
\begin{equation} \label{eq:C-limit}
    \lim_{\lambda\to 0} \lambda \mathcal{C} (\lambda, b \log\frac{1}{\lambda}, L) =0 ,
\end{equation} 
for any fixed $b>0$. This estimate controls the normalized difference between the exact and semi-classical integrals as $\lambda$ tends to zero, first for fixed $B$ and then on a logarithmically growing interval. Indeed, using
\begin{equation}
    \int_{I}\mathrm d\omega\,e^{-\delta_{\eta}|\omega|}\leq\frac{2}{\delta_{\eta}},
\end{equation}
we obtain 
\begin{equation}
    \frac{1}{\mathcal{N}}\int_{I}\mathrm d\omega\,|e^{\tau\omega}||K(\omega,\lambda)-P^{{\rm sc}}_{0}(\omega,\lambda)|\leq\lambda\frac{2\mathcal{C}}{\delta_{\eta}}.
\end{equation}
Setting $B=b \log\frac{1}{\lambda}$ and taking $\lambda\to0$, we obtain \eqref{eq:integral-bound} from \eqref{eq:C-limit}.

To prove \eqref{eq:bound-pointwise-I}, we use the triangle inequality to separate the two sources of error:
\begin{equation} \label{eq:KP-difference}
    \left|K(\omega,\lambda)-P^{\mathrm{sc}}_{0}(\omega,\lambda)\right|\leq|K(\omega,\lambda)-P_{0}(\omega,\lambda)|+|P_{0}-P^{{\rm sc}}_{0}|.
\end{equation}
The first term measures the difference between the exact 6j-symbol and the dominant residues at finite $\lambda$. The second measures the difference between those residues and their semi-classical expression on the right-hand side of \eqref{eq:6j-symbol-limit}.

For $\omega\in I$, the bounds \eqref{eq:bound-1}--\eqref{eq:bound-3} give
\begin{equation} \label{eq:bound-KP}
    \begin{aligned}
\left|K -P_0\right| & \leq \sum_{n \geq 1}\left|P_n\right|+\sum_{n \geq 0, \ell \neq 0}\left|P_{n, \ell}\right|+\sum_{k=1,3} \sum_{n, \ell}\left|T_{n, \ell}^{(k)}\right| \\
& \leq\left(C_{1,1} (1+|B|)^{k_1} \lambda^2+C_2 e^{-c_2 / \lambda}+C_3 e^{-c_3 / \lambda}\right)\left(\left|P_0\right|+\left|R \right|\right) \\
& \leq \widetilde{C}_B  \lambda^2\left(\left|P_0\right|+\left|R_\lambda \right|\right),
\end{aligned}
\end{equation} 
with $\widetilde{C}_B>0$ independent of $\lambda$ and depending on $|B|$ only through a polynomial of finite degree, so that when we take $B=b \log\lambda^{-1}$, $\lim_{\lambda\to 0} \lambda \widetilde{C}_B =0$. 

It remains to bound $|P_0|$ and $|R_\lambda|$. Their exact expressions at finite $\lambda$ are
\begin{align}
P_{0}(\omega,\lambda)	&=\mathcal{N}\frac{e^{i\omega L}\mathcal{A}_{\lambda,+}(\omega)-e^{-i\omega L}\mathcal{A}_{\lambda,-}(\omega)}{i\mathit{\omega}},\\
R_\lambda (\omega)	&=\mathcal{N}e^{i\omega L}\mathcal{A}_{\lambda,+},
\end{align}
where $L=\log\kappa_\theta$ and the amplitudes $\mathcal{A}_{\lambda,\pm}$ are obtained from the residue \eqref{eq:f-residue}. The corresponding semi-classical expressions are
\begin{align}
P_0^{\rm sc }(\omega)&=\mathcal{N}\frac{e^{i \omega L} \mathcal{A}_{0,+}(\omega)-e^{-i \omega L} \mathcal{A}_{0,-}(\omega)}{i \omega}, \\
R^{\rm sc} (\omega)&=\mathcal{N} e^{i \omega L} \mathcal{A}_{0,+},
\end{align}
where the $\lambda=0$ limits of the amplitudes $\mathcal{A}_{\lambda,\pm}$ are
\begin{equation}
    \mathcal{A}_{0,+}(\omega)=e^{i\chi+i\vartheta_{+}}\Gamma(1+i\omega),\quad\mathcal{A}_{0,-}=e^{i\chi+i\vartheta_{-}}\Gamma(1-i\omega).
\end{equation}
The norms $|\ma_{0,\pm}|$ and the derivatives of the amplitudes are bounded functions of $\omega$ on the closure of $I$. More precisely, they satisfy the weighted estimate\footnote{This follows from \begin{equation}
    \left|\mathcal{A}_{0, \pm}\right|+\left|\mathcal{A}_{0, \pm}^{\prime}\right| \leq C(1+|\omega|)^{3 / 2} e^{-\pi|\omega| / 2} ,
\end{equation}
for a constant $C>0$. One possible choice of $C$ is
\begin{equation}
    C=3\left[4+\left|\psi\left(\Delta_1\right)\right|+\left|\psi\left(\Delta_2\right)\right|+\frac{1+\left|\alpha_{13}\right|+\left|\alpha_{24}\right|}{2}\left(\psi^{\prime}\left(\Delta_1\right)+\psi^{\prime}\left(\Delta_2\right)\right)\right],
\end{equation} where $\psi(x)$ is the digamma function.  }
\begin{equation}
    \begin{split}
        & |e^{\tau \omega}|\left( \left|\mathcal{A}_{0,\pm}(\omega)\right|+\left|\mathcal{A}'_{0,\pm}(\omega)\right| \right)\leq C_A \, e^{-\delta_\eta |\omega|} \quad\forall\omega\in I, \\
        & C_A = C (1 + B)^{\frac{3}{2}},
    \end{split}
\end{equation}
for a constant $C>0$ independent of $B$ and $\lambda$. Applying the divided-difference argument used in \eqref{eq:uniform-bound-h-differnece} then gives the semi-classical bounds
\begin{equation} \label{eq:bound-PRsc}
    |e^{\tau \omega} P^{{\rm sc}}_{0}(\omega)|\leq C_A  \mathcal{N}(1+L) e^{-\delta_\eta |\omega|}, \quad
    |e^{\tau \omega} R^{{\rm sc}}(\omega)|\leq C_A \mathcal N e^{-\delta_\eta |\omega|}
\end{equation}
Next, we estimate the differences $| P_0 - P_0^{\rm sc} |$ and $|R - R^{\rm sc}|$ uniformly for $\omega\in I$.
First, define
\[
x=\Delta_1-\Delta_2+i\alpha_{13},\qquad h_\pm=\frac{\alpha_{13}\pm\alpha_{24}}{2}.
\]
The finite-$\lambda$ amplitudes $\ma_{\lambda,\pm}$ are related to $\ma_{0,\pm}$ by
\begin{equation}
    \ma_{\lambda, \pm } = \ma_{0,\pm} e^{\mathcal{E}_{\pm}},
\end{equation}
where
\begin{equation} \label{eq:me-def}
    \begin{aligned}\mathcal{E}_{\pm}= & \lambda z_{2,\pm}+\Phi_{\lambda}+G_{\lambda,\pm}+\delta_{\lambda}(x)+\delta_{\lambda}(1-x)+\delta_{\lambda}(\pm i\omega)\\
 & -\sum^{2}_{j=1}\left[\delta_{\lambda}\left(1-z_{j,\pm}\right)+\frac{\delta_{\lambda}\left(z_{j,\pm}\right)+\delta_{\lambda}\left(\overline{z_{j,\pm}}\right)}{2}+\frac{\delta_{\lambda}\left(z_{j,\mp}\right)-\delta_{\lambda}\left(\overline{z_{j,\mp}}\right)}{2}\right].
\end{aligned}
\end{equation}
The arguments $z_{j,\pm},\, j=1,2$ are of order $\mo(1)$ for fixed frequency and are defined, together with the shifts, by
\begin{equation}
\begin{aligned}
& z_{1,+}=\Delta_1-i \alpha_{32}, \quad z_{2,+}=\Delta_2-i \alpha_{12}, \quad  s_{+}=\frac{\omega}{2}-\alpha_{24} \\
& z_{1,-}=\Delta_1+i \alpha_{14},\quad  z_{2,-}=\Delta_2+i \alpha_{34}, \quad s_{-}=-\frac{\omega}{2}-\alpha_{24}
\end{aligned}
\end{equation}
The functions $\Phi_\lambda, G_\lambda , \delta_\lambda$ are defined by
\begin{equation}
    G_{\lambda,\pm}=\sum_{\sigma=\pm}\log\frac{\Gamma_{q}\left(\sigma\frac{2i\theta}{\lambda}+is_{\pm}\right)}{\Gamma_{q}\left(\sigma\frac{2i\theta}{\lambda}\right)}-2is_{\pm}L,
\end{equation}
\begin{equation}
    \Phi_{\lambda}=\frac{1}{2}\sum_{\sigma=\pm}\left[\log\frac{\Gamma_{q}\left(\sigma\frac{2i\theta}{\lambda}+\Delta_{1}-ih_{-}\right)\Gamma_{q}\left(\sigma\frac{2i\theta}{\lambda}+\Delta_{2}+ih_{+}\right)}{\Gamma_{q}\left(\sigma\frac{2i\theta}{\lambda}+\Delta_{1}+ih_{-}\right)\Gamma_{q}\left(\sigma\frac{2i\theta}{\lambda}+\Delta_{2}-ih_{+}\right)}\right]-2i\alpha_{24}L,
\end{equation}
\begin{equation}
    \delta_\lambda (z)=\log\frac{\Gamma_{q}(z)}{\Gamma(z)}.
\end{equation}
The decomposition of $\me_\pm$ in \eqref{eq:me-def} allows us to bound $\me_{\pm}$ and its first derivative with respect to $\omega$~\footnote{For fixed $\Delta_{1,2}$ and $\theta$, the constraint $\sum_{i=1}^4 \alpha_i =0$ leaves three independent kinematic variables. We choose $\omega, \alpha_{13}$ and $\alpha_{24}$, so derivatives with respect to $\omega$ are taken with $\alpha_{13}$ and $\alpha_{24}$ held fixed.} uniformly on $I$. In the parameter range under consideration, $\delta_\lambda$ is smooth and satisfies the uniform $\mo(\lambda)$ estimate
\begin{equation}
\left|\delta_\lambda(z)\right| + \left|\delta^\prime_\lambda (z)\right| \leq 2 \lambda(2+M)^2,\quad |z| \leq M,  \quad 0<\lambda \leq \frac{1}{2 M+2}, 
\end{equation}
where $M=1+\Delta_1+\Delta_2+B+|\alpha_{13}|+|\alpha_{24}|$ is chosen so that the disk of this radius contains all arguments of $\delta_\lambda$ in \eqref{eq:me-def}. 

Analogous uniform $\mo(\lambda)$ bounds for $G_{\lambda,\pm}$ and $\Phi_\lambda$ and their first derivatives follow from the asymptotic expansion of $q$-Gamma function ratios with $\mathcal{O}(\lambda^{-1})$ arguments:
\begin{equation}
    \log\frac{\Gamma_{q}\left(\sigma\frac{2i\theta}{\lambda}+a\right)}{\Gamma_{q}\left(\sigma\frac{2i\theta}{\lambda}+b\right)}=(a-b)\log\frac{1-e^{-\sigma2i\theta}}{\lambda}+O(\lambda), \quad \sigma = \pm 1.
\end{equation}
Tracking the dependence on $a,b$ of the $O(\lambda)$ remainder and its derivative with respect to $\omega$, and combining the estimates for $\delta_\lambda$, $G_\lambda$ and $\Phi_\lambda$, we obtain, uniformly for $\omega\in I$,
\begin{equation} \label{eq:bound-me}
    \left|\mathcal{E}_{ \pm}(\omega)\right|+\left|\mathcal{E}_{ \pm}^{\prime}(\omega)\right| \leq C_\me \lambda(1+|\omega|)^2 ,
\end{equation}
where $C_\mathcal{E}>0$ is a constant independent of $\lambda$ and $B$.
This bound implies
\begin{equation}
\begin{aligned}
\left|e^{\tau \omega}\right|\left|R-R^{\mathrm{sc}}\right| & =\mathcal{N}\left|\mathcal{A}_{0,+} e^{\tau \omega}\right|\left|e^{\mathcal{E}_{+}}-1\right| \\
& \leq \mathcal{N}\left|\mathcal{E}_{+}\right| e^{\left|\mathcal{E}_{+}\right|} e^{\eta|\omega|}|\Gamma(1+i \omega)| \\
& \leq \lambda C_{\mathcal{E}} \mathcal{N} e^{\eta|\omega|+\lambda C_{\mathcal{E}}\left(1+|B|^2\right)}(1+|\omega|)^2|\Gamma(1+i \omega)| .
\end{aligned}
\label{eq:bound-R-differnece}
\end{equation}
Applying the Gamma-function estimate
\begin{equation}
    |\Gamma(1+i\omega)|\leq3(1+|\omega|)^{\frac{1}{2}}e^{-\frac{\pi}{2}|\omega|}
\end{equation}
to the last line gives 
\begin{equation} \label{eq:bound-R-difference}
    |e^{\tau\omega}||R_\lambda -R^{{\rm sc}}|\leq\lambda\mathcal{N}e^{-\delta_{\eta}|\omega|}C_{R},
\end{equation}
with 
\begin{equation}
    C_{R}=3C_{\me}e^{\lambda C_{\me}(1+|B|^{2})}(1+|B|)^{\frac{5}{2}}.
\end{equation}
A similar argument gives a constant $C_P>0$~\footnote{For example, one may choose $C_P$ as 
 \begin{equation}
    C_{P}=C_{\mathcal{E}}e^{C_{\mathcal{E}}\lambda\left(1+B^{2}\right)}\max\left\{ 4e^{\pi/2}\left(1+|L|+C_{\ma}\right),2\sqrt{2\pi}(1+B)^{3/2}\right\},
 \end{equation} where $C_{\ma}=\operatorname{max}_{\pm}\sup_{|\omega|\leq1}\left|\mathcal{A}^{\prime}_{0,\pm}(\omega)\right|$. } such that  
\begin{equation} \label{eq:bound-P-difference}
|e^{\tau\omega}||P_{0}-P^{{\rm sc}}_{0}|\leq\lambda\mathcal{N}e^{-\delta_{\eta}|\omega|}C_{P}.
\end{equation}
The coefficients $C_A$, $C_P$ and $C_R$ depend on $\lambda, B$. When $B=B_\lambda = b \log\frac{1}{\lambda}$ grows logarithmically with $\lambda^{-1}$, all three coefficients satisfy
\begin{equation} \label{eq:C-limit-2}
    \lim_{\lambda\to 0} \lambda C_A |_{B=B_\lambda}= \lim_{\lambda\to 0}  \lambda  C_P |_{B=B_\lambda} = \lim_{\lambda \to 0}  \lambda  C_R |_{B=B_\lambda} =0 .
\end{equation}

Combining \eqref{eq:bound-KP}, \eqref{eq:bound-R-differnece}, \eqref{eq:bound-P-difference}, and \eqref{eq:bound-PRsc}, we bound the difference in \eqref{eq:KP-difference} by
\begin{equation} \label{eq:pointwise-bound-proof}
    \begin{aligned}
\frac{1}{\mathcal{N}}\left|e^{\tau \omega}\right|\left|K-P_0^{\mathrm{sc}}\right| & \leq \frac{\left|e^{\tau \omega}\right|}{\mathcal{N}}\left(\widetilde{C}_B  \lambda^2\left(\left|P_0-P_0^{\mathrm{sc}}\right|+\left|R_\lambda -R^{\mathrm{sc}}\right|+\left|P_0^{\mathrm{sc}}\right|+\left|R^{\mathrm{sc}}\right|\right)\right) \\
& +\frac{1}{\mathcal{N}}\left|e^{\tau \omega}\right|\left|P_0-P_0^{\mathrm{sc}}\right| \\
& \leq \lambda e^{-\delta_\eta|\omega|}\left(\widetilde{C}_B \lambda\left(\lambda C_P+\lambda C_R+C_A(L+2)\right)+C_P\right).
\end{aligned}
\end{equation}
We therefore choose 
\begin{equation}
    \mathcal{C}=\widetilde{C}_{B}\lambda(\lambda C_{P}+\lambda C_{R}+C_{A}(L+2))+C_{P}.
\end{equation}
This coefficient satisfies \eqref{eq:C-limit} by \eqref{eq:C-limit-2}, so \eqref{eq:pointwise-bound-proof} gives the desired estimate \eqref{eq:bound-pointwise-I}.

\paragraph{Controlling the large-$\omega$ tail.} To establish the tail bound \eqref{eq:bound-tail}, we must control the integral outside $I$. We do so by estimating the exponential decay of the exact kernel $K (\omega,\lambda)$ at large frequency $\omega$, with explicit control of the dependence on $\lambda$. Specifically, we prove the following bound on $K$:
\begin{equation} \label{eq:bound-K-tail}
    \frac{\left|K(\omega, \lambda)\right|}{\mathcal{N}}=\left|\frac{N_{R}}{\mathcal{N}}\right|\left|\int_{\mathcal{C}}\frac{du}{2\pi i}q^{u-\Delta_{2}}f_{q}(u)\right|\leq C\lambda^{-M}e^{-\frac{\pi(1+v)}{4}|\omega|},\quad\omega\in D_{\lambda},
\end{equation}
where $v \in(0,1)$ is related to the thermal saddle point by $\theta=\theta_\beta=\pi(1+v) / 2$.  We use the reflected product representation \eqref{eq:f-simpler} and choose a fixed $0<c<\min \left(1, \Delta_1, \Delta_2\right)$ with $\Delta_j-c \notin \mathbb{Z}$, so that the sine denominators do not vanish on the contour. Define the even, $2 \pi$-periodic function $g$ by
\begin{equation}
    g(\phi)=\frac{\pi|\phi|}{2}-\frac{\phi^2}{4}, \quad|\phi| \leq 2 \pi.
\end{equation}
For every fixed real $a>0$, the product definition gives
\begin{equation}
\log \left|\left(q^a e^{i \phi} ; q\right)_{\infty}\right|=\frac{g(\phi)-\pi^2 / 6}{\lambda}+\mo (\lambda^0 , \log (1 / \lambda)),
\end{equation}
uniformly for real $\phi$. 
For a fixed numerator shift $a\leq0$, choose an integer $J\geq0$ with $a+J>0$ and write
\begin{equation}
(q^ae^{i\phi};q)_\infty
=(q^ae^{i\phi};q)_J(q^{a+J}e^{i\phi};q)_\infty.
\end{equation}
The modulus of $(q^ae^{i\phi};q)_J$ is bounded uniformly in $\phi$ and sufficiently small $\lambda$, so the same exponential upper bound applies.
We parameterize the contour $\mathcal{C}$ as
\begin{equation}
 u=c+\frac{i\omega}{2}+\frac{iy}{\lambda},
 \qquad y\in\mathbb R,
 \qquad x=\frac{\lambda|\omega|}{2}.
\end{equation}
The fixed parameters $\alpha_{13}$ and $\alpha_{24}$ displace the product phases by $O(\lambda)$. Since $g$ is Lipschitz continuous, these displacements change the exponential bound only by a constant factor. The inverse sine factors are bounded by $C e^{-2 \pi|y| / \lambda}$ for some constant $C$, with the finite shifts of their imaginary parts absorbed into $C$.   Thus, for fixed constants $C, M_0>0$ independent of $\omega$ and sufficiently small $\lambda$,
\begin{equation}\label{eq:f-behavior}
    \left|f_q(u)\right| \leq C \lambda^{-M_0} e^{E(y, x) / \lambda},
\end{equation}
where 
\begin{equation}
    \begin{aligned} E(y, x)= & -2 \pi|y|-g(y-x)-g(y+x) \\ & -g(y-2 d)-g(y+2 d)+4 g(y), \quad d=\min (\theta, \pi-\theta) .\end{aligned}
\end{equation}
On the physical domain, $0\leq x\leq2d\leq\pi$. For $|y|\leq\pi$, the piecewise-quadratic expression for $g$ gives
\begin{equation}
    \begin{aligned} E(y, x) &=  -2 d(\pi-d)-\pi x+\frac{x^2}{2} -\pi[\max \{|y|-x, 0\}+\max \{|y|-2 d, 0\}] \\ 
    & \leq -2 \theta(\pi-\theta)-\frac{\pi \lambda|\omega|}{2}+\frac{\lambda^2 \omega^2}{8}.
\end{aligned}
\end{equation}
To bound the full contour integral, write $y=y_0+2\pi k$, with $|y_0|\leq\pi$ and $k\in\mathbb Z$. Periodicity gives
\begin{equation}
    E\left(y_0+2 \pi k, x\right)=E\left(y_0, x\right)-2 \pi\left(\left|y_0+2 \pi k\right|-\left|y_0\right|\right).
\end{equation}
The central and two neighboring periods are bounded by the same maximum. The more distant periods are suppressed by at least $e^{-4 \pi^2(|k|-1) / \lambda}$, so their sum is uniformly bounded.  Together with $d u=i d y / \lambda$ and $\left|q^{u-\Delta_2}\right|=q^{c-\Delta_2}$, this gives fixed constants $C_f, M_1>0$ such that
\begin{equation} \label{eq:fint-bound}
 \left|\int_{\mathcal{C}} \frac{d u}{2 \pi i} q^{u-\Delta_2} f_q(u)\right| \leq C_f \lambda^{-M_1} e^{-2 \theta(\pi-\theta) / \lambda} e^{-\pi|\omega| / 2+\lambda \omega^2 / 8} .
\end{equation}
The power $M_1$ includes the logarithmic errors in \eqref{eq:f-behavior} and the factor $1 / \lambda$ from the contour measure.  We next combine this estimate with the exact normalization in \eqref{eq:NR-def}. With $\mathcal{N}=1 /(2 \pi \lambda \rho(\theta))$, the density  \eqref{eq:E-def} gives
\begin{equation}
    \frac{N_R}{\mathcal{N}}=\frac{\lambda}{1-q} \frac{Q_\lambda}{\Gamma_q(2 i \theta / \lambda) \Gamma_q(-2 i \theta / \lambda)}.
\end{equation}
Here $Q_\lambda$ denotes the quotient of gamma functions in \eqref{eq:NR-def}.
Consequently, there are $\lambda$-independent constants $C_N,M_2>0$ such that
\begin{equation}\label{eq:Nratio-bound}
 \left|\frac{N_R}{\mathcal N}\right|
 \leq C_N\lambda^{-M_2}e^{2\theta(\pi-\theta)/\lambda}.
\end{equation}
The $\theta$-dependent exponential cancels the corresponding suppression in \eqref{eq:fint-bound}. For $\omega\in D_\lambda$, we have $|\omega|\leq4d/\lambda=2\pi(1-v)/\lambda$, so
\begin{equation}
 -\frac{\pi|\omega|}{2}+\frac{\lambda\omega^2}{8}
 \leq-\frac{\pi|\omega|}{2}+\frac{d|\omega|}{2}
 =-\frac{\pi(1+v)}4|\omega|.
\end{equation}
Combining this estimate with \eqref{eq:fint-bound} and
\eqref{eq:Nratio-bound}, and setting $C_K=C_fC_N$ and $M=M_1+M_2$, proves \eqref{eq:bound-K-tail}.

Let $\eta=|\operatorname{Re} \tau|$ and recall $\delta_\eta=\pi / 2-\eta$. For the tail estimate assume $\delta_\eta>d / 2$, equivalently $\eta<\pi(1+v) / 4. $ Integrating \eqref{eq:bound-K-tail} gives
\begin{equation}
    \begin{aligned} \frac{1}{\mathcal{N}} \int_{D_\lambda \backslash I_\lambda} d \omega\left|e^{\tau \omega} K(\omega, \lambda)\right| & \leq C \lambda^{-M} e^{-\left(\delta_\eta-d / 2\right) B} \\ & =C \lambda^{b\left(\delta_\eta-d / 2\right)-M}, \quad B=b \log (1 / \lambda)\end{aligned}
\end{equation}
Choose the fixed constant $b$ so that $b\left(\delta_\eta-d / 2\right)>M$. This proves \eqref{eq:bound-tail}.

\section{Convergence of the partition function and entropy}
\label{app:disorder-averaged-partition-function}

In this appendix, we establish an analytic result used implicitly in the main text. The chord partition function is not merely a formal generating function for the moments in \eqref{eq:conv-moments}; it arises as the double-scaled limit of the disorder-averaged finite-$N$ partition function defined with respect to the normalized trace. In addition to the convergence of each fixed moment, the essential ingredient is a uniform Wick-contraction bound that allows us to interchange the double-scaled limit with the infinite series expansion in $\beta/\sqrt\lambda$.\footnote{See also \cite{Feng:2018spectrum} for rigorous convergence results for the SYK spectral distribution, including the double-scaled regime.}

Let $H_N$ be the finite-$N$ Hamiltonian defined in \eqref{eq:HN-def}, and define the corresponding thermal partition function by
\begin{equation}
Z_N(\beta)
:=
\tr e^{-\beta H_N}.
\end{equation}
We will show that
\begin{equation} \label{eq:claim-ZN}
\lim_{\rm DS}
\overline{Z_N(\beta)}
=
\langle0| e^{-\beta H}|0\rangle,
\end{equation}
where $H$ is the double-scaled SYK Hamiltonian used in the main text and $|0\rangle$ is the zero-chord state. Our strategy is to show that the expansion of the left-hand side of \eqref{eq:claim-ZN} in powers of $\beta/\sqrt{\lambda}$ converges uniformly in $N$ at fixed $\beta$ and $\lambda$. This allows us to exchange the double-scaled limit with the infinite sum, after which the convergence of the individual moments yields the right-hand side of \eqref{eq:claim-ZN}.

\subsection{Uniform Wick bound}

We begin by deriving the following uniform bound on the moments $m_{2k}^{(N)}$ for any values of $N$ and $p$:
\begin{equation} \label{eq:uniform-bound-2k}
    |m_{2k}^{(N)}| \leq \lambda^{-k}  (2k-1)!!.
\end{equation}

Since the couplings $J_I$ are Gaussian random variables, Wick's theorem gives
\begin{equation}
\overline{J_{I_1}\cdots J_{I_{2k}}}
=
\sum_{\pi\in \operatorname{Pair}(2k)}
\prod_{(a,b)\in\pi}
\overline{J_{I_a}J_{I_b}}.
\end{equation}
It follows that
\begin{align}
m_{2k}^{(N)}
&=
i^{kp}\sum_{I_1,\dots,I_{2k}}
\overline{J_{I_1}\cdots J_{I_{2k}}}\,
\tr\left(\Psi_{I_1}\cdots \Psi_{I_{2k}}\right)
\nonumber \\
&=
i^{kp}\sum_{\pi\in \operatorname{Pair}(2k)}
\sum_{I_1,\dots,I_{2k}}
\left(\prod_{(a,b)\in\pi}
\overline{J_{I_a}J_{I_b}}\right)
\tr\left(\Psi_{I_1}\cdots \Psi_{I_{2k}}\right).
\end{align}
The Majorana strings $\Psi_I$ have unit operator norm,
\begin{equation}
\left|
\Psi_I
\right|=1.
\end{equation}
The normalized trace therefore satisfies
\begin{equation}
\left|
\tr\left(\Psi_{I_1}\cdots \Psi_{I_{2k}}\right)
\right|
\leq 1.
\end{equation}
This estimate is uniform in the microscopic parameters: its right-hand side is independent of both $N$ and $p$. Consequently, the absolute value of each moment is bounded entirely by the corresponding Wick contractions, without any additional enhancement from the fermion strings. Taking absolute values, we obtain
\begin{align}
\left|m_{2k}^{(N)}\right|
&\leq
\sum_{\pi\in \operatorname{Pair}(2k)}
\sum_{I_1,\dots,I_{2k}}
\prod_{(a,b)\in\pi}
\left|
\overline{J_{I_a}J_{I_b}}
\right| .
\end{align}
Because the couplings $J_I$ are independent,
\begin{equation}
\overline{J_IJ_J}
=
\delta_{IJ} \overline{J_I^2}.
\end{equation}
For each Wick pairing $\pi$, the sums over the indices therefore collapse pairwise, giving
\begin{equation}
\sum_{I_1,\dots,I_{2k}}
\prod_{(a,b)\in\pi}
\left|
\overline{J_{I_a}J_{I_b}}
\right|
=
\left(
\sum_I \overline{J_I^2}
\right)^k .
\end{equation}
Moreover,
\begin{equation}
\sum_I \overline{J_I^2}
=
\overline{\tr H_N^2}
=
m_2^{(N)}.
\end{equation}
We therefore find
\begin{equation} \label{eq:bound-2k}
\left|m_{2k}^{(N)}\right|
\leq  (m_2^{(N)})^k |\mathrm{Pair}(2k)| = \lambda^{-k} (2k-1)!!.
\end{equation}
This establishes \eqref{eq:uniform-bound-2k}.
Note that the bound is uniform in both $N$ and $p$, and is therefore independent of the precise manner in which the large-$N$ limit is taken. Since the double-scaling parameter $\lambda=2p^2/N$ is held fixed, the same bound remains valid after taking the double-scaled limit at any fixed value of $\lambda$.

\subsection{Dominated convergence for the thermal series}

We now show that the averaged partition function $\overline{Z_N(\beta)}$ converges to the partition function of double-scaled SYK defined via $\langle 0 | e^{-\beta H } | 0\rangle $.  We next expand the disorder-averaged partition function as
\begin{equation}
\overline{Z_N(\beta)}
=
\overline{\tr e^{-\beta H_N}}
=
\sum_{L=0}^{\infty}
\frac{(-\beta)^L}{L!}m_L^{(N)} .
\end{equation}
Since all odd moments vanish, this reduces to
\begin{equation}
\overline{Z_N(\beta)}
=
\sum_{k=0}^{\infty}
\frac{\beta^{2k}}{(2k)!}m_{2k}^{(N)} .
\end{equation}
Using the uniform Wick bound \eqref{eq:bound-2k}, we obtain
\begin{align}
\left|
\frac{\beta^{2k}}{(2k)!}m_{2k}^{(N)}
\right|
&\leq
\frac{|\beta|^{2k}}{(2k)!}
(2k-1)!!\,\lambda^{-k}
=
\frac{1}{k!}
\left(
\frac{\beta^2}{2\lambda}
\right)^k.
\end{align}
The corresponding dominating series is
\begin{equation}
\sum_{k=0}^{\infty}
\frac{1}{k!}
\left(
\frac{\beta^2}{2\lambda}
\right)^k
=
\exp\left(\frac{\beta^2}{2\lambda}\right),
\end{equation}
which is finite for every fixed real value of $\beta$.

The dominated convergence theorem states that if each term in a sequence converges pointwise and the sequence is bounded by a single summable sequence independent of the limiting parameter, then the limit may be interchanged with the infinite sum. In the present setting, the limiting parameter is $N$, with $p$ determined as a function of $N$ so that $\lambda$ remains fixed. Pointwise convergence follows from the convergence of each fixed moment $m_{2k}^{(N)}$, while the summable dominating sequence is
\begin{equation}
\frac{1}{k!}
\left(
\frac{\beta^2}{2\lambda}
\right)^k.
\end{equation}
We may therefore interchange the double-scaled limit with the thermal series:
\begin{equation}
\lim_{\rm DS}
\overline{Z_N(\beta)}
=
\sum_{k=0}^{\infty}
\frac{\beta^{2k}}{(2k)!}
\lim_{\rm DS}
m_{2k}^{(N)}
=
\sum_{k=0}^{\infty}
\frac{\beta^{2k}}{(2k)!}
m_{2k}^{\rm chord}
=
\sum_{k=0}^{\infty}
\frac{\beta^{2k}}{(2k)!}
\langle 0 | H^{2k} |0\rangle .
\end{equation}
Restoring the odd moments, which vanish on both sides, confirms that
\begin{equation}
\lim_{\rm DS} \overline{Z_N(\beta)}
=
\langle0|e^{-\beta H}|0\rangle.
\end{equation}

\subsection{Convergence of the entropy}
\label{subsec:entropy-convergence}

The same argument establishes the convergence of the semi-quenched Rényi entropy defined in \eqref{eq:semi-quenched-renyi-def}. Let $\rho_{R,N}$ denote the finite-$N$ reduced density matrix constructed using the normalized trace and the finite-$N$ matter operators $O_{N,i}$. For each fixed positive integer $n$, the two quantities entering the semi-quenched entropy are
\begin{equation}
A_N^{(n)}
:=\overline{\tr_R\rho_{R,N}^n},
\qquad
B_N^{(n)}
:=\overline{(\tr_R\rho_{R,N})^n}.
\end{equation}
Both quantities are finite linear combinations of disorder-averaged, normalized Euclidean correlation functions involving $H_N$ and the matter operators $O_{N,i}$. More explicitly, the numerator is obtained by cyclically gluing the right endpoints of the $n$ replicas and therefore takes the schematic form
\begin{equation}
A_N^{(n)}
=
\sum_{i_1,\ldots,i_n,j_1,\ldots,j_n}
\widetilde\rho_{i_1j_1}\cdots\widetilde\rho_{i_nj_n}
\overline{\tr \left(
e^{-\beta_L H_N}O_{N,i_1}
e^{-\beta_R H_N}O_{N,j_2}
\cdots
e^{-\beta_L H_N}O_{N,i_n}
e^{-\beta_R H_N}O_{N,j_1}
\right)} .
\label{eq:semi-quenched-numerator-correlator}
\end{equation}
The denominator is the disorder average of a product of the corresponding one-replica normalized correlators:
\begin{equation}
B_N^{(n)}
=
\sum_{i_1,\ldots,i_n,j_1,\ldots,j_n}
\widetilde\rho_{i_1j_1}\cdots\widetilde\rho_{i_nj_n}
\overline{\prod_{a=1}^{n}
\tr \left(
e^{-\beta_L H_N}O_{N,i_a}
e^{-\beta_R H_N}O_{N,j_a}
\right)} .
\label{eq:semi-quenched-denominator-correlator}
\end{equation}
The above two equations contain the essential information required to establish convergence. Expanding each Euclidean evolution operator in powers of $H_N$ expresses each coefficient as a mixed moment involving finitely many Hamiltonian strings and finitely many matter strings. The uniform convergence of the relevant mixed chord moments, which can be proved by an argument similar to that in the last subsection,  guarantees the pointwise double-scaled convergence of each such coefficient.

Furthermore, the uniform Wick bound derived above is modified only by a multiplicative constant depending on the fixed number of matter insertions and the finite dimension of the flavor space. Each Majorana monomial has unit operator norm, and Wick-contracting the fixed number of matter insertions contributes only an $N$-independent factor. The factorial bound on the Hamiltonian insertions is therefore unchanged. Dominated convergence therefore again allows the double-scaled limit to pass through the thermal expansions.

It follows that the numerator converges to the DSSYK thermal $2n$-point correlator with the cyclic replica ordering:
\begin{equation}
\lim_{\rm DS}A_N^{(n)}
=
Z_{2n}^{\rm DSSYK},
\end{equation}
where $Z_{2n}^{\rm DSSYK}$ is the replicated DSSYK amplitude given in \eqref{eq:Z2n-result-exact}. Similarly, the denominator converges to the DSSYK correlator associated with the replicated normalization. Contractions connecting different normalized traces obey additional Majorana-index constraints and are therefore suppressed in the double-scaled limit. Consequently, only the disconnected disk contributions survive:
\begin{equation}
\lim_{\rm DS}B_N^{(n)}
=
\left(\widetilde\Tr\widetilde\rho\right)^n
\left[Z_2^{\rm DSSYK}(\beta_L,\beta_R;\Delta)\right]^n .
\end{equation}
For positive Euclidean evolution times, this denominator is nonzero. Hence the ratio appearing in \eqref{eq:semi-quenched-renyi-def} has a well-defined double-scaled limit:
\begin{equation}
\lim_{\rm DS}S^{(n)}(\rho_{R,N})
=
\frac{1}{1-n}
\log
\frac{Z_{2n}^{\rm DSSYK}}
{\left(\widetilde\Tr\widetilde\rho\right)^n
\left[Z_2^{\rm DSSYK}(\beta_L,\beta_R;\Delta)\right]^n} .
\end{equation}
This is the statement used in the main text in \eqref{eq:entropy-DS}. In other words, the normalized finite-$N$ traces entering the semi-quenched entropy converge to correlation functions of the double-scaled algebra evaluated in the chord-vacuum tracial state $\langle0|\cdot|0\rangle$. The only contribution not captured by this convergent normalized trace is the state-independent microscopic shift obtained by restoring the full Hilbert-space trace.

Finally, the preceding result concerns quantities defined using the normalized trace. If instead one uses the full Hilbert-space trace, the finite-$N$ partition function
\begin{equation}
Z_N^{\rm full}(\beta)
:=
\Tr_{\mathcal H_N}e^{-\beta H_N}
\end{equation}
is related to the normalized partition function by
\begin{equation}
\overline{Z_N^{\rm full}(\beta)}
=
2^{N/2}\overline{Z_N(\beta)}
=
2^{N/2}
\left[\langle0|e^{-\beta H}|0\rangle+o(1)\right] .
\end{equation}
Consequently,
\begin{equation}
\log \overline{Z_N^{\rm full}(\beta)}
=
\frac{N}{2}\log 2
+
\log \langle0|e^{-\beta H}|0\rangle
+
o(1).
\end{equation}
Thus, the double-scaled Hamiltonian captures the finite, normalized contribution to the disorder-averaged partition function, while the extensive microscopic entropy
\begin{equation}
\frac{N}{2}\log 2
\end{equation}
arises from restoring the full finite-$N$ Hilbert-space trace.

\section{Limitations of the fake-disk interpretation for higher-point functions} \label{app:Limitations}

In this appendix, we revisit the fake-disk interpretation of semiclassical two-point functions \cite{Lin_2023} and show that it does not, in general, extend to a single assignment of fake coordinates compatible with all Wick contractions in a higher-point function.

We write the two-point function as
\begin{equation}
    \la\mo(\tau_{1})\mo(\tau_{2})\ra \equiv G_2^{\Delta}(\tau_{12},\beta-\tau_{12})=\left(\frac{\cos\frac{\pi v}{2}}{\cos\frac{\pi v}{\beta}(\tau_{12}-\frac{\beta}{2})}\right)^{2\Del}, \quad \tau_1 >\tau_2. 
\end{equation}
Here $\tau_{12}\equiv \tau_1 - \tau_2 \in [0,\beta]$, and $\mathcal{O}(\tau) = e^{\tau H} \mathcal{O}e^{-\tau H}$. The correlator is evaluated in a thermal state and given by the normalized trace:
\begin{equation}
    \langle \mo(\tau_1) \mo(\tau_2) \rangle = \frac{1}{Z(\beta)}\tr (e^{-\beta H} \mo(\tau_1) \mo(\tau_2)).
\end{equation}
This parametrization of the two-point function makes the KMS condition transparent:
\begin{equation}
    \la\mo(\tau_{2}+\beta)\mo(\tau_{1})\ra=\left(\frac{\cos\frac{\pi v}{2}}{\cos\frac{\pi v}{\beta}(\tau_{21}+\frac{\beta}{2})}\right)^{2\Del}=\la\mo(\tau_{1})\mo(\tau_{2})\ra.
\end{equation}
To make the geometric meaning of these coordinates more transparent, we rescale $\tau_{12}$ by defining
\begin{equation}
    \theta_{12}=\frac{2\pi}{\beta}\tau_{12}.
\end{equation}
We denote the corresponding two-point function by
\begin{equation}
\la\mo(\te_{1})\mo(\te_{2})\ra=\left(\frac{\cos\frac{\pi v}{2}}{\sin\frac{v\te_{12}+(1-v)\pi}{2}}\right)^{2\Del}.
    \end{equation}
Thus, in order to express this correlator as an ordinary conformal two-point function on the fake circle, the fake angular separation must satisfy
\begin{equation}
    \phi_{12}
    =
    v\theta_{12}+\pi(1-v),
\end{equation}
up to the usual periodicity and reflection ambiguities of the sine function. The additive term $\pi(1-v)$ represents the fake region separating the two insertions. Importantly, this shift is naturally associated with a pair of insertions rather than with a globally defined map from each physical position $\theta_i$ to a fake position $\phi_i$.

To see the resulting obstruction explicitly, consider the four-point function 
\begin{equation}
    G_4 = \langle \mo(\theta_1) \mo(\theta_2) \mo(\theta_3) \mo(\theta_4) \rangle , \quad 2\pi \geq \te_1 \geq \te_2 \geq \te_3 \geq \te_4 \geq  0.
\end{equation}
Now let us consider four insertions and the assignment
\begin{equation}
    \begin{aligned}
        \phi_1-\frac{3\pi}{4}
        &=
        v\left(\theta_1-\frac{3\pi}{4}\right),
        &
        \phi_2-\frac{\pi}{4}
        &=
        v\left(\theta_2-\frac{\pi}{4}\right),
        \\
        \phi_3+\frac{\pi}{4}
        &=
        v\left(\theta_3+\frac{\pi}{4}\right),
        &
        \phi_4+\frac{3\pi}{4}
        &=
        v\left(\theta_4+\frac{3\pi}{4}\right).
    \end{aligned}
\end{equation}
This assignment gives
\begin{equation}
    \phi_{13}
    =
    v\theta_{13}+\pi(1-v),
    \qquad
    \phi_{24}
    =
    v\theta_{24}+\pi(1-v).
\end{equation}
Consequently, the following crossed contraction can be written as a product of conformal two-point functions on the fake circle:
\begin{equation}
  \la\mo(\te_{1})\mo(\te_{3})\ra\la\mo(\te_{2})\mo(\te_{4})\ra
    =
    \left(
    \frac{\cos\frac{\pi v}{2}}
    {\sin\frac{\phi_{13}}{2}}
    \right)^{2\Delta}
    \left(
    \frac{\cos\frac{\pi v}{2}}
    {\sin\frac{\phi_{24}}{2}}
    \right)^{2\Delta}.
\end{equation}
The same fake-coordinate assignment, however, does not correctly represent the other contraction channels. For example,
\begin{equation}
    \phi_{12}
    =
    v\theta_{12}+\frac{\pi}{2}(1-v),
\end{equation}
and therefore
\begin{equation}
    \frac{\pi v}{\beta}
    \left(
    \frac{\beta}{2}-\tau_{12}
    \right)
    =
    \frac{\pi}{4}(1+v)-\frac{\phi_{12}}{2}.
\end{equation}
It follows that
\begin{equation}
    \langle \mo (\theta_1) \mo(\theta_2) \rangle
    =
    \left[
    \frac{\cos\frac{\pi v}{2}}
    {\cos\left(
    \frac{\pi}{4}(1+v)-\frac{\phi_{12}}{2}
    \right)}
    \right]^{2\Delta},
\end{equation}
which is not an ordinary conformal two-point function proportional to $\sin^{-2\Delta}(\phi_{12}/2)$.

More generally, the incompatibility follows immediately from the additivity of coordinate differences. If one demands that the pairs $(1,2)$ and $(2,3)$ both be separated by a complete fake region, then
\begin{equation}
    \phi_{12}
    =
    v\theta_{12}+\pi(1-v),
    \qquad
    \phi_{23}
    =
    v\theta_{23}+\pi(1-v).
\end{equation}
These relations imply
\begin{equation}
    \phi_{13}
    =
    \phi_{12}+\phi_{23}
    =
    v\theta_{13}+2\pi(1-v).
\end{equation}
On the other hand, representing $\langle \mo(\theta_1)\mo(\theta_3)\rangle $ on the fake circle would require
\begin{equation}
    \phi_{13}
    =
    v\theta_{13}+\pi(1-v).
\end{equation}
The two conditions are incompatible for generic $0<v<1$. The obstruction disappears at $v=1$, precisely when the fake region itself vanishes.

This becomes relevant because the leading semiclassical four-point function obeys factorization. For identical probe operators, one has schematically
\begin{equation}
    G_4 = \la\mo_{1}\mo_{2}\ra\la\mo_{3}\mo_{4}\ra+\la\mo_{1}\mo_{3}\ra\la\mo_{2}\mo_{4}\ra+\la\mo_{1}\mo_{4}\ra\la\mo_{2}\mo_{3}\ra.
\end{equation}
A fake-disk configuration may be chosen to represent any one of these Wick-contraction channels. Different channels, however, require different and mutually incompatible assignments of the fake regions. Therefore, the full factorized four-point function cannot be represented by placing all four insertions on a single fake disk.

The same obstruction persists for general $2n$-point functions. At leading semiclassical order,
\begin{equation}
    \left\langle
    \mathcal{O}_1\cdots\mathcal{O}_{2n}
    \right\rangle
    =
    \sum_{\mathcal{P}}
    \prod_{(ij)\in\mathcal{P}}
    G_2(\tau_{ij})
    +
    O(\lambda),
\end{equation}
where $\mathcal{P}$ runs over the allowed Wick pairings. Each two-point factor admits the fake-disk representation described above, but there is generally no single assignment of fake coordinates that represents all pairings simultaneously.

\bibliographystyle{JHEP.bst}
\bibliography{ref.bib}

\end{document}